\documentclass[sigconf]{acmart}

\AtBeginDocument{%
  }

\setcopyright{acmlicensed}
\copyrightyear{2026}
\acmYear{2026}
\setcopyright{cc}
\setcctype{by}
\acmConference[IUI '26]{31st International Conference on Intelligent User Interfaces}{March 23--26, 2026}{Paphos, Cyprus}
\acmBooktitle{31st International Conference on Intelligent User Interfaces (IUI '26), March 23--26, 2026, Paphos, Cyprus}
\acmPrice{}
\acmDOI{10.1145/3742413.3789124}
\acmISBN{979-8-4007-1984-4/2026/03}

\acmSubmissionID{9004}

\usepackage{bm}
\usepackage{amsmath}
\usepackage{enumitem}
\usepackage{multirow}
\usepackage{colortbl} 
\usepackage{makecell}
\usepackage{cleveref}
\usepackage{balance}

\usepackage{soul}
\soulregister\cite7 % allow highlight with \cite inside
\soulregister\ref7
\soulregister\textit7 
\soulregister\rev7 
\newcommand{\ns}[1]{{\color{purple}#1}}

\newcommand{\hlc}[2][yellow]{{\sethlcolor{#1}\hl{#2}}}
\definecolor{revisionColor}{HTML}{FFFFFF}
\newcommand{\rev}[1]{\hlc[revisionColor]{#1}} % for revision highlight

\begin{document}

\title[Transferable XAI: Relating Understanding Across Domains with Explanation Transfer]{Transferable XAI: Relating Understanding Across Domains with~Explanation Transfer}

\author{Fei Wang}
\affiliation{%
  \department{Department of Computer Science}
  \institution{National University of Singapore}
  \city{Singapore}
  \country{Singapore}
}
\email{wang-fei@nus.edu.sg}

\author{Yifan Zhang}
\affiliation{%
 \department{Department of Computer Science}
 \institution{National University of Singapore}
 \city{Singapore}
 \country{Singapore}
}
\email{yifan.zhang\_@u.nus.edu}

\author{Brian Y Lim}
\authornote{Corresponding author}
\affiliation{%
  \department{Department of Computer Science}
  \institution{National University of Singapore}
  \city{Singapore}
  \country{Singapore}
}
\email{brianlim@nus.edu.sg}

\renewcommand{\shortauthors}{Wang, Zhang, and Lim}

\begin{abstract}
  Current Explainable AI (XAI) focuses on explaining a single application, but when encountering related applications, users may rely on their prior understanding from previous explanations. This leads to either overgeneralization and AI overreliance, or burdensome independent memorization. Indeed, related decision tasks can share explanatory factors, but with some notable differences; e.g., body mass index (BMI) affects the risks for heart disease and diabetes at the same rate, but chest pain is more indicative of heart disease. Similarly, models using different attributes for the same task still share signals; e.g., temperature and pressure affect air pollution but in opposite directions due to the ideal gas law. Leveraging transfer of learning, we propose Transferable XAI to enable users to transfer understanding across related domains by explaining the relationship between domain explanations using a general affine transformation framework applied to linear factor explanations. The framework supports explanation transfer across various domain types: \textit{translation} for data subspace (subsuming prior work on Incremental XAI), \textit{scaling} for decision task, and \textit{mapping} for attributes. Focusing on task and attributes domain types, in formative and summative user studies, we investigated how well participants could understand AI decisions from one domain to another. Compared to single-domain and domain-independent explanations, Transferable XAI was the most helpful for understanding the second domain, leading to the best decision faithfulness, factor recall, and ability to relate explanations between domains. This framework contributes to improving the reusability of explanations across related AI applications by explaining factor relationships between subspaces, tasks, and attributes.
\end{abstract}





\maketitle

\section{Introduction}

\begin{figure}
    \centering
    \includegraphics[width=\linewidth]{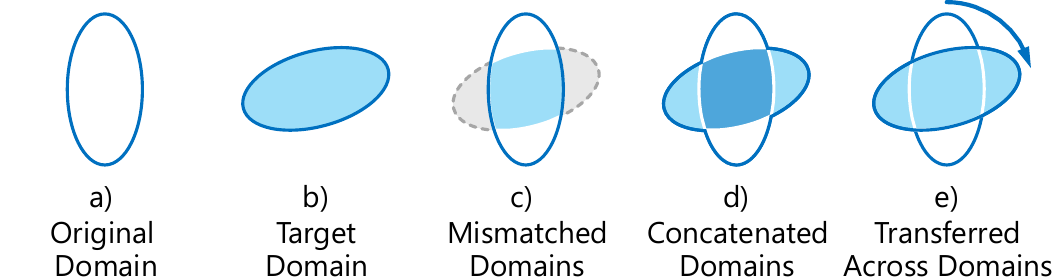}
    \caption{
    Concept of Transferable XAI for cross-domain understanding.
    a) Original Domain: Initial understanding of AI decisions.    
    b) Target Domain: Understanding goal for the second domain.    
    c) Mismatched Domains: Applying original understanding to the target domain results in knowledge gaps (gray) and potential errors.    
    d) Concatenated Domains: Learning both domains independently is cognitively demanding and covers shared knowledge redundantly (dark blue).    
    e) Transferred Across Domains: Leveraging original understanding, the user can transfer understanding across domains for efficient understanding of the target domain.
    }
    \Description{The figure contains five schematic panels labeled a) through e), each showing simplified shapes representing domains and their relationships. 
    Panel a), labeled ``Original Domain'', shows a single vertical oval shape. 
    Panel b), labeled ``Applied Original Domain'', shows the same vertical oval overlapping with a tilted dashed oval, indicating partial overlap between the two shapes. 
    Panel c), labeled ``Target Domain'', shows a single tilted horizontal oval shape. 
    Panel d), labeled ``Original + Target Domains'', shows a vertical oval and a horizontal oval overlapping, with a combined region in the center. 
    Panel e), labeled ``Transferable XAI'', shows a horizontal oval overlapping with a vertical oval, accompanied by a curved arrow above the shapes indicating transformation or adaptation.}
    \label{fig:intro_conceptual_graph}
\end{figure}

Artificial Intelligence (AI) systems have become prevalent in various domains to support human decision-making. Driven by the needs such as transparency and fairness, explainable AI (XAI) methods are developed to explain the reasoning behind AI decisions~\cite{dwivedi2023explainable,karimi2022survey}.
Despite some success of existing XAI methods, most of the studies have been focusing on explaining AI applications in isolation.
However, experiencing a variety of AI usage is not uncommon for people today.
For example, large enterprises sometimes segment markets based on regions or user types~\cite{wang2022efficient,rezaeinia2016recommender}, and operators may care about how the decisions of related AI systems differ across these submarkets. Similar demands also exist in scenarios, such as AI making predictions for different but relevant decision tasks, and providing explanations with different sets of attributes.
This requires users to \textbf{understand AI decisions across diverse \textit{domains}}, i.e., applications contexts with varying datasets, decision tasks, and input attributes.
Hence, toward AI literacy~\cite{long2020ai}, it is important for users to be able to apply knowledge across domains (see conceptual goal in Fig.~\ref{fig:intro_conceptual_graph}).
Without proper guidance, users may rely on their prior understanding from previous explanations, which leads to either overgeneralization and AI overreliance~\cite{bauer2023expl} (Fig.~\ref{fig:intro_conceptual_graph}c), or suffer from burdensome independent memorization of each domain (Fig.~\ref{fig:intro_conceptual_graph}d). 
Therefore, our research goal is to help domain users, with reasonable terminology knowledge and numeracy but limited knowledge in AI,
i) to leverage their understanding of an AI in one domain,
ii) to understand another related AI in a target domain.

Inspired by education literature on the transfer of learning~\cite{perkins1992transfer,bossard2008transfer} and cumulative human learning~\cite{shuell1986cognitive}, we hypothesize that helping users to \textit{transfer} between different domains of related AI applications can improve their understanding when using multiple applications (Fig.~\ref{fig:intro_conceptual_graph}e). 
The explanation transfer can be done by first explaining one, and showing how changes to that explanation can fit for the new domain.
Bo et al. had proposed Incremental XAI~\cite{bo2024incremental} to show the incremental difference between explanation instances in general (e.g., pricing for all houses) and for special cases (for large houses).
This explains the difference between instances from different \textit{subspaces} or distributions.
In this work, we propose to extend this toward Transferable XAI to cover more types of explanation transfer.
Specifically, we examine how similar AI systems can have related but different prediction \textit{tasks}, and unshared input \textit{attributes} (features).
For example, although heart disease and diabetes are different diseases, they can share the same risk indicators (e.g., being overweight) and varied ones (e.g., chest pain for heart disease, blood glucose for diabetes).
Another example is explaining disease risk to children with the attributes of body weight and height,
but explaining to health-savvy users with the more sophisticated attribute of body mass index (BMI).
With linear explanations, users learn the importance weight of how each attribute influences the AI decision, 
but the weights differ across domains for different tasks or attributes.
Transferable XAI aims to teach the users how to \textit{relatably modify} the weights from one domain to another, 
instead of learning the weights independently for both domains. This makes the user's knowledge of the original weights transferable to the new domain via the modification.

To implement Transferable XAI, we introduce an \textit{interpretable} Affine Transformation framework for explanation transfer.
We focus on linear factor explanations where each attribute is multiplied with a factor (weight) to constitute partial sums (contributions) and summed for a decision, and define the factors as a vector.
The transformation consists of a matrix multiplication and an addition on the explanatory factor vectors to shift explanations from the original domain to a target domain for the three aforementioned domain types.
Specifically, 
i) addition represents \textit{translation} to transfer the explanation from one subspace to another,
ii) vector multiplication represents \textit{scaling} to transfer across decision tasks, and
iii) matrix multiplication represents linear \textit{mapping} to transfer across different sets of attributes.
To manage the cognitive load of understanding relational differences, we apply sparsity regularization on the affine transformation parameters to minimize the number of terms users need to recall.
In summary, our contributions are:
\begin{itemize}
    \item \textbf{Transferable XAI} formulation for explaining across domains 
    (subspace\footnote{Subspace transfer is conceptually unchanged from Bo et al. for Incremental XAI~\cite{bo2024incremental}, but Transferable XAI subsumes it under a general framework.}, task, attributes).
    \item \textbf{Affine transformation} general interpretable framework for explanation transfer via \textit{translation} for subspace, \textit{scaling} for task, and \textit{mapping} for attributes.
\end{itemize}

We evaluated Transferable XAI in two experiments\footnote{Subspace transfer had been investigated by Bo et al.~\cite{bo2024incremental}, so we omit this redundant insight.} on task and attributes transfer across applications (health risk, air pollution) and domains (heart disease and diabetes; small PM2.5 and coarse PM10 particles).
We examined whether users could apply correct domain knowledge when deciding on instances from two related domains.
We measured user understanding, user alignment to the relevant explanation domain, and perceived usefulness across domains. 
In general, though higher in cognitive load, we found that Transferable XAI best helped users understand AI decisions and correctly report the relationship of explanatory factors across domains.

\section{Related Work}
We summarize how AI explanations have focused on single domains, 
yet AI is applied across domains, and discuss the prospect of transfer learning for appropriate generalization of AI explanations.

\subsection{Surrogate Explanations of AI}
Explainable AI (XAI) is widely investigated to support the understanding of AI systems for the needs of transparency~\cite{gyevnar2023bridging}, trust~\cite{naiseh2023different,cau2025influence}, fairness~\cite{kusner2017counterfactual,schoeffer2024explanations}, etc.
Due to the trade-off between accuracy and explainability, many users prefer post-hoc explanations that preserve AI model accuracy and are model agnostic~\cite{madsen2022post}.
Techniques for post-hoc explanations vary, including surrogate explanations that show attribute importances~\cite{ribeiro2016should,bove2022contextualization}, attribution~\cite{lundberg2017unified}, or distilled rules~\cite{van2021evaluating} in AI predictions, saliency-based explanations that show attribute contributions~\cite{sundararajan2017axiomatic}, example-based explanations that provide reflective examples (e.g., counterfactual~\cite{wachter2017counterfactual} and contrastive examples~\cite{miller2019explanation}).
Additive models such as linear factor models are considered intuitive in understanding, and their linearity and additivity make it easy to spot spurious associations and to justify AI decisions~\cite{caruana2015intelligible}.

However, current XAI mainly focuses on explaining individual applications, without considering the challenges that arise when users learn from explanations across multiple AI applications. Studies by Bauer and colleagues~\cite{bauer2023expl} identified spillover effects to other decision domains in explanation-driven mental models. This implies an overgeneralization or overreliance on previous explanations. Therefore, investigating AI explanation in multiple domains is necessary.

\subsection{AI for Multiple Domains}

Instead of modeling one focused aspect, recent AI models exploit commonalities and relationships across domains to predict for multiple objectives.
We discuss this for the domains of study in our work: subspace, tasks, and attributes.

\subsubsection{Subspace} 
Due to the complexity of AI systems, different groups of instances---in different subspaces---can have very different relationships between their inputs and their predictions.
Training on a subset of cases but inferring on very different cases can lead to the out-of-distribute problem~\cite{moreno2012unifying,yang2024generalized}.
Subspace models have been proposed to train sub-models for the instances in each subspace~\cite{natesan2020model, bjorklund2019sparse, chowdhury2022equi}, but the relationship across subspaces remain unclear.
Transfer learning methods have been developed to learn the relationship between the different subspaces to mitigate this problem~\cite{wang2022efficient, rezaeinia2016recommender}.
Recently, Bo et al. proposed Incremental XAI to relate explanatory factors across subspaces with sparse differences~\cite{bo2024incremental}.
In this work, we propose a framework to subsume this and extend this work for other domains.

\subsubsection{Task} 
The richness of datasets enables training AI to predict different decisions or labels, e.g., diagnosing various diseases for the same patient.
Yet, these different decision tasks can be related if they use the same attributes similarly.
Multi-task learning~\cite{caruana1997multitask, ma2018modeling, zhang2021survey} and multi-label learning~\cite{liu2021emerging, barutcuoglu2006hierarchical} exploit this shared information by co-training multiple tasks with the same set of attributes.
While these models improve AI accuracy, there has been no investigation into how users can understand the relationship between multiple tasks.
Hence, we study explanatory task transfer in this work.

\subsubsection{Attributes} 
The attributes used for an AI task could differ due to data collection methods, user explanation preferences, etc.
For example, Ames Housing~\cite{de2011ames} and California Housing~\cite{pace1997sparse} are two datasets for house price prediction, of which Ames provides property-level structural attributes, whereas California contains aggregated regional socioeconomic attributes.
AI applications may expose different input attributes to different users---for example, Isabel Healthcare\footnote{\url{https://www.isabelhealthcare.com/}} provides detailed patient data for professionals but only basic demographics and symptoms for patients. Likewise, stakeholders vary in the attributes they emphasize in interpretation~\cite{Tomsett2018InterpretableToWhom,zhang2022effects}.
While models based on different attribute sets are typically trained separately, their decision processes can still be related. Explaining them independently may overlook cross-model attribute relationships, such as equivalent information represented by different attributes (e.g., temperature represented in Celsius or Fahrenheit) or correlated attributes that contribute differently across models.
Therefore, explaining the mapping between attributes helps with AI understanding, which is studied in this work.

While these works shows the need for AI systems to vary across Domain Type dimensions, they did not study the different domains together.
In this work, we investigate how users can transfer (or struggle to adapt) their understanding from one domain to another, and propose a method to aid them.

\subsection{Knowledge Transfers in Human Learning and AI Models}
Human learning is cumulative in nature~\cite{shuell1986cognitive}.
Transfer of learning occurs when people apply their prior knowledge or skills to the learning or performance in a new context~\cite{perkins1992transfer,haskell2000transfer}. Thorndike~\cite{thorndike1923influence} concluded that transfer depends on identical elements between two performances. Indeed, transfer of learning is important but nontrivial for humans, which requires abilities such as abstraction and active self-motivation~\cite{perkins1992transfer}.
Such difficulty in transfer may hinder users’ understanding of new AI systems: insufficient transfer fails to ease learning, whereas excessive transfer causes misunderstanding through overgeneralization of prior AI explanations.
Therefore, proper guidance of transferable explanations is necessary.
In this work, we propose to support this with Transferable XAI methods and user interfaces.

Knowledge transfer strategies have also been widely applied in AI model design and training. Representative methods include domain adaptation~\cite{fang2024source} and transfer learning~\cite{zhuang2020comprehensive}, which transfer model knowledge across data spaces or tasks; incremental learning~\cite{wang2024comprehensive} and life-long learning~\cite{zhou2024continual}, which update AI knowledge incrementally with new data; and federated learning~\cite{kairouz2021advances}, which aggregates AI knowledge across subsystems. 
These methods aim to improve the AI reasoning and performance, but we focus on improving the human understanding and performance to transfer across domains.
Style transfer shares methodological similarity with our approach in that it uses an affine transformation to linearly scale and shift the data representations of the content image to match the statistical characteristics of the style image~\cite{li2017universal,huang2017arbitrary}. By contrast, we apply affine transformation in a more general sense, focusing on the transfer of explanations rather than data representations.

\section{Technical Approach} \label{sec:technical-approach}

We establish 
the background on the three types of domain shifts we investigated in this work, and
key concepts for AI prediction and explanation
for a single domain.
Then, we introduce our key contribution, Transferable XAI, a general framework of affine transformation for explanation transfer.

In this work, we focus on regression tasks to predict numeric labels that are more detailed than classification of categorical labels.
Given an AI model $f$ that predicts label $\hat{y}$ for instance $\bm{x}$, we want to explain its prediction with a proxy explainable AI (XAI) model $g$ that estimates the label $\tilde{y}$.
For a single domain we could rely on just one explanation model $g_O$, but for different contexts (shifted to another domain), we should use specific explanation model $g_T$ for the target domain to avoid overgeneralization and XAI overreliance.
With Transferable XAI, we aim to teach users a mapping function $h$ to obtain target explanation $g_T$ by adapting (transferring) knowledge from the original explanation $g_O$.

\begin{figure*}[t]
    \centering
    \includegraphics[width=0.8\linewidth]{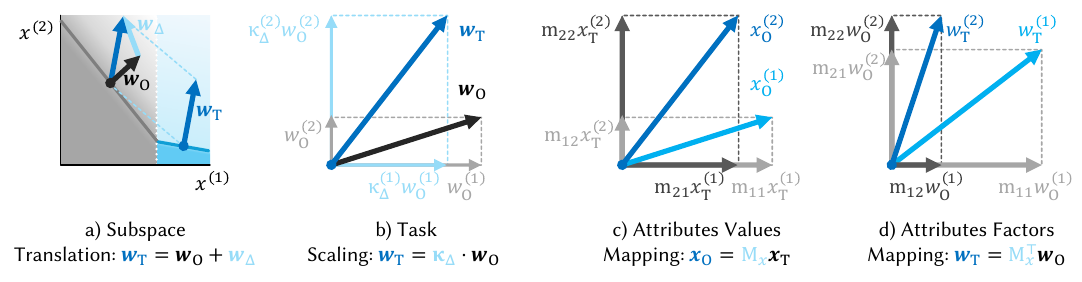}
    \caption{Conceptual example of affine transformations across different domain types, shown with 2-dimensional data (two attributes) for simplicity. See technical details in Section~\ref{sec:affine-transformation}.}
    \Description{This figure conceptually illustrates four types of affine transformations used to transfer explanatory relationships between domains, shown with two-dimensional data for simplicity. 
    The figure consists of four sub-figures (a–d), each of which shows a conceptual diagram of vectors---arrows that represent how factors (or explanatory variables) relate to outcomes. The arrows show transformations from the Original Domain (in black or gray) to the Target Domain (in blue).
    (a) Subspace Translation, where the Target vector is shifted from the Original one but keeps a similar direction. 
    This represents cases where the same explanatory factors apply but with an offset baseline.
    (b) Task Transfer via Scaling. Here, the blue Target arrow is longer or shorter than the black Original arrow, indicating that a factor may have a stronger, weaker, or even opposite influence in a related task.  
    Scaling helps capture these strength or direction differences, while sparsity keeps unchanged factors ignored to reduce cognitive load.
    (c) Attribute Value Mapping. Different sets of attributes may represent the same concept for different users. The mapping shows how one set of input variables can be linearly transformed into another to maintain interpretability across models.
    (d) shows Attribute Factor Mapping, which transfers the factors of attributes between domains. Each factor in the Target Domain is expressed as a combination of factors from the Original Domain. This allows explanations to remain meaningful when features differ across models. Sparse mappings preserve shared attributes while minimizing unnecessary changes.
    Together, these transformations demonstrate how explanatory factors can be shifted, scaled, or remapped across domains, maintaining interpretability and faithfulness when adapting models to new application contexts.}
    \label{fig:conceptual_fig}
\end{figure*}

\subsection{Multiple Domains}

To support appropriate generalization of XAI across applications, we identified various \textit{domains} for which contextual shift needs to be accounted, namely: subspace, task, and attributes.
We consider a simplified case of users learning about a domain first, then extending their understanding to another domain. 
% Heretofore, we call these the \textit{Original} and \textit{Target} domains, respectively. Explanations for the original domain are defined as $g_O$ and for the target domain as $g_T$.
Heretofore, we call these \textit{Original} and \textit{Target} domains, with explanations denoted as $g_O$ and $g_T$, respectively.

\subsubsection{Subspace}
As investigated by Bo et al. on Incremental XAI~\cite{bo2024incremental}, 
different groups of instances can have different explanations for their prediction outcome.
For example, small houses are priced at \$63k per ksqft of living area, but large houses (> 2.5ksqft) have a higher rate of \$240k/ksqft.
Thus, the feature space of instances should be partitioned into \textit{subspaces} due to their different explanatory contexts.
After understanding one subspace, users should understand other target subspaces relative to the original.

\subsubsection{Task}
Even for the same instances, different decisions can be made.
In machine learning, these are called prediction tasks.
For example, a patient's health indicators can be informative to predict the patient's risk for multiple diseases;
each disease risk is a separate decision task.
Yet, the health factors could explain various disease, with some notable factors emphasizing one disease over another.
Regarding what constitutes a discrete task, different prediction tasks cannot be synonymous (almost perfect correlation), otherwise, there is little need to transfer learning, and yet the tasks need to be sufficiently related to share some inductive bias where some features have similar effects on both tasks~\cite{caruana1997multitask}.
Thus, we define \textit{task} as another domain type.

\subsubsection{Attributes}
Even with the same task on the same instance, one could use different attributes to predict or explain the decision.
For example, a doctor could consider body mass index (BMI) for disease risk, while a lay patient may prefer to understand in terms of height and weight.
Yet, BMI is related to height and weight, so it is meaningful to understand how explanations for the same task relate across these attributes.
Thus, \textit{attributes}---also known as feature sets---are another domain for explanation transfer.

\subsection{Single-Domain Explanation}
\label{sec:app-single-domain}
For simplicity, we focus on linear factor explanations which are accessible to lay users. 
The explanation for a prediction on instance $\bm{x} = (x^{(1)}, \dots, x^{(n)})^\top$, with $n$ attributes, is:
\begin{equation}
    \label{xai_linear}
    \tilde{y} = \bm{w}^\top \bm{x} + b = \sum_{0<r<n} w^{(r)} x^{(r)} + w^{(0)},
\end{equation}
where $\bm{w} = (w^{(1)}, \dots, w^{(n)})^\top$ are the linear factors (weights) for each attribute (feature), and $w^{(0)}$ is the bias term.

\subsubsection{Preprocessing: Relative Attribute Values}
\label{sec:relative-attribute-values}
The bias term $w^{(0)}$ in linear factor explanations can be confusing to lay users to interpret. 
Users are either told that it just exists as an arbitrary adjustment or is the label value when $x$ is zero, but the latter explanation is jargon if $x = 0$ is meaningless.
Hence, to simplify interpretation, we eliminate the bias term by zeroing the attribute values to \textit{relative} attribute values.

Considering the mean of all instances $\bar{\bm{x}} = \frac{1}{n} \sum_i \bm{x}_i$, where $n$ is the number of instances, the XAI model label at this point is the centroid label:
\begin{equation}
    \label{xai_linear_mean}
    % \tilde{y}_{\bar{x}} = \bm{w}^\top \bar{\bm{x}} + w^{(0)}.
    \bar{\tilde{y}} = \bm{w}^\top \bar{\bm{x}} + w^{(0)}.
\end{equation}
Subtracting Eq. \ref{xai_linear_mean} from Eq. \ref{xai_linear} to eliminate $w^{(0)}$, and rearranging, we get the same factors in terms of relative values $\bm{\chi} = \bm{x} - \bar{\bm{x}}$:
\begin{equation}
    \label{xai_linear_relative}
    % \tilde{y} = \bm{w}^\top (\bm{x} - \bar{\bm{x}}) + \tilde{y}_{\bar{x}}
    \tilde{y} = g(\bm{\chi}) = \bm{w}^\top \bm{\chi} + \bar{\tilde{y}}.
\end{equation}
Thus, the bias term $w^{(0)}$ in Eq. \ref{xai_linear} is the difference between the centroid label and weighted mean instance, i.e., $w^{(0)} = \bar{\tilde{y}} - \bm{w}^\top\bar{\bm{x}}$.

\subsubsection{Original Domain}
\label{sec:app-single-org-domain}
Given an initial domain (Original $O$), the explanation for instance $\bm{x}$ is simply:
\begin{equation}
    \label{xai_linear_domain_original}
    % \tilde{y}_O = g_O(\bm{x}) = \bm{w}_O^\top \bm{x} + b_O.
    % \tilde{y}_O = g_O(\bm{\chi}) = \bm{w}_O^\top \bm{\chi} + \tilde{y}_{\bar{x}}.
    \tilde{y}_O = g_O(\bm{\chi}) = \bm{w}_O^\top \bm{\chi} + \bar{\tilde{y}}_O.
\end{equation}
$\bm{w}_O$ are the XAI model parameters, which are trained such that the XAI label $\tilde{y}_O$ is faithful to the AI prediction $\hat{y}_O$, by minimizing the MSE loss $L_O(\tilde{y}_O, \hat{y}_O)$.

\subsubsection{Target Domain}
\label{sec:app-single-target-domain}
For the second domain (Target $T$), we denote its explanation as:
\begin{equation}
    \label{xai_linear_domain_target}
    % \tilde{y}_T = g_T(\bm{x}) = \bm{w}_T^\top \bm{x} + b_T.
    % \tilde{y}_T = g_T(\bm{\chi}) = \bm{w}_T^\top \bm{\chi} + \tilde{y}_{\bar{x}}.
    \tilde{y}_T = g_T(\bm{\chi}) = \bm{w}_T^\top \bm{\chi} + \bar{\tilde{y}}_T.
\end{equation}
Parameters $\bm{w}_T$ can be determined \textit{independently} by minimizing the MSE loss $L_T(\tilde{y}_T, \hat{y}_T)$, 
or \textit{relatively} with transfer learning, which we discuss in the next section.

\subsection{Affine Transformation Framework for Explanation Transfer}
\label{sec:affine-transformation}

We extend the work of Bo et al. on Incremental XAI~\cite{bo2024incremental} beyond subspace domain transfer to the additional transfers of task and attributes domains.
Specifically, we seek an operation to map $h: \bm{w}_O \longmapsto \bm{w}_T$.
Retaining the simplicity of linear explanations, we found that affine transformations are a suitable framework to support multiple types of domain transfer.
An affine transformation consists of a linear mapping with matrix $A$ on the original vector---which can include operations like scaling, sheering and rotation---and translation defined by the direction of $\bm{b}$, i.e.,
\begin{equation}
    \label{eq:xai_affine_transformation}
    \bm{w}_T = h(\bm{w}_O) = A \bm{w}_O + \bm{b}
\end{equation}

Next, we discuss how choosing different values for $A$ and $\bm{b}$ allows us to transfer explanatory factors $\bm{w}_O$ to $\bm{w}_T$ for different domain types: subspace, task, and attributes.
We summarize and illustrate each type of explanation transfer in Fig.~\ref{fig:conceptual_fig}.
For all domain types, the explanations for the Original domain $O$ are the same as defined in Eq. \ref{xai_linear_domain_original}.
We only describe the explanations for the Target domain $T$.

For each domain type, we co-train XAI models for both domains together with sparsity constraints to maximize faithfulness and minimize cognitive load, i.e.,
\begin{equation}
    \mathcal{L} = L_O(\tilde{y}_O, \hat{y}_O) + L_T(\tilde{y}_T, \hat{y}_T) + \lambda L_s(A, \bm{b}),
    \label{eq:loss_functions}
\end{equation}
where the sparsity regularization function $L_s$ depends on the domain type and hyperparameter $\lambda$ controls the regularization strength.

\subsubsection{Subspace Transfer via Translation}
Although this transformation has been defined by Bo et al. on Incremental XAI~\cite{bo2024incremental} as an additive change in factors $\bm{w}_{\Delta}$,
here we subsume it as an affine transformation with only a translation, i.e., $A = I$ the identity matrix, and $\bm{b} = \bm{w}_{\Delta}$.
Thus, the transferred explanation for the Target domain $T$ is:
\begin{equation}
    \label{eq:xai_affine_transformation_subspace}
    \bm{w}_T = \bm{w}_O + \bm{w}_{\Delta}
    % \label{eq:transform_subspace}
\end{equation}
To limit cognitive load, a sparsity regularization with L1 norm is applied to make most factor changes to be negligible, i.e., $L_s(\bm{b}) = ||\bm{w}_{\Delta}||_1$.

\subsubsection{Task Transfer via Scaling}
\label{sec:app-task-transfer-via-scaling}
Here, we consider the relationship between two related decision tasks (such as predicting the risks of different diseases).
A factor could have stronger (or weaker) effect on one task, but weaker (or stronger) effect on the other.
Moreover, a factor could also have an opposite effect on one task than the another (e.g., body mass index increases the risk of diabetes, but decreases the risk of osteoporosis).

Scaling can account for these factorial and direction changes by multiplying a scale $\bm{\kappa}_\Delta$ on the explanatory factors $\bm{w}$.
Interpreting each scale: 
$\kappa_\Delta \approx 1$ indicates no difference between domains;
$|\kappa_\Delta| > 1$ indicates higher influence in the Target task than the Original task, and vice versa ($|\kappa_\Delta| < 1$);
$\kappa_\Delta < 0$ indicates influence in the opposite direction, i.e., increasing (or decreasing) effect in the Original task but decreasing (or increasing) effect in the Target task.

In terms of the affine transformation, $A = D_\kappa = \text{diag}(\bm{\kappa}_\Delta)$ a diagonal matrix which scales each factor $w^{(r)}$ by a corresponding amount $\kappa_\Delta^{(r)}$, and $\bm{b} = \bm{0}$, i.e.,
\begin{equation}
    \label{eq:xai_affine_transformation_task}
    \bm{w}_T = D_\kappa \bm{w}_O % = \bm{\kappa}_\Delta^\top \bm{w}_O
    % \label{eq:transform_task}
\end{equation}
To limit cognitive load, we apply L1 norm sparsity regularization on non-unity ($\neq 1$) scales, i.e., $L_s(A) = ||\bm{\kappa}_{\Delta} - \bm{1}||_1$.
Hence, most scales will be 1 and can be ignored, and the user can consider those factors as being the same across tasks.

\subsubsection{Attributes Transfer via Mapping}
\label{sec:app-attributes-transfer-via-mapping}
Even for the same prediction tasks, one may develop different XAI models to accommodate different target users (e.g., lay users vs. domain experts).
This often involves different attributes (feature sets) that are more precise or accessible to the different users.
For example, health experts are comfortable with body mass index (BMI = weight / height\textsuperscript{2}) as a health factor, while lay users may prefer to interpret in terms of height and weight.
While independently interpretable, this may cause disjoint understanding across both attribute sets.

However, if both sets of attributes are predictive, then they must be correlated even if the attributes are different.
We hypothesize that each attribute of the Target domain $\bm{\chi}_T$ can be expressed as a linear combination of the attributes of the Original domain $\bm{\chi}_O$.
For example, BMI could be approximated as $\text{BMI} = m_w \times \text{weight} - m_h \times \text{height}$.
For multiple attributes, this can be expressed as a matrix mapping that weighs each attribute from one domain to compute each attribute in the other domain. 
However, this mapping is for attribute values, but we aim to map factors across domains.

Consider the \textit{reverse} mapping of the Target attributes to the Original relative attributes $\bm{\chi}_T \longmapsto \bm{\chi}_O$, i.e., 
\begin{equation}
    \label{eq:xai_affine_transformation_attributes_values}
    \bm{\chi}_O = M_{\chi} \bm{\chi}_T.
\end{equation}
Conveniently, we can determine the factor mapping from the attribute mapping with the following steps:
\begin{enumerate}
    \item[i)] Assert the equivalence of XAI prediction across attribute domains: $y = \bm{w}_T^\top \bm{\chi}_T + \bar{\tilde{y}}_T = \bm{w}_O^\top \bm{\chi}_O + \bar{\tilde{y}}_O$.
    \item[ii)] Since both domains are predicting the same task, $\bar{\tilde{y}}_T = \bar{\tilde{y}}_O$, and this term can be canceled out.
    \item[iii)] Substitute Eq. \ref{eq:xai_affine_transformation_attributes_values} for $\bm{\chi}_O$: $\bm{w}_T^\top \bm{\chi}_T = \bm{w}_O^\top (M_{\chi} \bm{\chi}_T)$.
    \item[iv)] Factor out $\bm{\chi}_T$: $\bm{w}_T^\top = \bm{w}_O^\top M_{\chi}$
    \item[v)] Transpose: $\bm{w}_T = (\bm{w}_O^\top M_{\chi})^\top = M_{\chi}^\top \bm{w}_O$.
\end{enumerate}
Hence, $M = M_w = M_{\chi}^\top$ maps Original attribute factors $\bm{w}_O$ to Target attribute factors $\bm{w}_T$, which is a reverse mapping (transpose) of the attribute value mapping $M_{\chi}$.
In terms of the affine transformation, $A = M_{\chi}^\top$, and $\bm{b} = \bm{0}$, i.e., 
\begin{equation}
    \label{xai_affine_transformation_attributes}
    \bm{w}_T = M_{\chi}^\top \bm{w}_O.
\end{equation}

Interpreting the mapping from attribute-to-attribute is more intuitive than from factor-to-factor, since the former is done in everyday tasks when thinking of unit or variable conversion (e.g., BMI $\leftrightarrow$ (height, weight)).
Although the direction of the matrix is transposed for attributes or factor mapping, this can be presented in the same visual format to users, which we describe later (see Section~\ref{sec:experiment-apparatus-target-attributes} and Fig.~\ref{fig:ui_transferable_xai_attributes}a, b).

To limit cognitive load, we can also apply sparsity regularization on the elements in $M_{\chi}$, but this depends on whether the attributes are shared $\bm{\chi}^\oplus$ or unshared $\bm{\chi}^\ominus$ between domains.
Consider two sets of attributes with shared and unshared attributes: $\bm{\chi}_O = (\bm{\chi}^{\oplus}, \bm{\chi}_O^{\ominus})^\top$ and $\bm{\chi}_T = (\bm{\chi}^{\oplus}, \bm{\chi}_T^{\ominus})^\top$.
The attribute mapping can be defined as:
\begin{equation}
    \label{xai_affine_transformation_attributes_sharedunshared}
    M_{\chi} = \begin{pmatrix}
        M_{\chi}^{\oplus} & M_{\chi}^{\oplus \ominus} \\
        M_{\chi}^{\ominus \oplus} & M_{\chi}^{\ominus}
    \end{pmatrix},
\end{equation}
where $M_{\chi}^{\oplus}$ maps the shared attributes (ideally minimally), $M_{\chi}^{\ominus}$ maps the unshared attributes,
and $M_{\chi}^{\oplus \ominus}$ (or $M_{\chi}^{\ominus \oplus}$) maps the unshared (or shared) Original attributes to the shared (or unshared) Target attributes.
We prefer all sub-matrices to be sparse, but especially expect $M_{\chi}^{\oplus} \approx I$ so as not to change shared attributes, unless required to accommodate latent effects that are changed from one domain to the other.
Hence, the sparsity loss is $L_s(A) = ||M_{\chi}^{\oplus} - I||_1 + ||M_{\chi}^{\ominus}||_1 + (||M_{\chi}^{\oplus\ominus}||_1 + ||M_{\chi}^{\ominus\oplus}||_1)$.

\subsubsection{Implementation Details}
We implemented Single-domain explanation and Transferable XAI using SciPy\footnote{\url{https://scipy.org/}} 1.11.4 and adopted the Broyden-Fletcher-Goldfarb-Shanno (BFGS)~\cite{fletcher1970new}, a standard optimization algorithm implemented in SciPy. The random normalization distribution $\mathcal N(0, 0.01)$ was used to initialize the model parameters: $(\bm{w}_O, \bm{w}_T)$ in Singe-domain explanation and $(\bm{w}_O, A, \bm b)$ in Transferable XAI. 

\begin{figure*}[t]
    \centering
    \includegraphics[width=0.85\linewidth]{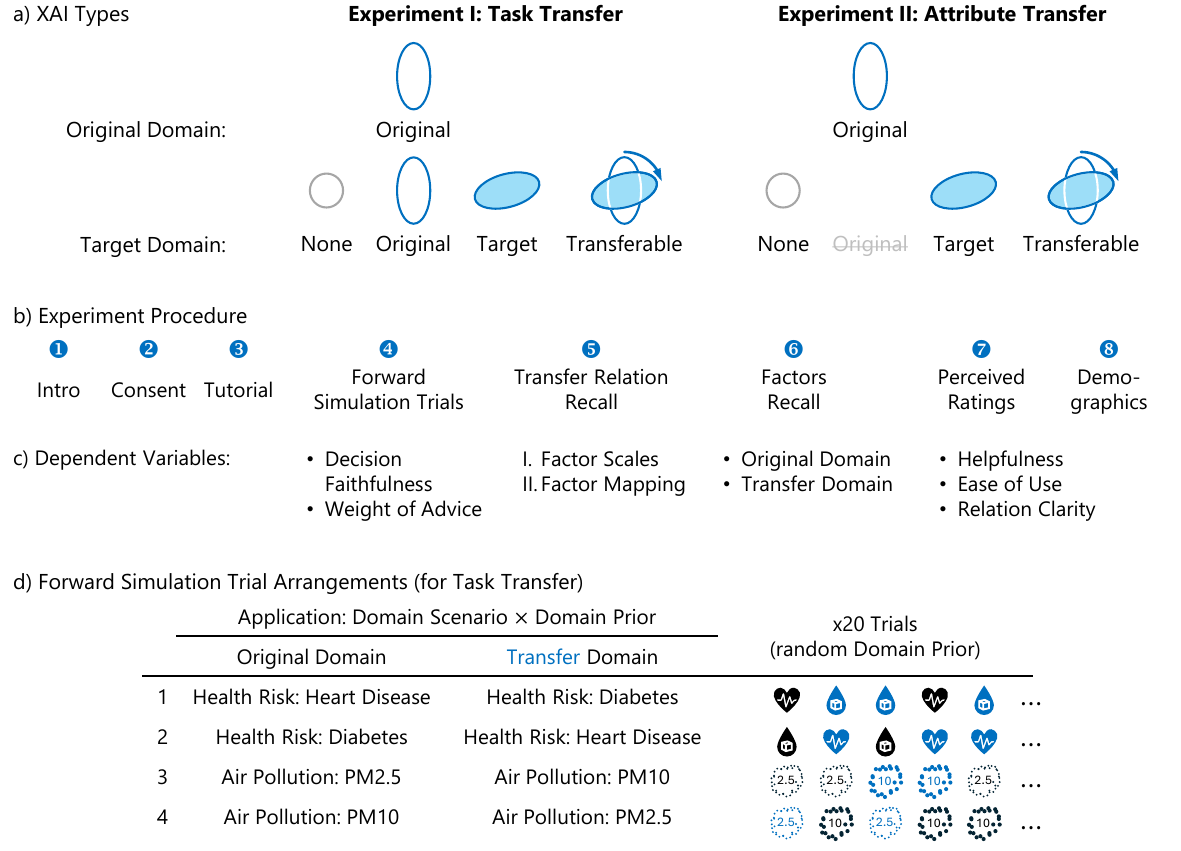}
    \caption{Overview of evaluations across two experiments, Experiment I: Task Transfer and Experiment II: Attributes Transfer.
    a) Both experiments share the same Base XAI with tabular UI for instances of the Original domain.
    But for Attributes Transfer, we omit Original XAI for Target domain XAI Types.
    b) Both experiments share the same experiment procedure and
    c) in each step, mostly share the same dependent variables.
    But for Task Transfer, we measure Factor Scale recall, and for Attributes Transfer, we measure Factor Mapping recall.
    d) For forward simulation trials, there are four between-subjects arrangements that participants see based on assigned Application, Domain Scenario.
    This determines which are their Original or Target domains.
    Arrangements shown for Task Transfer; attribute sets vary for Attributes Transfer.
    Note that the experiment design is the same for the formative and the summative user studies in each experiment.
    }
    \Description{The figure provides an overview of the evaluations across two experiments in four labeled sections, a) through d). 
    Section a), labeled ``XAI Types'', illustrates explanation configurations across experiments. 
    Two columns correspond to ``Experiment I: Task Transfer'' and ``Experiment II: Attributes Transfer''. 
    In both columns, the Original Domain is shown at the top with a vertical oval icon labeled ``Original''. 
    Below, the Target Domain row shows four XAI Types labeled ``None'', ``Original'', ``Target'', and ``Transferable'', each represented by distinct icons. 
    In Experiment I, all four Target Domain options are visually active, while in Experiment II the ``Original'' option is visually muted. 
    Section b), labeled ``Experiment Procedure'', presents a horizontal sequence of eight numbered stages. 
    The stages are labeled ``Intro'', ``Consent'', ``Tutorial'', ``Forward Simulation Trials'', ``Transfer Relation Recall'', ``Factors Recall'', ``Perceived Ratings'', and ``Demographics''.
    Section c), labeled ``Dependent Variables:'', lists outcome measures grouped by procedure stage. 
    Under Forward Simulation Trials, the listed variables are ``Decision Faithfulness'' and ``Weight of Advice''. 
    Under Transfer Relation Recall, the listed variables are ``Factor Scales'' and ``Factor Mapping''. 
    Under Factors Recall, the listed variables are ``Original Domain'' and ``Transfer Domain''. 
    Under Perceived Ratings, the listed variables are ``Helpfulness'', ``Ease of Use'', and ``Relation Clarity''.
    Section d), labeled ``Forward Simulation Trial Arrangements'', shows a table describing trial configurations. 
    The header reads ``Application: Domain Scenario × Domain Prior''. 
    Two columns are labeled ``Original Domain'' and ``Transfer Domain''. 
    Four rows list application pairs: 
    ``Health Risk: Heart Disease'' to ``Health Risk: Diabetes''; 
    ``Health Risk: Diabetes'' to ``Health Risk: Heart Disease''; 
    ``Air Pollution: PM2.5'' to ``Air Pollution: PM10''; 
    ``Air Pollution: PM10'' to ``Air Pollution: PM2.5''. 
    To the right of the table, a note indicates ``×20 Trials (random Domain Prior)'', accompanied by repeated heart and droplet icons representing randomized trial instances.}
    \label{fig:evaluation_diagram}
\end{figure*}

\section{Evaluation Method}

We conducted two experiments on two domain types: Task transfer and Attributes transfer.
We omitted evaluating Subspace transfer as it had been studied by Bo et al.~\cite{bo2024incremental}.
The experiments were conducted separately because of the large difference in cognitive complexity for each domain type transfer; Attributes transfer requires combinatorial mapping and a more sophisticated explanation user interface than Task transfer.
Nevertheless, for both experiments, we had the same application scenarios, AI models, experiment design, base apparatus, procedure, and mixed methods user studies (qualitative formative and quantitative summative). Fig.~\ref{fig:evaluation_diagram} shows the overview of evaluations across two experiments, and we describe the shared methods here.

\subsection{Experiment Design}
\label{sec:all_exp_design}

To investigate the transfer of understanding, we focus on the participant's understanding of AI decisions in the Target domain.
We adopted a 4$\times$2 factorial mixed-design experiment with primary independent variable (IV) as \textbf{XAI type} in the Target domain, and secondary IV as \textbf{Domain Prior}. 
We conducted two experiments across Domain Types, and varied Domain Scenarios for generality.

\subsubsection{Domain Type}
To evaluate the usefulness of Transferable XAI across Domain Types---Task and Attributes---in different experiments.
These were kept separate due to the large differences in reasoning when performing decisions across these domains, and very different explanation user interfaces.

\subsubsection{Applications and Domain Scenarios}
For generalization, we evaluated multi-domain understanding on two Applications: Health risk and Air pollution.
For each Domain Type, we further split each application into different Domain Scenarios.
We selected these applications due to the availability of appropriate datasets that are amenable to our conditions of varying Tasks and Attribute sets.

In the Health risk application, we trained models on the National Health and Nutrition Examination Survey (NHANES) dataset\footnote{\url{https://wwwn.cdc.gov/nchs/nhanes/}} of patient health indicators and risks~\cite{curtin2013national,chen2018national,johnson2013national}. 
We used the data collected from 2005 to 2015 and sampled a relatively balanced subset, containing 3,310 instances. 
For \textit{Task transfer}, we define the tasks as predicting the health risks of Heart disease and Diabetes.
We trained a multi-task AI model (neural network) on attributes: Age, High blood pressure, Chest pain, BMI, Average blood sugar (HbA1c), and Relative has diabetes. 
For \textit{Attributes transfer}, we selected two overlapping sets of attributes to specify two domains and trained separate neural network models on each attribute set.
One attribute set comprises: Age, High blood pressure, BMI, Average blood sugar (HbA1c), and Relative has diabetes.
The other attribute set replaces High blood pressure and BMI with Weight and Height.

In the Air pollution application, we used the Beijing Multi-site Air Quality dataset\footnote{\url{https://archive.ics.uci.edu/dataset/501/beijing+multi+site+air+quality+data}}~\cite{zhang2017cautionary} of hourly measurements of various air pollutants, and particulate matter (PM) readings. 
For \textit{Task transfer}, we define the tasks as predicting the PM readings of particles of different sizes in the air: PM2.5 (2.5$\mu$m) and PM10 (10$\mu$m).
We trained a multi-task AI model (neural network) on attributes: SO$_2$, NO$_2$, CO, O$_3$, and Wind speed.
For \textit{Attributes transfer}, we selected two overlapping sets of attributes to specify two domains and trained separate neural network models on each attribute set.
One attribute set comprises: SO$_2$, NO$_2$, CO, Wind speed, and Pressure.
The other attribute set replaces Pressure with Temperature.

We trained the Transferable XAI with a different loss hyperparameter $\lambda$ based on application. 
For Task transfer, $\lambda = 0.1$ for Health Risk and $\lambda = 10$ for Air Pollution.
For Attributes transfer, $\lambda = 100$ for Health Risk and $\lambda = 30$ for Air Pollution.
See Appendix~\ref{sec:app_modeling_performance} for AI and XAI modeling performance.

\subsubsection{Domain Prior}
We investigate the transfer of learning from the Original domain to the Target domain. These levels make up the Domain Prior independent variable.
Each participant is assigned a random Application (between-subject) and a random Domain Scenario (between-subject) for the Original domain, and the other scenario for the Target domain. 
Note that this is not sequential in the sense of experiment procedure phases of exposing participants to different conditions.
Instead, this represents the sequence of prior knowledge, i.e., which domain is known first, and which is learned next or transferred to.
See the between-subjects trial arrangement for the Task Transfer experiment in Fig.~\ref{fig:evaluation_diagram}d.

\subsubsection{XAI Types for Target Domain}
For instances in the Original domain, all instances are provided with explanations of linear factors in the Original domain $\bm{w}_O$.
For instances in the Target domain, we show different XAI Types to investigate Transferable XAI against baselines of no explanation (None), explanations only for the Original domain, and independent explanations of the Target domain. 
We provide them \textit{between-subject} to avoid participant confusion to mix up between different explanations and user interfaces. 
XAI Types are:
\begin{enumerate}
    \item[i)] \textit{None} that does not provide any explanation. Participants rely on their understanding of the Original domain to infer decisions in the Target domain.
    \item[ii)] \textit{Original} domain ($\bm{w}_O$) (only for Task transfer) that explains the AI prediction for the Original task, but the participant will have to adapt this to infer the decision for the Target task.
    This XAI Type is not applicable for Attributes transfer since providing the Original attributes for an instance with only Target attributes is data leakage.
    \item[iii)] \textit{Target} domain ($\bm{w}_T$) that explains the AI prediction for the Target domain.
    \item[iv)] \textit{Transferable} XAI. ($h(\bm{w}_O)$) that explains the Target domain by including the Original domain factors $\bm{w}_O$ and contextualizing it for the Target domain via the affine transformations described in Section \ref{sec:affine-transformation}.
    Transforming information depends on Domain Type: scales $\bm{\kappa}_\Delta$ for Task, and mapping matrix $M_\chi$ for Attributes.
\end{enumerate}

In the user studies, after tutorial training, we present participants with instances randomly chosen from the Original or Target domain for each trial.
Explanation shown depends on the Domain Prior, Domain Scenario, and XAI Type for the Target domain.

\subsubsection{Measures}
\label{sec:all_exp_design_measures}
Following Incremental XAI~\cite{bo2024incremental}, we investigate how well participants understand, recall, and apply explanations. 
In addition, we pay attention to participants' understanding and memorability of the factor relationships across the domain:
\begin{itemize}
    % \item \textit{Decision Faithfulness} 
    \item \textit{Decision Faithfulness} 
    measures the participant's \textbf{understanding} of AI behavior, by the participant's label $\mathring{y}$ was correct in forward simulating~\cite{doshi2017towards} the AI Explainer (XAI) prediction $\tilde{y}$. Taking the labeling error $|\mathring{y} - \tilde{y}|$, we apply a log transform since the errors were skewed with a long tail:
    \begin{equation}
        \operatorname{Log} \mathrm{Unfaithulness} = \log(|\mathring{y} - \tilde{y}|).
    \end{equation}
    
    \item \textit{Weight of Advice}~\cite{bailey2023meta} 
    measures the \textbf{alignment} of the participant's decision label $\mathring{y}$ is closer to the XAI of the aligned domain ($y_{=}$ when XAI label is of the same domain as instance), or the misaligned domain ($y_{\neq}$).
    The standard Weight of Advice (WoA) ratio is defined as:
    \begin{equation}
        \text{WoA} = \Big( \frac{|\mathring{y} - y_{=}|}{|\mathring{y} - y_{\neq}|} \Big)^{-1},
    \end{equation}
    where we inverted the expression such that WoA is higher when the participant's label $\mathring{y}$ is closer to the aligned domain $y_{=}$.
    However, ratio is a skewed metric\footnote{Consider the case of $\text{numerator} = 10 \times \text{denominator}$, then the $\text{ratio} = 10/1$ is 9 away from 1; 
    but for the case of $\text{denominator} = 10 \times \text{numerator}$, then the $\text{ratio} = 1/10$ is 0.9 away from 1, which is a much smaller difference ($0.9 \ll 9$).}, 
    so we apply a log transform to get the log-ratio which is symmetric, i.e.,
    \begin{equation}
     \operatorname{Log} \mathrm{WoA} = \log\big(|\mathring{y} - y_{\neq}|\big) - \log\big(|\mathring{y} - y_{=}|\big).
    \end{equation}
    
    \item \textit{Factors Recall} evaluates how accurately participants can reproduce each factor for both the Original domain, $w_O^{(r)}$, and the Target domain, $w_T^{(r)}$, by typing them out. 
    We measure this with the Absolute Percentage Error (APE) to account for kurtosis (larger values tend to have higher error), and applied a log transform to handle the skewed responses, i.e.,
    \begin{equation}
        \operatorname{Log} \mathrm{APE}^{(r)} = \log \Big( \frac{\lvert \mathring{w}^{(r)} - w^{(r)} \rvert}{\lvert w^{(r)} \rvert}\times 100\% \Big).
    \end{equation}
    
    \item \textit{Transfer Relation} measures participants’ understanding of the relationship between XAI explanations of the Original and the Target domain.
    Section~\ref{sec:task_exp_design} details how we measure the scaling relationship between factors across tasks, and 
    Section~\ref{sec:attributes_transfer_exp_design} details how we measure the mapping relationship across attributes.

    \item \textit{Perceived ratings} of Helpfulness, Ease of Use, and Relation Clarity between domains for the XAI Type measured on a 7-pt Likert scale: Strongly Disagree ($-3$), $\dots$, Neither Agree nor Disagree ($0$), $\dots$, Somewhat Agree ($+3$). 

\end{itemize}
We measured all metrics for both the Original and Target domains.

\begin{figure*}[t]
    \centering
    \includegraphics[width=0.96\linewidth]{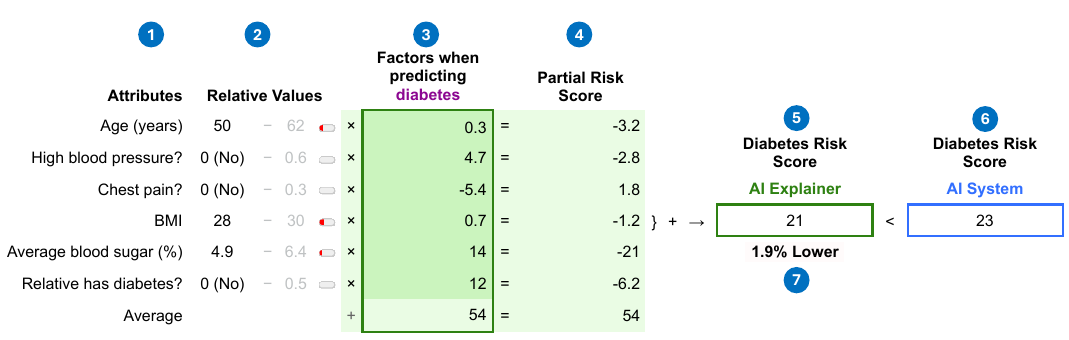}
    \caption{User interface (UI) of AI System with Base XAI of the Original domain showing: 
    % linear factor explanation showing: 
    1) attributes used in the prediction task, 
    2) instance relative values, computed as the actual value minus the average value for each attribute, i.e., $x^{(r)}$ {\color{lightgray}- $\bar{x}^{(r)}$}, 
    3) factors $w^{(r)}$ of each attributes, 
    4) partial contributions $w^{(r)}x^{(r)}$ of each attributes, 5) estimation $\tilde{y} = \sum_r \tilde{y}^{(r)}$ from the AI Explainer, 
    6) prediction $\hat{y}$ from the AI System, 
    7) indicator of how different the AI Explainer estimation is from the AI System.}
    \Description{This figure shows the user interface of an AI system that explains its diabetes risk prediction using a linear factor model. 
    The interface consists of seven main components, each numbered from 1 to 7.
    Component 1 lists the attributes used in the prediction, including age, blood pressure, chest pain, body mass index (BMI), average blood sugar, and whether a relative has diabetes.
    Component 2 shows the relative values of these attributes, calculated as the person’s actual value minus the average value in the dataset. 
    The values are as follows: 
    Age is 50 years, compared to an average of 62 (a difference of -12); 
    High blood pressure is 0 (No), compared to an average of 0.2; 
    Chest pain is 0 (No), compared to an average of 0.3; 
    BMI is 28, compared to an average of 30; 
    Average blood sugar is 4.9\%, compared to an average of 6.4\%; and Relative has diabetes is 0 (No), compared to an average of 0.5. 
    The “Average” row is also shown for reference.
    These relative values help identify which attributes deviate from the norm.
    Component 3 displays the model’s learned factors for predicting diabetes. The factors are: 0.3 for Age, 4.7 for High blood pressure, -5.4 for Chest pain, 0.7 for BMI, 14 for Average blood sugar, and 12 for Relative has diabetes. The “Average”, which is a biased term value, is 54.
    Each factor represents the strength and direction of influence of an attribute on diabetes risk.
    Component 4 shows the partial risk scores, obtained by multiplying each attribute’s relative value by its factor weight. 
    Each contribution is color-coded: green cells indicate factors that lower risk, while red cells indicate factors that increase it.
    Component 5 presents the AI Explainer’s estimated overall diabetes risk score, shown as 21. 
    Component 6 shows the AI System’s official prediction score, which is 23. 
    Component 7 highlights that the Explainer’s estimate is 1.9\% lower than the System’s score.
    Together, these components visually explain how each attribute contributes to the predicted diabetes risk and how closely the human-readable explanation matches the AI model’s prediction.}
    \label{fig:ui_org_domain}
\end{figure*}
\subsection{Base Experiment Apparatus}
\label{sec:apparatus_org}
To convey the linear factor explanation for each AI prediction, we adapted the tabular user interface (UI) from Incremental XAI~\cite{bo2024incremental}.
For generalizability across domains (described in Section \ref{sec:relative-attribute-values}), we show the relative attribute values instead of the raw attribute values, making the bias term more interpretable.
By explicitly presenting values and weights with numbers and text, this UI enables users to precisely calculate steps to arrive at the explained prediction.
Although, users require some numeracy to read the UI, we found that participants in our formative study could understand it.

Fig.~\ref{fig:ui_org_domain} shows the explanation UI for all instances in the Original domain, and for instances in the Target domain with baseline XAI Types (Original, Target). 
We describe the explanation UI for Transferable XAI for each Domain Type later in Sections \ref{sec:experiment-apparatus-target-task} and \ref{sec:experiment-apparatus-target-attributes}.
During training, we show the \% error between the AI Explainer and AI System to make this salient and accelerate learning about explanation faithfulness.
We implemented our survey in Qualtrics and embedded the user interface. 

\subsubsection{User Interface Design Process}
\rev{We assume users have numeracy skills and can readily read tables.
Starting with reference XAI user interfaces for linear factor explanations~\cite{bo2024incremental, poursabzi2021manipulating}, we employed a user-centered design process~\cite{norman1986user} to iteratively refine our explanation UI for Transferable XAI.
Specifically, we performed heuristic evaluation~\cite{nielsen2005ten} and cognitive walkthrough~\cite{lewis1997cognitive}, and elicited feedback from pilot users for both domain types (Transfer, Attribute).

We applied the following heuristics.
\textit{Match to real world} with simplified names to familiar terms, avoiding mathematical, technical, or domain jargon.
\textit{Consistency and standards} across XAI and domain variants, using green for XAI and blue for AI.
\textit{Aesthetic and minimalist design} for scannability with left-to-right information and calculation flow.
\textit{Recognition rather than recall} by showing explicit working steps in the table and matrix layout, and in tooltips.
\textit{Effective use of cues} to contextualized numbers using meters for attribute values and red text for negative scales.
\textit{Separation of concerns} to reduce cognitive load by separating the matrix mapping from the tabular interface.

With cognitive walkthroughs, the co-authors and several members from the same HCI research lab used each version of the explanation UI to perform forward simulation task to use the explanations to estimate the AI predictions in the Original and Target tasks.
We noted when information could be misunderstood or misused, and made redesigns to mitigate these issues.
Subsequently, we conducted pilot studies with the set up for the formative user studies to ascertain remaining usability or interpretability issues.
By the end of our design iterations, this included 10 pilot participants with a similar profile as the reported formative user studies.
See Sections~\ref{sec:experiment-apparatus-target-task} and ~\ref{sec:experiment-apparatus-target-attributes} for the design narrative and design rationale of the Task Transfer and Attributes Transfer XAI UIs, respectively.
}

\subsection{Experiment Procedure}
\label{sec:all-exp_procedure}
Participants are recruited separately for different Domain Types (Task or Attributes).
We first conducted a formative study focusing on qualitatively learning usage patterns and interpretations, 
then conducted a larger summative user study to formally evaluate benefits and effects across XAI Types.
Both user studies followed the same experiment procedure, but the formative study has fewer trials across longer duration than the summative study.
This allows us to elicit utterances from participants while observing their usage and interactions.

For the formative study, 
we used the think aloud protocol to elicit the participant's thought processes as they used the explanations. 
With participant consent, we audio and screen recorded participant vocalizations and interactions with the UI.

\begin{figure*}[tbp]
    \centering
    \includegraphics[width=.95\linewidth]{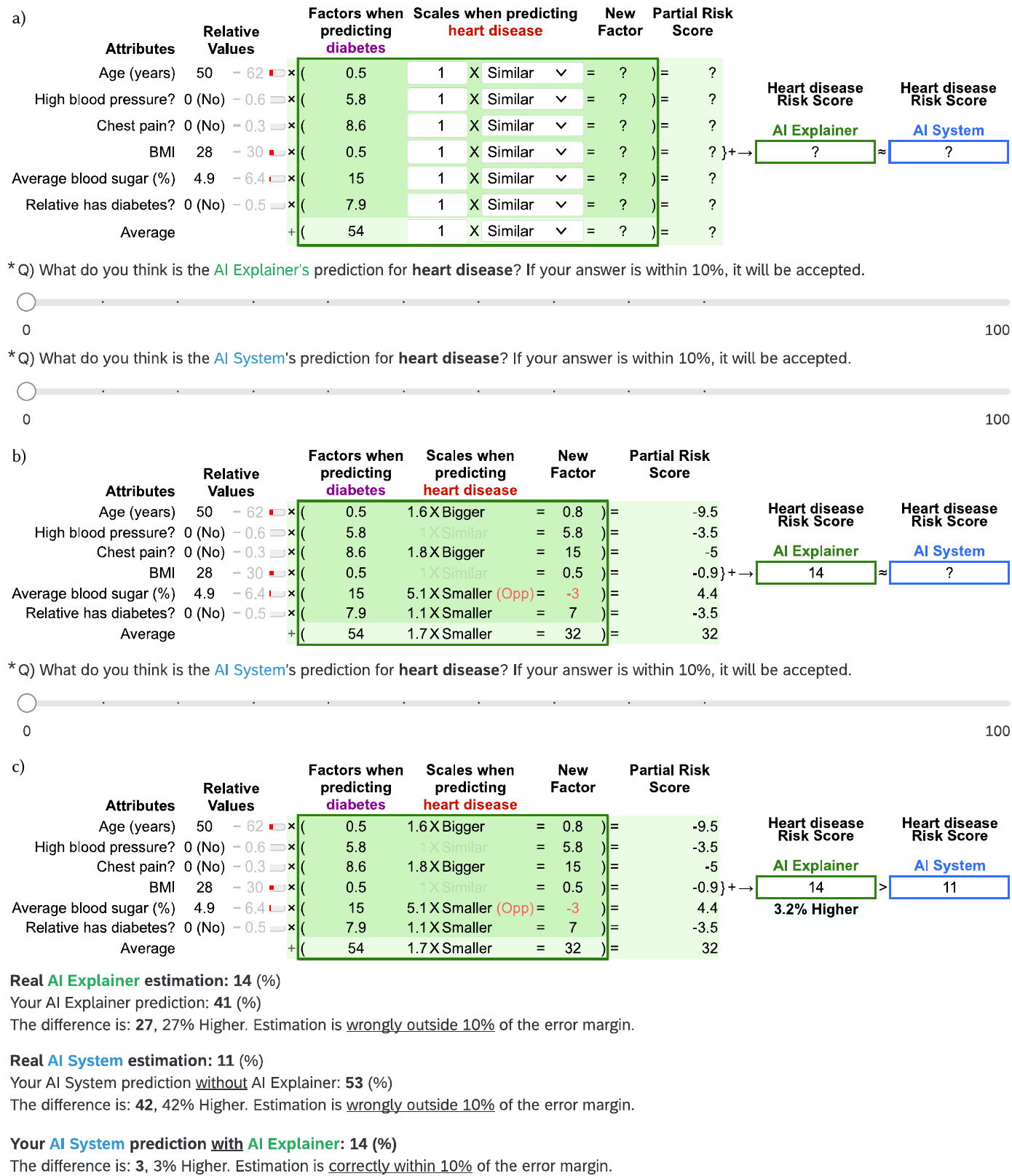}
    \caption{
    Forward simulation task in summative user study main trial in three-page arrangement for each trial:
    a) Test page for user to estimate XAI and AI predictions,
    b) Explanation page to read XAI and update estimation of AI prediction,
    c) Learning page to review estimations with respect to actual XAI and AI predictions.
    Example shown is for an instance in the Target domain for the Task transfer experiment.
    }
    \label{fig:task_transfer_3pages}
    \Description{The figure consists of three vertically arranged subfigures labeled a), b), and c), each showing an interface for estimating heart disease risk using an AI Explainer and an AI System. 
    All three subfigures share a common table structure with columns labeled ``Attributes'', ``Relative Values'', ``Factors when predicting diabetes'', ``Scales when predicting heart disease'', ``New Factor'', and ``Partial Risk Score''. 
    The rows list the same attributes: Age (years), High blood pressure? (Yes), Chest pain? (No), BMI, Average blood sugar (\%), Relative has diabetes? (Yes), and an ``Average'' row.
    In subfigure a), the columns ``Scales when predicting heart disease'', ``New Factor'', and ``Partial Risk Score'' display dropdowns set to ``1× Similar'' and question marks, indicating unspecified values. 
    On the right, two boxes labeled ``Heart disease Risk Score'' show placeholders for ``AI Explainer'' and ``AI System''. 
    Below the table, two horizontal slider bars are shown, asking for the user’s estimates of the AI Explainer’s prediction and the AI System’s prediction for heart disease, each ranging from 0 to 100.
    Subfigure b) shows the same table layout as a), but the ``Scales when predicting heart disease'' column now contains specific values such as ``1.6× Bigger'', ``1.8× Bigger'', ``5.1× Smaller (Opp)'', and ``1.1× Smaller''. 
    The ``New Factor'' and ``Partial Risk Score'' columns are filled with numeric values. 
    On the right, the ``AI Explainer'' box displays a numeric heart disease risk score of 37, while the ``AI System'' box still shows a placeholder. 
    Below the table, only a single horizontal slider bar is shown, asking for the AI System’s prediction for heart disease.
    Subfigure c) presents the same table content as b). 
    On the right, both the ``AI Explainer'' and ``AI System'' boxes now display numeric values, with the AI Explainer showing 37 and the AI System showing 24, accompanied by a label indicating ``13\% Higher''. 
    Below the table, the slider bar is replaced by three blocks of textual feedback. 
    These blocks report the real AI Explainer estimation, the real AI System estimation, and the user’s AI System prediction with AI Explainer, each followed by difference values and statements indicating whether the estimation is within or outside the 10\% error margin.}
\end{figure*}

For the summative study, each participant was engaged in the following procedure (see Appendix Figs.~\ref{fig:task_transfer_survey_scenarios}--\ref{fig:task_transfer_survey_factor_recall} for screenshots of the survey):
\begin{enumerate}
    \item[1)] Introduction to the study.
    Each participant is randomly assigned to an application (Health Risk or Air Pollution), 
    and randomly assigned a Domain Scenario as the Original domain condition.
    \item[2)] Consent to participate. This study was approved by the university institutional review board (IRB).
    \item[3)] Tutorial on the applications and user interface (UI). 
    UI Components are introduced depending on the assigned conditions, with screening questions to ensure participants have understood the tasks and can use the explanation correctly.
    \item[4)] Forward simulation session of 6 trials with reflection questions to help participants understand and 14 regular trials. 
    All trials are randomly assigned to the Original or Target domain.
    Following~\cite{bo2024incremental}, we have three pages for each trial:
    \begin{enumerate}
        \item[i)] On the test page \rev{(Fig.~\ref{fig:task_transfer_3pages}a)}, % (Appendix Fig.~\ref{fig:task_transfer_survey_test_page1}),
        participants saw a UI showing only values and were asked to forward-simulate the outputs of the AI Explainer ($\tilde{y}$) and AI System ($\hat{y}$), with factors and consequent calculations hidden. The UI is interactive, which allows them to enter different factor values and calculate automatically while attempting to estimate the AI outputs. We used two sliders % (bottom of Fig.~\ref{fig:task_transfer_survey_test_page1})
        to capture their estimates. Reflection questions %(middle of Fig.~\ref{fig:task_transfer_survey_test_page1})
        were posed in the first 6 trials to aid understanding without prolonging the survey.
        \item[ii)] On the explanation page \rev{(Fig.~\ref{fig:task_transfer_3pages}b)}, % (Appendix Fig.~\ref{fig:task_transfer_survey_test_page2}), 
        explanatory factors and calculations were provided to participants based on the XAI type condition. A slider was used to record participants' estimates of AI predictions.
        \item[iii)] On the learning page \rev{(Fig.~\ref{fig:task_transfer_3pages}c)}, % (Appendix Fig.~\ref{fig:task_transfer_survey_test_page3}), 
        participants review their answers with the actual AI Explainer prediction and AI System prediction. Through frequent review, participants can learn from mistakes, improve their attention to the factors, and reinforce their understanding across trials. 
    \end{enumerate}
    \item[5)] Transfer relation recall session.
    % (Appendix Fig.~\ref{fig:task_transfer_survey_relation_question}). 
    Participants recalled and judged factor transfer relationships across the two domains.
    % , considering relative magnitudes (ignoring sign) and whether their effects on predictions were the same or opposite.
    Specific questions depend on the transfer type experiment (Task or Attribute).
    \item[6)] Factor recall session.
    % (Appendix Fig.~\ref{fig:task_transfer_survey_factor_recall}). 
    Participants recalled the factors based on the AI Explainer they saw. No specific instances were shown, but participants can use the interactive UI to examine their attempts and help them recall.
    \item[7)] Answer rating questions on \textit{Perceived helpfulness}, \textit{Perceived ease-of-use}, and \textit{Perceived clarity of factor relationships}.
    \item[8)] Answer demographics questions.
\end{enumerate}
We provided an incentive bonus of £0.03 for each accurate response, defined as cases where the participant’s selected range contained the correct value and the relative error of the range width was within 10\% (max £1.8), and a max of £1.1 for Factor relationship recall task and Factor recall task based on the mean relative error.

\begin{figure*}[t]
    \centering
    \includegraphics[width=.95\linewidth]{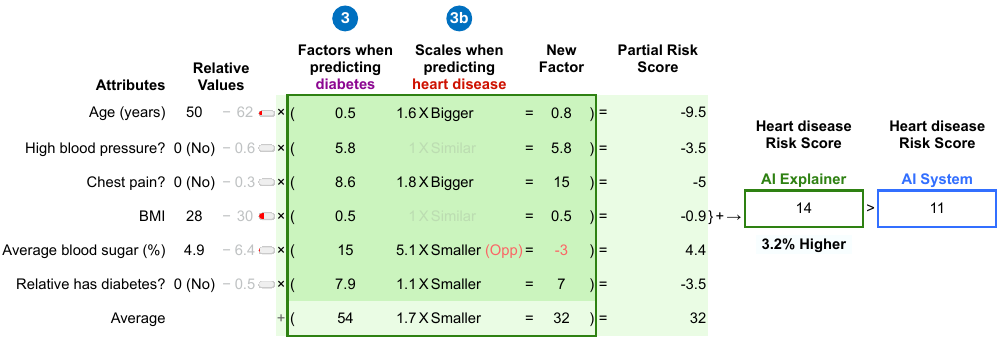}
    \caption{User interface (UI) of the Transferable explanation in \textbf{Task transfer}, showing: 
    3) factors from the Original Domain, $w_O^{(r)}$ (diabetes risk prediction task), 
    3b) the corresponding scaled factors for the Target domain (heart disease risk prediction task). For example, BMI retains a similar influence across tasks, Age becomes 1.6× more influential, and Average blood sugar (\%) reverses its effect (5.1× smaller) when transferred to heart disease prediction.}
    \Description{This figure shows the user interface of the Transferable AI Explainer in the Task transfer setting, 
illustrating how explanatory factors are adapted from the diabetes risk prediction task to a new heart disease prediction task. 
The interface consists of seven main components, numbered 1 to 7, similar to the previous figure.
Component 1 lists the attributes used for prediction: 
Age (years), High blood pressure, Chest pain, Body Mass Index (BMI), Average blood sugar (\%), and whether a Relative has diabetes. 
Component 2 shows the instance’s relative values, calculated as the difference between the actual and average values in the dataset. 
The relative values are: 
Age is 50 years versus an average of 62 (–12 difference); 
High blood pressure is 0 (No) versus an average of 0.6; 
Chest pain is 0 (No) versus an average of 0.3; 
BMI is 28 versus an average of 30; 
Average blood sugar is 4.9\% versus an average of 6.4\%; 
and Relative has diabetes is 0 (No) versus an average of 0.5. 
The “Average” row is also displayed.
Component 3 presents the original factors used when predicting diabetes: 
0.5 for Age, 
5.8 for High blood pressure, 
8.6 for Chest pain, 
0.5 for BMI, 
15 for Average blood sugar, 
and 7.9 for Relative has diabetes. 
The Average (bias term) value is 54.
Component 3b shows the scaling factors applied when transferring to the heart disease prediction task. 
These indicate how much stronger or weaker each attribute’s influence becomes. 
The scaling values are: 
1.6× Bigger for Age, 
1× Similar for High blood pressure, 
1.8× Bigger for Chest pain, 
1× Similar for BMI, 
5.1× Smaller (Opposite direction) for Average blood sugar, 
and 1.1× Smaller for Relative has diabetes. 
The Average scaling factor is 1.7× Smaller. 
After scaling, the New Factors for the heart disease task are: 
0.8 for Age, 
5.8 for High blood pressure, 
15 for Chest pain, 
0.5 for BMI, 
–3 for Average blood sugar (reversed direction), 
and 7 for Relative has diabetes. 
The Average new factor value is 32.
Component 4 shows the resulting Partial Risk Scores, obtained by multiplying the relative values by the new factors. 
The partial scores are: 
–9.5 for Age, 
–3.5 for High blood pressure, 
–5 for Chest pain, 
–0.9 for BMI, 
4.4 for Average blood sugar, 
and –3.5 for Relative has diabetes. 
The Average partial score is 32.
Component 5 displays the AI Explainer’s estimated Heart Disease Risk Score, which is 14. 
Component 6 shows the AI System’s true Heart Disease Risk Score, which is 11. 
Component 7 highlights that the Explainer’s estimate is 3.2\% higher than the AI System’s prediction.
Overall, this figure demonstrates how the Transferable AI Explainer adapts explanatory factors between related prediction tasks. 
For example, Age becomes 1.6 times more influential when moving from diabetes to heart disease, 
while Average blood sugar not only becomes 5.1 times smaller in effect but also reverses its influence direction.}
    \label{fig:ui_transferable_xai_task}
\end{figure*}
\section{Task Transfer Experiment}

We evaluated Transferable explanations against three baselines (None, Original, and Target) to examine:
i) usage strategies to interpret the AI decisions across different prediction tasks in \rev{a qualitative formative user study}, 
ii) impact on decision faithfulness, explanation recall, and relations understanding in \rev{a quantitative summative user study}.

\subsection{Experiment Design for Task Transfer}
\label{sec:task_exp_design}
The study follows the experiment design described in Section~\ref{sec:all_exp_design} and the procedure outlined in Section~\ref{sec:all-exp_procedure}.
but with additional specific measures.
\rev{Specifically, we clarify the transfer-specific measures of Transfer Relation for task transfer.}

\subsubsection{Measures on Transfer Relation (Factor Scaling)}

We measured participant understanding of how factors relate across the Original and Target domains with:
\begin{itemize}  
    \item \textit{Factor Relation Magnitude}
    of the scale between each $r$th factor across domains, i.e.,
    \begin{equation}
        |\mathring{\kappa}^{(r)}| \simeq \Big| w_T^{(r)}/w_O^{(r)} \Big|.
    \end{equation}
    To simplify participant answering, we asked them to indicate this in five ordinal levels:
    \textless{}$1.5\times$ smaller ($-2$), $1\times$–$1.5\times$ smaller ($-1$), similar (0), $1\times$–$1.5\times$ larger ($+1$), and more than $1.5\times$ larger ($+2$). 
    We calculate the (in)correctness of participant response as an Ordinal Error, treating differences between consecutive levels as 1.
    
    \item \textit{Factor Relation Direction}
    of whether the $r$th factor for both domains are in the same ($+1$) or opposite ($-1$) directions, i.e.,
    \begin{equation}
        % \mathbb{I}\!\left(\mathring{D}^{(r)} = \operatorname{sgn}(\rho^{(r)})\right).
        \operatorname{sgn}\Big(\mathring{\kappa}^{(r)}\Big)
    \end{equation}
\end{itemize}

\subsection{Experiment Apparatus for Task Transfer}
\label{sec:experiment-apparatus-target-task}

Fig.~\ref{fig:ui_transferable_xai_task} shows the explanation UI for Transferable XAI with Task transfer. 
\rev{As a generalization of Incremental XAI~\cite{bo2024incremental}, Transferable XAI uses the same UI with some changes.}
The UI for Transferable XAI has an additional ``New Factor'' column to show a modification on the factors.
For Incremental XAI, the column shows the sparse \textit{differences} ($+$) in factors, where unchanged factors have $\Delta\text{factor} = 0$.
For Transferable XAI, 
the column shows the sparse \textit{scales} ($\times$) in factors, where unchanged factors have $\text{scale} = 1$.

\rev{Through our cognitive walkthrough, we realized that users may just directly multiply the factor scale numbers without semantically understanding their implications or be confused.}
We noted the asymmetric interpretation of scale values above or below 1, which may lead to biased interpretations.
For example, $\kappa_\Delta^{(1)} = 0.2$ can be perceived as being closer to 1 than $\kappa_\Delta^{(2)} = 5$ with differences of 0.8 and 4, respectively.
This contradicting interpretation is more apparent when reversing the direction of the transformation from Target to Original, where the difference effect is reversed.
Yet, both scales are just inverse \textit{ratios} of each other, i.e., $\kappa_\Delta^{(1)} = w_T^{(1)}/w_O^{(1)} = 1/5$ and $\kappa_\Delta^{(2)} = w_T^{(2)}/w_O^{(2)} = 5/1$.
To mitigate this bias, we report scales $> 1$ as Bigger and $< 1$ as Smaller, with the \textit{ratio} numerator and denominator, respectively.
In our example, we would denote $\kappa_\Delta^{(1)} = 5 \times \text{Smaller}$ and $\kappa_\Delta^{(2)} = 5 \times \text{Bigger}$.

Furthermore, negative scales are rather unintuitive, and rote flipping of the factor sign is not sufficiently interpretive.
Instead, a negative scale means that the same attribute contributes in the \textit{opposite} direction for the Target task than for the Original task.
We convey this by appending a red ``(Opp)'' label.

\subsection{Formative User Study}

We examined how users use and interpret the explanations across different prediction tasks.

\subsubsection{Participants}
We recruited 16 university students (mean age = 23, 21 to 29; 10 female, 6 male) from diverse disciplines, including Data Science, Economics, Robotics, and Engineering Leadership. 

\subsubsection{Qualitative Findings}
We analyzed participants' interactions and utterances, and report key findings in terms of XAI Types for the Target domain.

Participants could well understand the Original domain in the \textbf{None} condition (No XAI), but would \textbf{overgeneralize} the explanations to the Target domain and \textbf{struggle to understand} the AI’s predictions. 
This was difficult if they had no domain knowledge, e.g., P9 felt that
\textit{``knowing the factors of heart disease [Target domain] would be useful''}. 
When deciding on a case in the Target domain, P2 \textit{``[did] mental math ... using the AI Explainer [from the Original domain], multiplying these attributes with respect to their factors...''}. 
Since the factors in the Target domain are actually different from those in the Original domain, this could be an overgeneralization.

When provided with independent Original and Target XAI Types, participants had to \textbf{recall} the explanations for both AI systems separately, and found this \textbf{burdensome}.
P14 felt that \textit{``[the factors] do not look similar, so it's hard to memorize them by their relations''}.
P8, after seeing the explanations of two domains, thought \textit{``some [factors] are about the same, and some are opposite, like one positive and one negative... but after a while, I probably couldn’t remember anymore''}.
Nevertheless, some participants still preferred the Original/Target XAI to Transferable XAI, because \textit{``it's \textbf{easier to understand}''} (P6, P12, P13) and \textit{``straightforward''} (P8).

In contrast, participants using Transferable XAI could \textbf{understand how factors change across domains}. 
P7 found the scales \textit{``useful when I want to know the risks of both diseases and make comparisons''}.
P9 learned that \textit{``different factors may be more important for heart disease, compared to diabetes... If it [the scale] is more than one, then it's more important for heart disease''}.
P14 felt that \textit{``changes in [factors of Transferable XAI] are not very large and have some identical factors, so I can remember them''}. 
However, some participants remarked that it \textit{``is just a\textbf{ bit complex} for me''} (P13).

\subsection{Summative User Study}
\label{sec:task_summative}
We conducted a summative study to qualitatively evaluate how well participants understand, recall, and apply explanations, as well as their understanding of factor relations across the prediction tasks.
\subsubsection{Participants}
We recruited participants from Prolific.co, of which 400 passed screening.
They had a median age of 38 years (range = 19--78; 41\% female). The survey took a median of 71 minutes, with £7.00 base pay and a median £1.24 bonus (up to £2.90).

\begin{table*}[t]
\caption{Hypotheses and findings from the \textbf{Task transfer} summative user study, examining various dependent variables across different XAI types: None (N), Original (O), Target (T), and Transferable ($\mathcal{T}$).}
\label{tab:task_transfer_hypo_and_finding}
\renewcommand{\arraystretch}{1} 
\begin{tabular}{@{}lrllll@{}}
\toprule
Measure & Domain & Hypothesis & Air Pollution & Health Risk & Evidence \\ 
\midrule
Decision Faithfulness & Original &  N $\approx$ O $\approx$ T $\approx$ $\mathcal{T}$ & N $\approx$ O $\approx$ T $<$ $\mathcal{T}$ & N $\approx$ O $\approx$ T $<$ $\mathcal{T}$ & Fig.~\ref{fig:summativestudy_task_forward}a,b (gray)  \\
\color{lightgray} &  Target   
&  N $<$ O $\approx$ T $<$ $\mathcal{T}$  
&  N $<$ O $\approx$ T $<$ $\mathcal{T}$  
& N \ns{$\approx$} O $\approx$ T $<$ $\mathcal{T}$ &  Fig.~\ref{fig:summativestudy_task_forward}a,b (blue)  \\
\midrule

Weight of Advice &  Original 
&  N $\approx$ O $\approx$ T $\approx$ $\mathcal{T}$  
&  N $\approx$ O $\approx$ T $<$ $\mathcal{T}$  
& N $\approx$ O $\approx$ T $\approx$ $\mathcal{T}$ & Fig.~\ref{fig:summativestudy_task_forward}c,d (gray) \\
\color{lightgray} &   Target  
&  N $<$ O $<$ T $<$ $\mathcal{T}$  
&  N \ns{$\approx$} O \ns{$\approx$} T $<$ $\mathcal{T}$  
& N \ns{$\approx$} O $<$ T \ns{$\approx$} $\mathcal{T}$   & Fig.~\ref{fig:summativestudy_task_forward}c,d (blue) \\
\midrule

Factors Recall  & Original &  N $\approx$ O $\approx$ T $\approx$ $\mathcal{T}$  &  N $\approx$ O $\approx$ T $\approx$ $\mathcal{T}$ & N $\approx$ O $\approx$ T $<$ $\mathcal{T}$ &  Fig.~\ref{fig:summativestudy_task_recall}a,b (gray)  \\
\color{lightgray} &  Target 
&  N $<$ O $\approx$ T $<$ $\mathcal{T}$  
&  N $<$ O $\approx$ T \ns{$\approx$} $\mathcal{T}$  
&  N $<$ O $\approx$ T \ns{$\approx$} $\mathcal{T}$  &Fig.~\ref{fig:summativestudy_task_recall}a,b (blue)  \\
\midrule

\makecell[l]{Factor Relation Magnitude} 
&  Target
&  N $<$ O $\approx$ T $<$ $\mathcal{T}$  
&  N \ns{$\approx$} O $\approx$ T $<$ $\mathcal{T}$  
&   N \ns{$\approx$} O $\approx$ T $<$ $\mathcal{T}$
& Fig.~\ref{fig:summativestudy_task_recall}c,d  \\
\makecell[l]{Factor Relation Direction} 
&  Target
&  N $<$ O $\approx$ T $<$ $\mathcal{T}$  
&  N $<$ O $\approx$ T \ns{$\approx$} $\mathcal{T}$  
& N $<$ O $\approx$ T $<$ $\mathcal{T}$ 

% Figs.~\ref{fig:summativestudy_task_recall}c,d
\\

\bottomrule
\end{tabular}
\end{table*}

\begin{table*}[t]
\caption{Statistical analysis of responses using linear mixed-effects models for \textbf{Task transfer}. 
Each row reports one effect, including fixed effects, random effects, and their interactions.
XAI Type denotes the explanation type applied in the Target domain.
$F$ and $p$ values are from ANOVA tests.}
\label{tab:task_transfer_summative_statistical_analysis}
\renewcommand{\arraystretch}{1}
\begin{tabular}{llrrrr}
\toprule
\multirow{2}{*}{Response} & \makecell[l]{Linear Effect Model} & \multicolumn{2}{c}{Air pollution} & \multicolumn{2}{c}{Health risk}  \\

 & \makecell[l]{(random effects: participants, case)} & $F$ & $p > F$ & $F$ & $p > F$  \\
\midrule
\multirow{11}{*}{Log UnFaithfulness}    
& XAI Type +  & 5.4 & .0016 & \color{lightgray}2.5 & \color{lightgray}n.s. \\
& Domain ID + & \color{lightgray}1.8 & \color{lightgray}n.s. & 20.5 & $<$.0001  \\
& w/ or w/o XAI + & 397.9 &  $<$.0001 & 155.7 & $<$.0001\\
& XAI Domain Gap + & 6.1  & .0160 & 13.2 & .0005 \\
& XAI Type $\times$ Domain ID & \color{lightgray}3.0 & \color{lightgray}.0290 & 5.9 & .0005 \\
& XAI Type $\times$ XAI Domain Gap & \color{lightgray}1.5 & \color{lightgray}n.s. & \color{lightgray}2.2 & \color{lightgray}n.s.  \\
& Domain ID $\times$ XAI Domain Gap & \color{lightgray}3.1  & 
\color{lightgray}n.s. &\color{lightgray}2.1  & \color{lightgray}n.s. \\
& XAI Type $\times$ Domain ID $\times$ XAI Domain Gap & \color{lightgray}1.5 & 
\color{lightgray}n.s.  & 3.3 & .0205 \\
& XAI Domain Gap $\times$ w/ or w/o XAI & \color{lightgray}0.1  & 
\color{lightgray}n.s.  & \color{lightgray}2.6 & \color{lightgray}n.s.  \\
& Log[Time Taken] & \color{lightgray}2.9  & 
\color{lightgray}n.s.  & 11.0 & .0009 \\
& XAI Domain Gap $\times$ Log[Time Taken] & 5.8  & .0160 & \color{lightgray}3.8 &  \color{lightgray}n.s. \\

\arrayrulecolor{lightgray}
\midrule
\arrayrulecolor{black}

\multirow{11}{*}{Log WoA}    
& XAI Type +  & 3.9 & .0108 & \color{lightgray}3.1 & \color{lightgray}.0304 \\
& Domain ID + & \color{lightgray}<.1 & \color{lightgray}n.s. & 27.7 & $<$.0001  \\
& w/ or w/o XAI + & 190.2 &  $<$.0001 & 145.9 & $<$.0001\\
& XAI Domain Gap + & \color{lightgray}4.8  & \color{lightgray}.0324 & 19.5  &  $<$.0001\\
& XAI Type $\times$ Domain ID & \color{lightgray} 0.5&  \color{lightgray}n.s.  & 9.8 & $<$.0001 \\
& XAI Type $\times$ XAI Domain Gap & \color{lightgray}3.2 & \color{lightgray}.0218 & \color{lightgray}2.6 & \color{lightgray}n.s. \\
& Domain ID $\times$ XAI Domain Gap & \color{lightgray}0.5  & 
\color{lightgray}n.s. & 9.4 & .0022\\
& XAI Type $\times$ Domain ID $\times$ XAI Domain Gap & \color{lightgray}1.9 & 
\color{lightgray}n.s.  & 3.6 & .0121\\
& XAI Domain Gap $\times$ w/ or w/o XAI & \color{lightgray}0.8  & 
\color{lightgray}n.s.  & 18.1 & $<$.0001 \\
& Log[Time Taken] & \color{lightgray}3.1  & 
\color{lightgray}n.s.  & 8.5 & .0035\\
& XAI Domain Gap $\times$ Log[Time Taken] & \color{lightgray}0.9  & \color{lightgray}n.s. & \color{lightgray}0.9 & \color{lightgray}n.s. \\

\midrule
Log APE of $\bm{w}_O$ Recall & XAI Type + & \color{lightgray}1.9  & \color{lightgray}n.s. & 4.2 & .0077 \\
Log APE of $\bm{w}_T$ Recall & XAI Type + & 6.0  & $<$.0001 & 14.4  & $<$.0001 \\
\arrayrulecolor{lightgray}
\midrule
\arrayrulecolor{black}
\makecell[l]{Magnitude Relation Accuracy} & XAI Type + & 15.9  & $<$.0001 & 11 & $<$.0001 \\
\makecell[l]{Direction Relation Accuracy}  & XAI Type +& 4.1  & .0081 & 14.4  & $<$.0001 \\
\midrule
Perceived Helpfulness (General) & XAI Type + & \color{lightgray}0.4 & \color{lightgray}n.s. & \color{lightgray}1.6 & \color{lightgray}n.s. \\
Perceived Ease-of-Use (General) & XAI Type  + & \color{lightgray}1.4 & \color{lightgray}n.s. & \color{lightgray}0.6 & \color{lightgray}n.s.\\
Perceived Helpfulness (Original) & XAI Type + & \color{lightgray}1.0 & \color{lightgray}n.s. & \color{lightgray}1.6 & \color{lightgray}n.s. \\
Perceived Ease-of-Use (Original) & XAI Type + & \color{lightgray}0.4 & \color{lightgray}n.s. & \color{lightgray}0.7 & \color{lightgray}n.s.\\
Perceived Helpfulness (Target) & XAI Type + & 4.7 & .0036  & \color{lightgray}1.4 & \color{lightgray}n.s.  \\
Perceived Ease-of-Use (Target) & XAI Type + & \color{lightgray}3.2 & \color{lightgray}.0251 & \color{lightgray}0.3 & \color{lightgray}n.s.\\
Perceived Relation Clarity & XAI Type + & 3.6 & .0146 & \color{lightgray}1.3 & \color{lightgray}n.s. \\

\arrayrulecolor{black}
\bottomrule
\end{tabular}
\end{table*}

\subsubsection{Statistical Analysis}
We fitted linear mixed-effects models for each measure as the response, with XAI type (for the Target domain), 
Domain ID (indicating whether the instance belongs to the Original or Target domain), 
XAI presence (with or without XAI), 
XAI Domain Gap\footnote{The difference between the predicted XAI labels of the Original and Target domains.
We found that the discrepancy in predictions between the Original domain XAI and Target domain XAI influenced whether participants considered the Target XAI. Thus, we modeled this as a confounder.
For simplicity, we binarize this to close ($\Delta \tilde{y}_{OT} \geq \text{Median}$) and far ($\Delta \tilde{y}_{OT} < \text{Median}$).} ($\Delta \tilde{y}_{OT} = |\tilde{y}_T - \tilde{y}_O| \geq \text{Median}$), and Time Taken as fixed effects. 
We also included interaction effects among some of these factors, while Participant ID and Case ID were modeled as random effects. We further performed post-hoc contrast tests for specific comparisons.
We applied the Benjamini–Hochberg procedure~\cite{Benjamini1995} to control the false discovery rate (FDR) at 0.05 across 70 comparisons. 
Accordingly, results with $p\le .0205$ were considered statistically significant. 
\begin{figure}[t]
    \centering
    \includegraphics[width=0.95\linewidth]{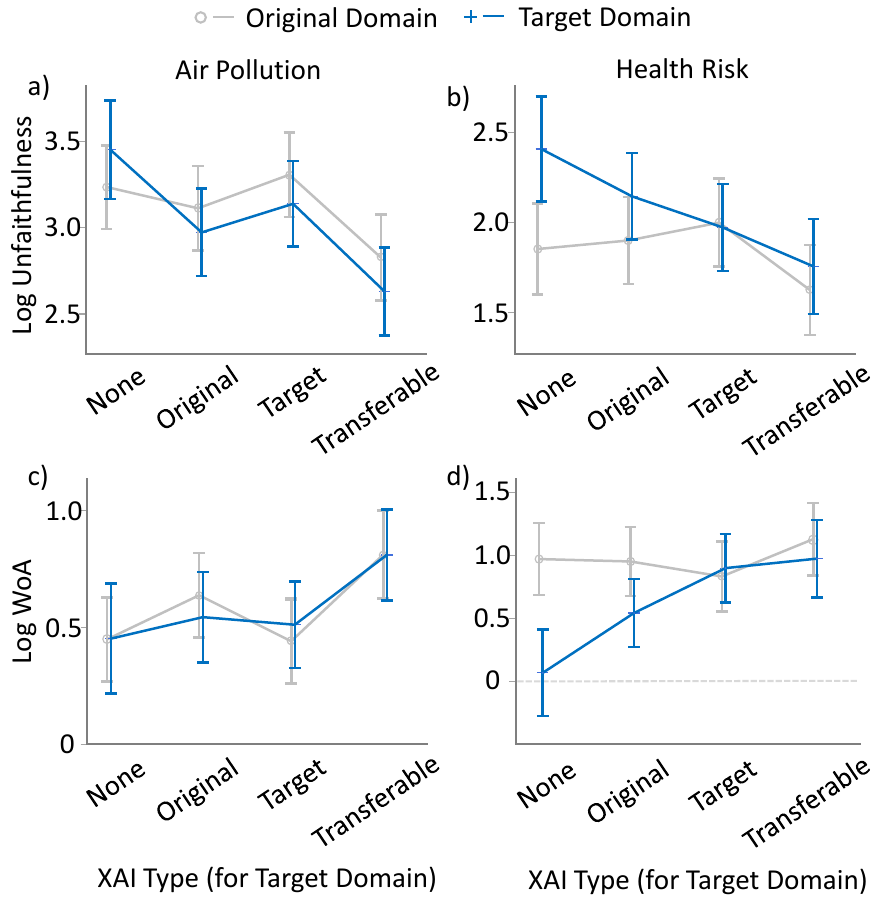}
    \caption{Results from \textbf{Task Transfer} forward simulation trials on Decision Faithfulness, evaluated by (a--b) users’ response unfaithfulness ($\downarrow$) in estimating the AI Explainer’s output $\tilde{y}$, and (c--d) Weight of Advice ($\uparrow$). Blue lines represent the Target domain, while gray lines represent the Original Domain. Error bars indicate 95\% confidence interval.}
    \Description{The figure shows four line plots arranged in a 2×2 grid, labeled a) through d). 
        A legend at the top indicates two conditions: ``Original Domain'', shown with a grey line and open circle markers, and ``Target Domain'', shown with a blue line and plus markers. 
        Subfigure a), titled ``Air Pollution'', plots ``Log Unfaithfulness'' on the y-axis against ``XAI Type (for Target Domain)'' on the x-axis. 
        The x-axis categories are ``None'', ``Original'', ``Target'', and ``Transferable''. 
        Both lines show a decrease from ``None'' to ``Original'', a slight increase at ``Target'', and a decrease at ``Transferable'', with vertical error bars at each point. 
        Subfigure b), titled ``Health Risk'', also plots ``Log Unfaithfulness'' on the y-axis against ``XAI Type (for Target Domain)''. 
        The same four x-axis categories are used. 
        Both Original Domain and Target Domain lines show an overall decreasing trend from ``None'' to ``Transferable'', with error bars displayed at all points. 
        Subfigure c) plots ``Log WoA'' on the y-axis against ``XAI Type (for Target Domain)''. 
        The x-axis categories are ``None'', ``Original'', ``Target'', and ``Transferable''. 
        Both lines increase overall from ``None'' to ``Transferable'', with small fluctuations between ``Original'' and ``Target'', and error bars shown for each condition. 
        Subfigure d) also plots ``Log WoA'' on the y-axis against ``XAI Type (for Target Domain)''. 
        The same four x-axis categories are used. 
        The Original Domain line starts higher at ``None'' and remains relatively stable across categories, while the Target Domain line starts lower at ``None'' and increases steadily toward ``Transferable''. 
        Error bars are shown for all data points.}
    \label{fig:summativestudy_task_forward}
\end{figure}

\begin{figure}[t] 
    \centering
    \includegraphics[width=.95\linewidth]{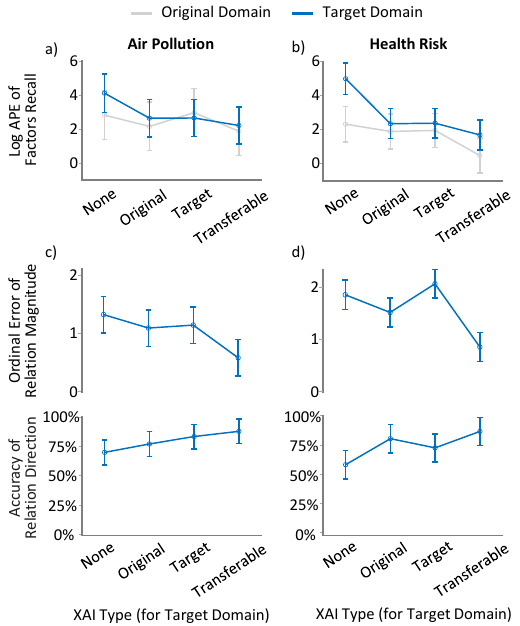}
    \caption{Results from \textbf{Task transfer} explanation recall trials. (a--b) Factor recall Absolute Percentage Error (APE $\downarrow$) for the Original Domain factors $\bm{w}_O$ (gray) and Target Domain factors $\bm{w}_T$ (blue). 
    (c--d) Factor Relation Magnitude, evaluated by Ordinal Error ($\downarrow$), and Accuracy of Relation Direction ($\uparrow$). Error bars indicate 95\% confidence intervals.} 
    \Description{This figure contains four line graphs labeled (a) through (d).
    The left column shows results for the Air Pollution application, and the right column shows results for the Health Risk application. Each graph compares performance across four explanation types on the x-axis: None, Original, Target, and Transferable. 
    Gray lines represent the Original Domain, and blue lines represent the Target Domain. Error bars indicate 95 percent confidence intervals.
    Graph (a): Air Pollution---Factors Recall. Y-axis label: Log Absolute Percentage Error (APE) of Factors Recall. Y-axis range: 0 to 6, increasing upward in steps of 2. Approximate mean values: 
    1) None: Original Domain 2.0, Target Domain 4.5 2) Original: Original Domain 2.2, Target Domain 2.5
    3) Target: Original Domain 2.3, Target Domain 2.5
    4) Transferable: Original Domain 2.0, Target Domain 2.2
    Error bars are approximately plus or minus 0.6 for all conditions.
    The blue Target Domain line decreases sharply from 4.5 to about 2.2, showing improvement (lower error) with Transferable explanations. The gray Original Domain line remains roughly flat near 2.0 across conditions.
    Graph (b): Health Risk---Factors Recall.
Y-axis label: Log Absolute Percentage Error (APE) of Factors Recall. Y-axis range: 0 to 6, increasing upward in steps of 2. Approximate mean values:
1) None: Original Domain 2.5, Target Domain 5.0
2) Original: Original Domain 2.0, Target Domain 2.5
3) Target: Original Domain 2.0, Target Domain 2.2
4) Transferable: Original Domain 1.5, Target Domain 2.0.
Error bars are approximately plus or minus 0.7 for all conditions.
The Target Domain (blue) shows a strong decrease from about 5.0 for None to about 2.0 for Transferable, indicating much better recall accuracy. The Original Domain (gray) stays lower and relatively stable between 1.5 and 2.5.
Graph (c): Air Pollution --- Factor Relation.
Y-axis label: Ordinal Error of Relation Magnitude (top) and Accuracy of Relation Direction (bottom axis).
The upper half of the y-axis represents ordinal error (0 to 2 in increments of 0.5), and the lower half represents direction accuracy in percentage (0\% to 100\% in increments of 25\%).
Approximate mean values for magnitude error (upper line, Target Domain only):
None 1.8, Original 1.5, Target 1.4, Transferable 1.0
Approximate mean values for direction accuracy (lower line, Target Domain only):
None 60\%, Original 70\%, Target 80\%, Transferable 90\%.
Error bars are approximately plus or minus 0.3 for both measures.
Both metrics show improvement: ordinal error decreases from 1.8 to 1.0, and direction accuracy increases from 60\% to 90\%, indicating that Transferable explanations improve users’ understanding of relation magnitude and direction.
---
Panel (d): Health Risk --- Factor Relation.
Y-axis label: Ordinal Error of Relation Magnitude (top) and Accuracy of Relation Direction (bottom axis).
The upper axis ranges from 0 to 2, the lower axis from 0\% to 100\%.
Approximate mean values for magnitude error (upper line, Target Domain only):
None 1.3, Original 1.6, Target 2.0, Transferable 1.2
Approximate mean values for direction accuracy (lower line, Target Domain only):
None 60\%, Original 55\%, Target 70\%, Transferable 80\%.
Error bars are approximately plus or minus 0.3 for all measures.
Magnitude error first increases and then drops again for Transferable, while direction accuracy consistently rises from 60\% to 80\%. This indicates that although intermediate explanations may confuse users, transferable explanations ultimately yield better relation understanding.}
    \label{fig:summativestudy_task_recall}
\end{figure}

Table~\ref{tab:task_transfer_hypo_and_finding} summarizes the model fits with respect to our hypotheses. We present the results for each measure. Table~\ref{tab:task_transfer_summative_statistical_analysis} summarizes the significant effects found.

\subsubsection{Quantitative Results}
In general, Transferable XAI best improved understanding and appropriate domain alignment.
Table~\ref{tab:task_transfer_hypo_and_finding} summarizes our hypotheses and corresponding results.
See Table~\ref{tab:task_transfer_summative_statistical_analysis} for effect test significances.

\begin{itemize}
    \item \textit{Decision Unfaithfulness.} Fig.~\ref{fig:summativestudy_task_forward}a,b shows that participants who received explanations estimated the AI Explainer’s output in Target domain more accurately ( lower Unfaithfulness). 
    \item \textit{Weight of Advice.}  Fig.~\ref{fig:summativestudy_task_forward}c,d shows that participants with \textit{Transferable} explanations made decisions more aligned with the explanations for the Target domain.
\end{itemize}

\begin{itemize}
    \item \textit{Factors Recall.} Fig.~\ref{fig:summativestudy_task_recall}a, b shows the results of recalling factors. 
    Participants with \textit{Transferable} explanations recalled factors better than with no explanation \textit{(None)} or with \textit{Original/Target} Domain XAI.
    \item \textit{Factor Relation.} Fig.~\ref{fig:summativestudy_task_recall}c, d shows that participants with \textit{Transferable} explanations achieved significantly better understanding, with lower ordinal error in relation magnitude and higher accuracy in identifying relational direction across prediction tasks.
\end{itemize}

\begin{figure*}[t] 
    \centering
    \includegraphics[width=0.92\linewidth]{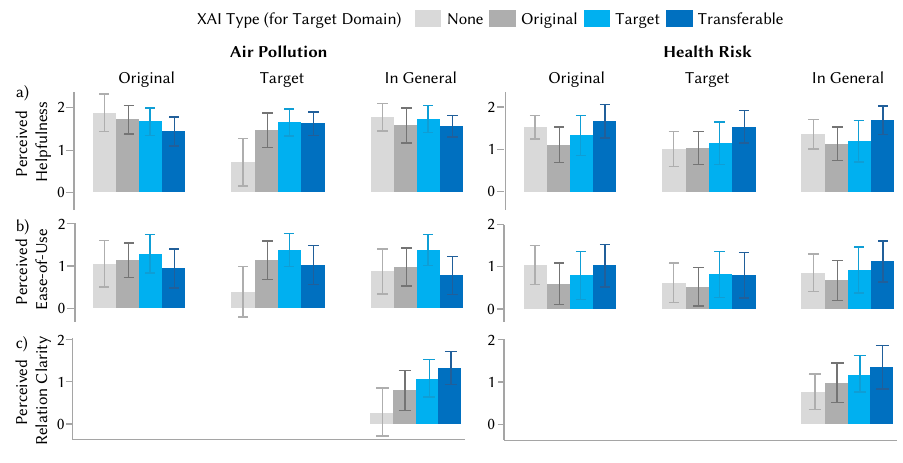}
    \caption{Results of a) Perceived Helpfulness, b) Perceived Ease-of-Use, and c) Perceived Relation Clarity across different XAI types (for Target Domain) in \textbf{Task transfer}, with ratings grouped by scope: Original (left), Target (middle), and In General (right). 
    Bars show mean ratings, and error bars represent 95\% confidence interval.}
    \Description{This figure contains six groups of bar charts arranged in two applications: Air Pollution on the left and Health Risk on the right. 
        Each application has three subplots labeled (a), (b), and (c). Each subplot contains three grouped sets of bars corresponding to rating scopes: Original, Target, and In General. Within each group, there are four colored bars representing different XAI types for the Target Domain: light gray for None, dark gray for Original, light blue for Target, and dark blue for Transferable.
        The y-axis values range from 0 to 2 in increments of 0.5. All bars show mean user ratings with vertical error bars representing 95 percent confidence intervals.
        a) Perceived Helpfulness. Y-axis label: Perceived Helpfulness. Across all groups, bar heights are approximately between 1.3 and 2.0.
        For Air Pollution application, 
        Original group: None 1.4, Original 1.8, Target 1.7, Transferable 1.8. 
        Target group: None 1.3, Original 1.6, Target 1.8, Transferable 1.8. 
        In General group: None 1.4, Original 1.7, Target 1.9, Transferable 2.0.
        For Health Risk, 
        Original group: None 1.2, Original 1.6, Target 1.8, Transferable 1.9. Target group: None 1.2, Original 1.5, Target 1.8, Transferable 2.0. In General group: None 1.3, Original 1.6, Target 1.8, Transferable 1.9. Error bars for all bars are approximately plus or minus 0.4.
        b) Perceived Ease-of-Use. Y-axis label: Perceived Ease-of-Use. Ratings are generally lower than in panel (a), with values between 0.6 and 1.5. 
        For the Air Pollution application, 
        Original group: None 0.8, Original 1.0, Target 1.2, Transferable 1.3. 
        Target group: None 0.6, Original 0.9, Target 1.1, Transferable 1.3. 
        In General group: None 0.8, Original 1.0, Target 1.2, Transferable 1.4. 
        For the Health Risk application, 
        Original group: None 0.6, Original 0.9, Target 1.1, Transferable 1.3. 
        Target group: None 0.7, Original 1.0, Target 1.2, Transferable 1.4. 
        In General group: None 0.8, Original 1.1, Target 1.2, Transferable 1.3. 
        Error bars are wide, approximately plus or minus 0.5.
        c) Perceived Relation Clarity. Y-axis label: Perceived Relation Clarity. Ratings range from 0.5 to 1.8. 
        For Air Pollution, 
        Original group: None 0.8, Original 1.0, Target 1.3, Transferable 1.6. 
        Target group: None 0.6, Original 0.9, Target 1.2, Transferable 1.7. 
        In General group: None 0.8, Original 1.1, Target 1.4, Transferable 1.8. 
        For Health Risk, 
        Original group: None 0.7, Original 1.0, Target 1.3, Transferable 1.7.
        Target group: None 0.8, Original 1.0, Target 1.4, Transferable 1.8. 
        In General group: None 0.9, Original 1.1, Target 1.4, Transferable 1.7. 
        Error bars are approximately plus or minus 0.4.}
    \label{fig:summative_tasktransfer_perceivedratings}
\end{figure*}

\begin{itemize}
    \item \textit{Perceived Helpfulness and Ease-of-Use.} 
% shows the results for perceived helpfulness and ease-of-use. 
    Fig.~\ref{fig:summative_tasktransfer_perceivedratings}a shows no significant differences between XAI Types when explaining instances in the Original domain with Original XAI. 
    For explaining Target domain instances in the Air Pollution application, showing any XAI Type was more helpful than None ($p = .0036$). 
    However, we found no significant differences for the Health Risk application, likely because its higher complexity made it challenging for participants to manage within a lab study in a limited time. 
    No significant differences across XAI Types for Perceived Ease-of-Use.
    % \item \textit{Perceived Ease-of-Use.}There were no significant differences across XAI Types.
    \item \textit{Perceived Relation Clarity.} Fig.\ref{fig:summative_tasktransfer_perceivedratings}c shows that Transferable XAI best helped to clarify the relation between domain factors, though only for the Air Pollution ($p =.0146$).
\end{itemize}

\section{Attributes Transfer Experiment}
We evaluated Transferable explanations against two baselines (None and Target) to examine:
i) usage strategies to interpret the AI decisions across different attributes in \rev{a qualitative formative user study}, 
ii) impact on decision faithfulness, explanation recall, and relations understanding in \rev{a quantitative summative user study}.
The Original condition is excluded to avoid data leakage from identical tasks and overlapping attributes.
Such a setup would not reflect a realistic situation.

\subsection{Experiment Design for Attributes Transfer}
\label{sec:attributes_transfer_exp_design}
The study follows the experiment design described in Section~\ref{sec:all_exp_design} and the procedure outlined in Section~\ref{sec:all-exp_procedure}. 
but with additional specific measures.
\rev{Specifically, we clarify the transfer-specific measures of Transfer Relation for attribution transfer.}

\subsubsection{Measures on Transfer Relation (Attributes Correlation)}

We evaluate how participants perceived the attributes correlation $M_{\chi}$ between the Target and the Original domains. 
Each element in $M_{\chi}$ represents the mapping relation between one attribute in the Target domain and one in the Original domain.   
To simplify responses, participants rated each relation on five ordinal levels: strongly positive, positive, neither positive nor negative, negative, and strongly negative. 
We quantified their understanding by computing the ordinal error between their responses $\mathring{M}_{\chi}$ and the actual correlation.

\subsection{Experiment Apparatus for Attributes Transfer}
\label{sec:experiment-apparatus-target-attributes}
\begin{figure*}[t] 
    \centering
    \includegraphics[width=1\linewidth]{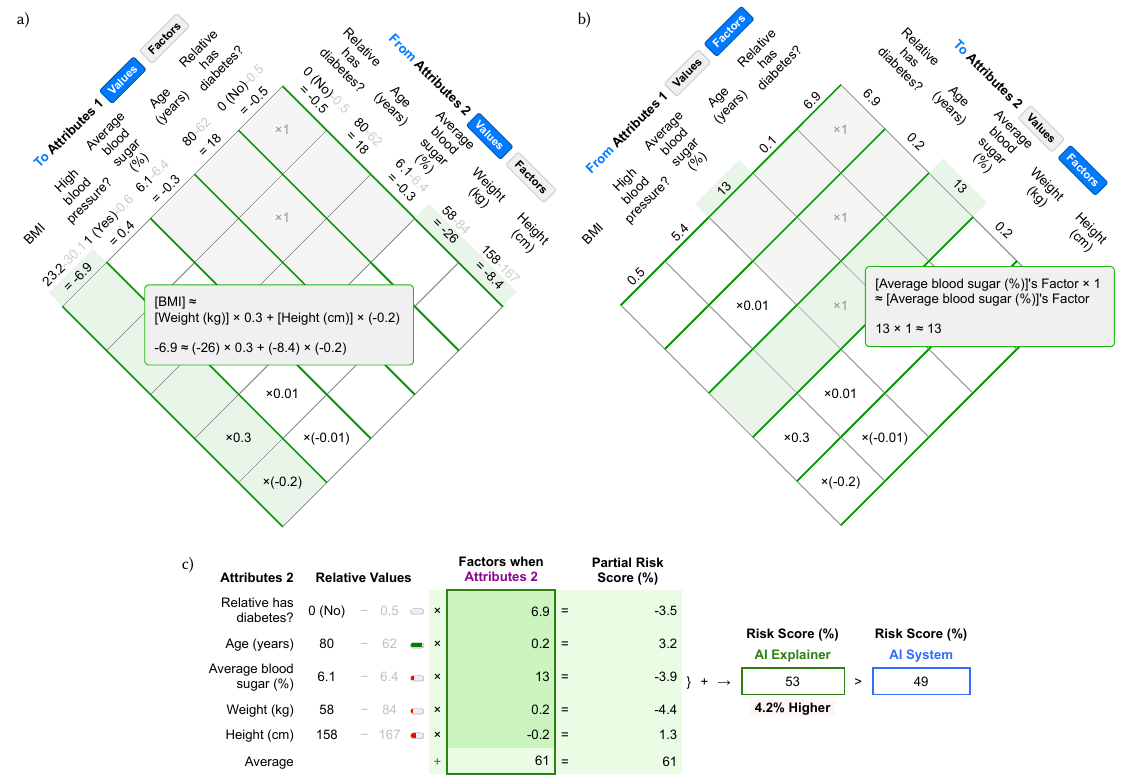}
    \caption{UI of the Transferable explanation in \textbf{Attributes transfer}. 
    a) Map Target attribute values to the corresponding Original attribute value. Matching row and attribute values are highlighted when the mouse hovers, with tooltips displaying the transformation formula (BMI in this case). 
    b) Map Original attribute factors to the corresponding Target attribute factor. Matching column and attribute factors are highlighted when the mouse hovers, with tooltips displaying the transformation formula (Average blood sugar in this case). 
    c) Tabular interface, similar to the one introduced previously for the Target domain.}
    \Description{This figure shows the user interface of the Transferable AI Explainer in the Attributes transfer. 
    It consists of three panels labeled a, b, and c.
    Panel a illustrates how the system maps Target attribute values to their corresponding Original attribute values. 
    The diamond-shaped matrix lists Target attributes along the left column (“To Attributes”) --- BMI, High blood pressure, 
    Average blood sugar, Age, Relative has diabetes, Weight, and Height --- and Original attributes along the top row (“From Attributes”). 
    Each cell shows a numeric mapping coefficient describing how one attribute contributes to another. 
    The diagonal cells have “×1” entries, showing one-to-one correspondences. 
    Off-diagonal values represent cross-attribute influences: for example, BMI depends on Weight and Height, with coefficients ×0.3 and ×(–0.2) respectively. 
    A tooltip appears when hovering over the BMI cell, displaying the explicit transformation formula: “BMI = Weight (kg) × 0.3 + Height (cm) × (–0.2)”. 
    Below the formula, an example substitution is shown: “–6.9 = (–26) × 0.3 + (–8.4) × (–0.2)”, illustrating how relative attribute differences combine to yield the predicted BMI deviation.
    Other small coefficients are shown, such as ×0.01, ×0.3, ×(–0.01), and ×(–0.2), capturing weaker or inverse effects.
    Panel b displays the reverse mapping, from Original attribute factors to Target attribute factors. The layout mirrors panel a but in transposed form: “From Attributes 1” (Original domain) are on the left, and “To Attributes 2” (Target domain) are on the top. 
    Diagonal values again show “×1” correspondences, while off-diagonal cells show transformation coefficients such as ×0.01, ×0.3, and ×(–0.2). 
    When hovering over a matching cell, a tooltip reveals the formula used in factor translation; for instance, the highlighted cell shows: “Average blood sugar (\%)’s Factor × 1 = Average blood sugar (\%)’s Factor,” confirming direct mapping. An example numerical annotation shows “13 × 1 = 13,” demonstrating the transformation from Original to Target factors.    
    Panel c presents a tabular interface summarizing the resulting Target-domain prediction using five attributes: 
    Relative has diabetes, Age, Average blood sugar, Weight, and Height. 
    The Relative Values column lists deviations from the population averages:
    Relative has diabetes = 0 (No) – 0.5, 
    Age = 80 – 62, Average blood sugar = 6.1 – 6.4, Weight = 58 – 84, Height = 158 – 167. The Factors column (corresponding to “Attributes 2”) lists: 6.9 for Relative has diabetes, 0.2 for Age, 13 for Average blood sugar, 0.2 for Weight, and –0.2 for Height. The resulting Partial Risk Scores (Relative × Factor) are: –3.5 for Relative has diabetes, 3.2 for Age, –3.9 for Average blood sugar, –4.4 for Weight, and 1.3 for Height. The Average factor sum is 61.
    On the right, the AI Explainer estimates a Risk Score of 53, while the AI System’s prediction is 49. The difference is highlighted as “4.2 \% Higher,” indicating that the explainer slightly overestimates the system’s score.
    Overall, the figure demonstrates how attribute mappings and factor transformations allow a model trained on one attribute set to generate interpretable explanations in another set, while maintaining transparency of both numerical relations and prediction accuracy.}
    \label{fig:ui_transferable_xai_attributes}
\end{figure*}

Fig.~\ref{fig:ui_transferable_xai_attributes} shows the explanation UI for Transferable XAI with Attributes transfer. 
\rev{It consists of two components: 
i) the main tabular UI (Fig.~\ref{fig:ui_transferable_xai_attributes}a) showing the linear factors explanation just like for Subspace (Incremental XAI~\cite{bo2024incremental}) and Task (Section~\ref{sec:experiment-apparatus-target-task}) transfer; and
ii) a matrix mapping attribute values from the Target domain to Original domain (Fig.~\ref{fig:ui_transferable_xai_attributes}a), or, in reverse, of attribute factors from the Original domain to Target domain (Fig.~\ref{fig:ui_transferable_xai_attributes}b).}

\subsubsection{\rev{User Interface Design Process}}
\rev{Due to the complexity of conveying the mapping relation between the attributes from the Original and Target domains, we went through a few design iterations with cognitive walkthrough and pilot feedback. 
Inspired by research on visualizing correlations~\cite{zhang2014visual}, we considered matrix and bipartite graph visualizations.
Specifically, we explored: embedded matrix, bipartite graph, rotated matrix.

First, we considered adding columns to the tabular UI, with new columns for each attribute in the Target domain. 
Together with each row for the Original domain's attributes, this presents the matrix embedded within the table. 
However, this made the the table confusing (since columns now also include attributes like the rows), and pilot users found the UI overwhelming. 
We then considered separating the matrix from the table.
This also helped with separation of concerns, where users can use the mapped attributes to determine partial scores in the Target domain without being burdened with how they are computed from the Original domain.

With a separate component for the attribute/factor mapping, we considered a streamline representation with a bipartite graph.
This shows the Original attributes in a column on the left, the Target attributes on the right, and edges connecting each attribute on the left to each on the right. An edge is thicker if attributes in a pair is more correlated.
However, due to many edges crisscrossing, this led to messy occlusion, and pilot users reported being unfamiliar and uncomfortable with such a visualization.
Hence, we focused on improving the matrix representation.

In Eq.~\ref{xai_affine_transformation_attributes}, we had determined that attribute value mapping is the transpose of attribute factor mapping.
Also, as found from our pilot users, it is more intuitive to understand value mapping than factor mapping.}
To facilitate reading the mapping in either direction from Original to Target or in reverse, we rotate the matrix $45^\circ$ in a diamond orientation.
\rev{To interpret the equivalence in mapping values and factors,} the user can select one of two modes to use the matrix: 
a) for mapping the Target attribute values to the Original attribute values, and
b) for mapping the Original factors to the Target factors.
\rev{However, some pilot users struggled with whether to read left-to-right or right-to-left.
Hence,} we implemented interactivity where hovering on a matrix cell will \rev{highlight a row of mapping} attribute values (or factors) from attributes of one domain to be combined in a weighted sum  to relate to an attribute value (or factor) in the other domain.
\rev{We include a tooltip to explicitly show the math of the weighted sum for further clarity.
This final UI (see Fig.~\ref{fig:ui_transferable_xai_attributes}) was adequate for our user study where all participants could understand and use the UI without great difficulty or confusion\footnote{\rev{We had conducted our formative and summative user studies face-to-face, such that the experimentalists could provide minor clarifications}.}.}

\subsection{Formative User Study}
We examined how users use and interpret the explanations across different sets of attributes.

\subsubsection{Participants}
We recruited 14 university students (mean age = 23, range = 18--32; 6 female, 8 male) from diverse disciplines, including computer science, public administration, nursing, and data science. 

\begin{table*}[htbp]
\caption{Statistical analysis of responses using linear mixed-effects models for \textbf{Attributes transfer}. 
Each row reports one effect, including fixed effects, random effects, and their interactions.
XAI Type denotes the explanation type applied in the Target domain.
$F$ and $p$ values are from ANOVA tests.}
\label{tab:attributes_transfer_summative_statistical_analysis}
\renewcommand{\arraystretch}{1}
\begin{tabular}{llrrrr}
\toprule
\multirow{2}{*}{Response} & \makecell[l]{Linear Effect Model} & \multicolumn{2}{c}{Air pollution} & \multicolumn{2}{c}{Health risk}  \\

 & \makecell[l]{(Random Effects: Participants, Case)} & $F$ & $p > F$ & $F$ & $p > F$  \\
\midrule
\multirow{11}{*}{Log UnFaithfulness}    
& XAI Type +  & 4.2 & .0193 & 3.4 & .0379 \\
& Domain ID + & 32.9 & $<$.0001 & \color{lightgray}2.7 & \color{lightgray}n.s. \\
& w/ or w/o XAI + & 490.9 &  $<$.0001 & 50.1 & $<$.0001\\
& XAI Domain Gap + & \color{lightgray}1.9  & \color{lightgray}n.s. & <.1 & \color{lightgray}n.s. \\
& XAI Type $\times$ Domain ID & 15.8 & $<$.0001 & 8.2 & .0003 \\
& XAI Type $\times$ XAI Domain Gap & \color{lightgray}0.4 & \color{lightgray}n.s. & \color{lightgray}0.3 & \color{lightgray}n.s.  \\
& Domain ID $\times$ XAI Domain Gap & \color{lightgray}2.1  & 
\color{lightgray}n.s. &\color{lightgray}<.1  & \color{lightgray}n.s. \\
& XAI Type $\times$ Domain ID $\times$ XAI Domain Gap & \color{lightgray}2.6 & 
\color{lightgray}n.s.  & \color{lightgray}0.8 & \color{lightgray}n.s. \\
& XAI Domain Gap $\times$ w/ or w/o XAI & 15.8  & 
$<$.0001  & \color{lightgray}0.3 & \color{lightgray}n.s.  \\
& Log[Time Taken] & \color{lightgray}3.6  & 
\color{lightgray}n.s.  & 26.2 & $<$.0001 \\
& XAI Domain Gap $\times$ Log[Time Taken] & \color{lightgray}3.7  & \color{lightgray}n.s. & \color{lightgray}0.6 &  \color{lightgray}n.s. \\
\arrayrulecolor{lightgray}\midrule
\arrayrulecolor{black}

\multirow{11}{*}{Log WoA}    
& XAI Type +  & \color{lightgray}2.7 & \color{lightgray}n.s. & \color{lightgray}0.8 & \color{lightgray}n.s. \\
& Domain ID + & \color{lightgray}<.1 & \color{lightgray}n.s. & \color{lightgray}0.1 & \color{lightgray}n.s. \\
& w/ or w/o XAI + & 27.2 &  $<$.0001 & 7.1 & .0076\\
& XAI Domain Gap + & 15.6  & $<$.0001 & 26.1  &  $<$.0001\\
& XAI Type $\times$ Domain ID & \color{lightgray} 0.7&  \color{lightgray}n.s.  & \color{lightgray}0.8 & \color{lightgray}n.s. \\
& XAI Type $\times$ XAI Domain Gap & 3.5 & .0302 & \color{lightgray}1.2 & \color{lightgray}n.s. \\
& Domain ID $\times$ XAI Domain Gap & \color{lightgray}0.3  & 
\color{lightgray}n.s. & \color{lightgray}0.3 & \color{lightgray}n.s.\\
& XAI Type $\times$ Domain ID $\times$ XAI Domain Gap & \color{lightgray}2.2 & 
\color{lightgray}n.s.  & \color{lightgray}0.2 & \color{lightgray}n.s.\\
& XAI Domain Gap $\times$ w/ or w/o XAI & \color{lightgray}3.3  & 
\color{lightgray}n.s.  & 8.4 & .0037 \\
& Log[Time Taken] & \color{lightgray}1.8  & 
\color{lightgray}n.s.  & \color{lightgray}1.3 & \color{lightgray}n.s.\\
& XAI Domain Gap $\times$ Log[Time Taken] & 6.3  & .0122 & \color{lightgray}1.2 & \color{lightgray}n.s. \\

\midrule
Log APE of $\bm{w}_O$ Recall & XAI Type + & \color{lightgray}2.3  & \color{lightgray}n.s. & \color{lightgray}0.3 & \color{lightgray}n.s \\
Log APE of $\bm{w}_T$ Recall & XAI Type + & 4.9  & .0105 & 14.0  & $<$.0001 \\
\arrayrulecolor{lightgray}\midrule
\arrayrulecolor{black}
\makecell[l]{Ordinal Error of\\ Attributes Correlation} & XAI Type + & 4.8  & .0106 & \color{lightgray}4.4 & \color{lightgray}.0148 \\
\midrule
Perceived Helpfulness (General) & XAI Type + & \color{lightgray}0.3 & \color{lightgray}n.s. & \color{lightgray}0.6 & \color{lightgray}n.s. \\
Perceived Ease-of-Use (General) & XAI Type + & \color{lightgray}0.2 & \color{lightgray}n.s. & \color{lightgray}0.4 & \color{lightgray}n.s.\\
Perceived Helpfulness (Original) & XAI Type + & \color{lightgray}0.1 & \color{lightgray}n.s. & \color{lightgray}3.2 & \color{lightgray}n.s. \\
Perceived Ease-of-Use (Original) & XAI Type + & \color{lightgray}0.4 & \color{lightgray}n.s. & \color{lightgray}<.1 & \color{lightgray}n.s.\\
Perceived Helpfulness (Target) & XAI Type + & 
\color{lightgray}2.6 & \color{lightgray}n.s.  & \color{lightgray}1.4 
& \color{lightgray}n.s.  \\
Perceived Ease-of-Use (Target) & XAI Type + & \color{lightgray}2.3 & \color{lightgray}n.s. & \color{lightgray}1.2 & \color{lightgray}n.s.\\
Perceived Relation Clarity & XAI Type + & \color{lightgray}2.2 & \color{lightgray}n.s. & \color{lightgray}0.5 & \color{lightgray}n.s. \\

\arrayrulecolor{black}
\bottomrule
\end{tabular}
\end{table*}
\subsubsection{Qualitative Findings}
We analyzed participants' interactions and utterances, and report key findings in terms of XAI Types for the Target domain.

With no explanations (None), participants tended to \textbf{duplicate}
attribute factors from the Original domain to the Target domain. 
P9 thought \textit{``[the factors] are similar to Attributes 1 [the Original domain]''} and duplicated the factors into the Target domain, including non-shared attributes, by copying the factors from the corresponding positions in the Original domain UI. P10 performed similarly but made small adjustments, \textit{``I’m trying to keep them... like 10\%... not a huge difference''}.
Participants inferred the relationship between the attribute values in the Original domain and the Target domain purely based on prior knowledge and guessing.

When presented with the Target domain XAI type, participants could \textbf{recognize similar factors} between corresponding attributes of the two domains, but had to \textbf{memorize unshared attributes independently} or used prior knowledge to help make sense. 
P3 memorized the factors in two domains independently and thought the differences were acceptable because they are \textit{``small, when you see the way it impacts the partial risk score [of the predictions]''}. 
Regarding the unshared attributes, P1 \textit{``knew how BMI relates to height and weight, but high blood pressure seems be irrelevant to other attributes''}. P13 noticed the unshared attribute (Temperature) and thought \textit{``its impact should be positive on PM2.5 [based on my knowledge]... but I don't have sufficient domain knowledge to know how much the attributes influence the results, so I just memorize them''}.
Some participants (P3, P4, P5) inferred the relationships based on the \textbf{factor similarity}, e.g., \textit{``if the factors are similar, then they are highly related''} (P3). 
However, this strategy is spurious because similar factors do not necessarily indicate related attributes. 
Yet, some participants preferred the Target domain XAI type to Transferable XAI, since \textit{``it's \textbf{simpler} for a general user''} (P9).

\begin{table*}[tp]
\caption{Hypotheses and findings from the Attributes transfer summative user study, examining various dependent variables across different XAI types: None (N), Target (T), and Transferable ($\mathcal{T}$), on Air Pollution and Health Risk applications.}
\label{tab:attributes_transfer_hypo_and_finding}
\renewcommand{\arraystretch}{1} 
\begin{tabular}{@{}lrllll@{}}
\toprule
Measure & Domain & Hypothesis & Air Pollution & Health Risk & Evidence \\ 
\midrule
Decision Faithfulness & Original 
&  N $\approx$ T $\approx$ $\mathcal{T}$ 
& N $\approx$ T $<$ $\mathcal{T}$ 
& N $\approx$ T $<$ $\mathcal{T}$ & Fig.~\ref{fig:summativestudy_attributes_forward}a,b (gray)  \\
\color{lightgray} &  Target   
&  N $<$ T $<$ $\mathcal{T}$  
&  N $<$ T \ns{$\approx$} $\mathcal{T}$  
&  N $<$ T \ns{$\approx$} $\mathcal{T}$ &  Fig.~\ref{fig:summativestudy_attributes_forward}a,b (blue)  \\
\midrule

Log WoA &  Original 
&  N  $\approx$ T $\approx$ $\mathcal{T}$  
&  N  $\approx$ T $<$ $\mathcal{T}$  
& N  $\approx$ T $\approx$ $\mathcal{T}$ & Fig.~\ref{fig:summativestudy_attributes_forward}c,d (gray) \\
\color{lightgray} &   Target  
&  N $<$ T $<$ $\mathcal{T}$  
&  N \ns{$\approx$} T \ns{$\approx$} $\mathcal{T}$  
& N \ns{$\approx$} T \ns{$\approx$} $\mathcal{T}$   & Fig.~\ref{fig:summativestudy_attributes_forward}c,d (blue) \\
\midrule

Factors Recall  & Original 
&  N $\approx$ T $\approx$ $\mathcal{T}$  
&  N $\approx$ T $\approx$ $\mathcal{T}$ 
&  N $\approx$ T $\approx$ $\mathcal{T}$ &  Fig.~\ref{fig:summativestudy_attributes_recall}a,b (gray) \\
\color{lightgray} &  Target 
&  N $\approx$ T $<$ $\mathcal{T}$  
&  N \ns{$\approx$} T \ns{$\approx$} $\mathcal{T}$  
&  N $<$ T \ns{$\approx$} $\mathcal{T}$  &Fig.~\ref{fig:summativestudy_attributes_recall}a,b (blue)  \\
\midrule

\makecell[l]{Attributes Correlation } 
&  
&  N $<$ T $<$ $\mathcal{T}$  
&  N \ns{$\approx$} T $<$ $\mathcal{T}$  
&   N \ns{$\approx$} T $<$ $\mathcal{T}$ &Fig.~\ref{fig:summativestudy_attributes_recall}c,d  \\
% \makecell[l]{Relation Direction \\ Understanding} 
% &  
% &  N $<$ T $<$ $\mathcal{T}$  
% &  N $<$ O $\approx$ T \ns{$\approx$} $\mathcal{T}$  
% & N $<$ O $\approx$ T $<$ $\mathcal{T}$  & Figs.~\ref{fig:summativestudy_attributes_recall}g,h  \\

\bottomrule
\end{tabular}
\end{table*}

\begin{figure}[tp] 
    \centering
    \includegraphics[width=0.95\linewidth]{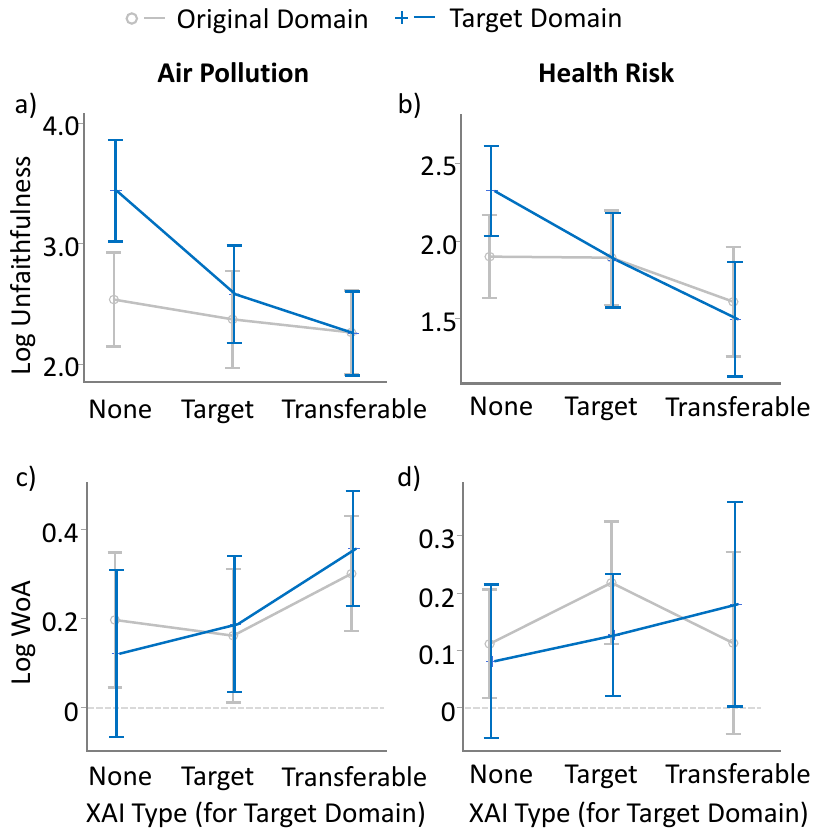}
    \caption{Results from \textbf{Attributes Transfer} forward simulation trials on Decision Faithfulness, evaluated by (a--b) users’ response unfaithfulness ($\downarrow$) in estimating the AI Explainer’s output $\tilde{y}$, and (c--d) Weight of Advice ($\uparrow$). Blue lines represent the Target domain, while gray lines represent the Original Domain. Error bars indicate 95\% confidence interval.}
    \Description{The figure contains four line plots arranged in a 2×2 grid, labeled a) through d). 
        A legend at the top indicates two conditions: ``Original Domain'' shown with a grey line and open circles, and ``Target Domain'' shown with a blue line and plus markers. 
        Subfigure a), titled ``Air Pollution'', plots ``Log Unfaithfulness'' on the y-axis against ``XAI Type (for Target Domain)'' on the x-axis. 
        The x-axis categories are ``None'', ``Target'', and ``Transferable''. 
        Both Original Domain and Target Domain lines decrease across the three XAI types, with vertical error bars shown at each point. 
        Subfigure b), titled ``Health Risk'', also plots ``Log Unfaithfulness'' on the y-axis against ``XAI Type (for Target Domain)'' on the x-axis, with the same three categories. 
        Both lines again show a decreasing trend from ``None'' to ``Transferable'', with error bars displayed for each condition. 
        Subfigure c) plots ``Log WoA'' on the y-axis against ``XAI Type (for Target Domain)'' on the x-axis. 
        The x-axis categories are ``None'', ``Target'', and ``Transferable''. 
        Both Original Domain and Target Domain lines increase across the XAI types, with error bars shown at each point. 
        Subfigure d) also plots ``Log WoA'' on the y-axis against ``XAI Type (for Target Domain)''. 
        The same three x-axis categories are used. 
        The Original Domain line shows a slight increase from ``None'' to ``Target'' and then decreases slightly, while the Target Domain line shows a gradual increase from ``None'' to ``Transferable''. 
        Error bars are shown for all data points.}
   \label{fig:summativestudy_attributes_forward}
\end{figure}

\begin{figure}[tp] 
    \centering
    \includegraphics[width=.97\linewidth]{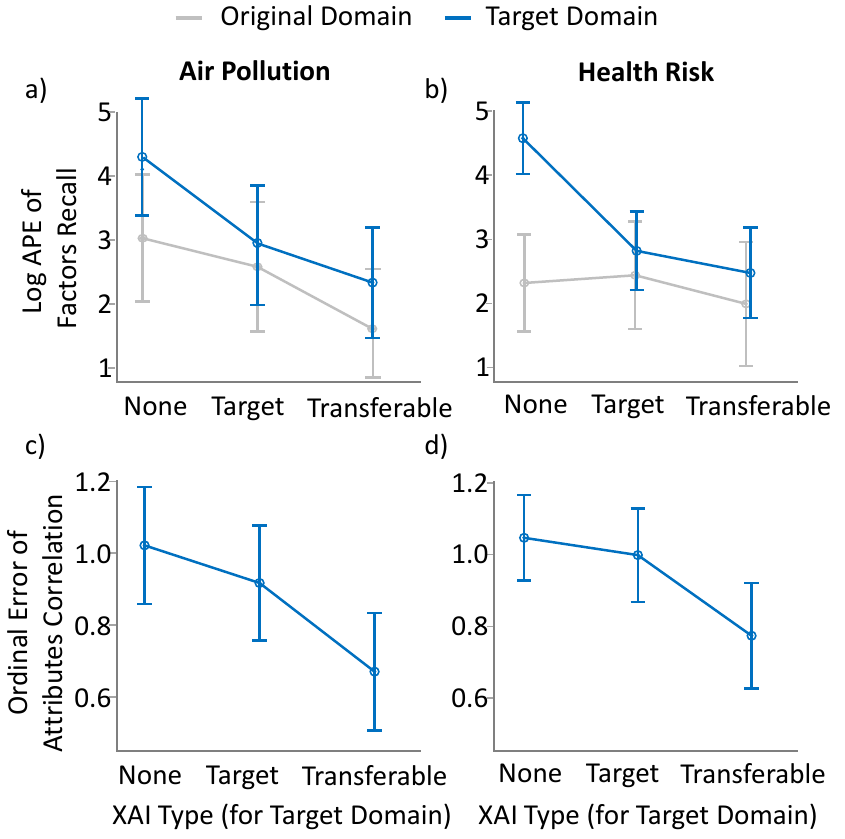}
    \caption{Results from \textbf{Attributes Transfer} explanation recall trials. 
    Factors recall (a–-b): Absolute Percentage Error (APE $\downarrow$) of the Original domain factors $\bm{w}_O$ (gray) and Target domain factors $\bm{w}_T$ (blue). 
    Factor Relation (c--d): Ordinal Error ($\downarrow$) of Attributes Correlation. Error bars indicate 95\% confidence interval.}
    \Description{This figure contains four line graphs (a) through (d), arranged in two columns for the Air Pollution (left) and Health Risk (right) applications. The top row (a–-b) presents Factor Recall results, and the bottom row (c–-d) presents Factor Relation results. 
    Each plot compares explanation types None, Target, and Transferable along the x-axis. The y-axis represents either Log APE (top) or Ordinal Error (bottom). 
    Gray lines with circle markers indicate the Original Domain, while blue lines with cross markers indicate the Target Domain. Error bars denote 95 percent confidence intervals.
    a) Air Pollution — Factors Recall. Y-axis label: Log Absolute Percentage Error (APE) of Factors Recall. Y-axis range: 1 to 5, increasing in steps of 1. X-axis categories: None, Target, Transferable. Approximate mean values:
    1) None: Original Domain 3.0, Target Domain 4.3
    2) Target: Original Domain 2.7, Target Domain 3.0
    3) Transferable: Original Domain 2.3, Target Domain 1.8. Error bars are approximately ±0.5.
    Trend: The Target Domain line (blue) decreases sharply from 4.3 to 1.8, showing improved factor recall with more transferable explanations. The Original Domain (gray) decreases slightly from 3.0 to 2.3.
    b) Health Risk — Factors Recall. Y-axis label: Log Absolute Percentage Error (APE) of Factors Recall. Y-axis range: 1 to 5. Approximate mean values:
    1) None: Original Domain 2.8, Target Domain 4.5
    2) Target: Original Domain 2.6, Target Domain 3.0
    3) Transferable: Original Domain 2.0, Target Domain 2.2. Error bars are approximately ±0.6.
    Trend: Similar to panel (a), the Target Domain line drops steeply from 4.5 to 2.2, indicating a reduction in recall error. The Original Domain line remains steady between 2.8 and 2.0. }
    \label{fig:summativestudy_attributes_recall}
\end{figure}
Six out of eleven participants who used Transferable XAI found the \textbf{relationships shown in the mapping matrix clear}, 
and noted that the identity matrix between shared attributes reduced the amount of numbers to memorize and helped them make estimations. 
For example, \textit{``...with the matrix, I can get the values from one side to another... understandable, for example, BMI is influenced by these two [Weight and Height] and but not by others} (P11), and \textit{``...these three (matrix numbers that equal 1) help to fix some of the big factors here''} (P3). 
Participants appreciated the relational information provided in Transferable XAI; e.g.,
P1 \textit{``prefer [Transferable XAI] when understanding and analyzing two systems, because it provides the numerical relationships, so I don't need to guess. It's more useful when I'm not familiar with the attributes''}.
However, the Transferable XAI can be \textbf{overwhelming} to non-technical participants. P7 (public administration) and P8 (nursing) did not understand the mapping matrix. P2 thought it was \textit{``a good system to predict, but more complicated''}, and P3 commented that \textit{``there are too many numbers on both sides [of the mapping matrix]... if the presentation of the diamond [matrix] can be a bit more tidy [with fewer numbers], I would definitely prefer [Transferable XAI]''}.

\begin{figure*}[t] 
    \centering
    \includegraphics[width=0.92\linewidth]{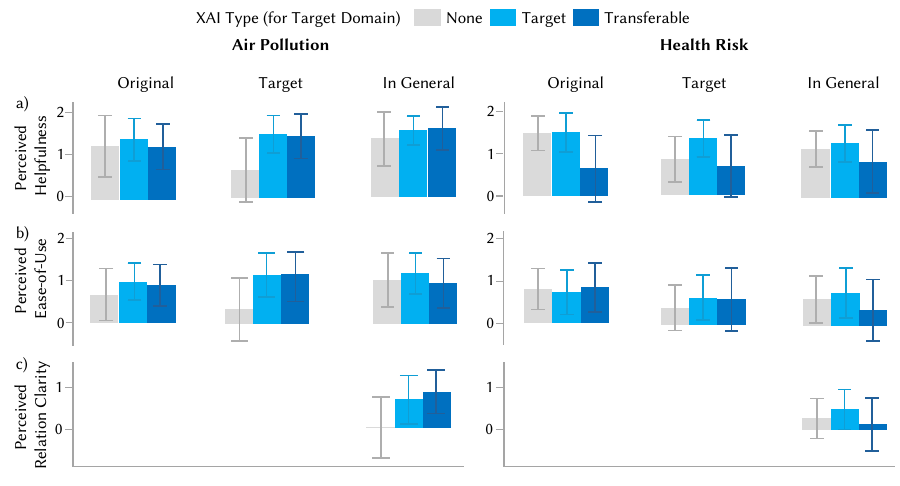}
    \caption{Results of participants’ a) Perceived Helpfulness, b) Perceived Ease-of-Use, and Perceived Relation clarity across different XAI types (for Target Domain) in \textbf{Attributes transfer}, with ratings grouped by scope: Original (left), Target (middle), and In General (right). 
    Bars show mean ratings, and error bars represent 95\% confidence interval.}
    \Description{This figure contains six groups of bar charts arranged in two applications: Air Pollution on the left and Health Risk on the right. Each domain has three subplots labeled (a), (b), and (c), corresponding to Perceived Helpfulness, Perceived Ease-of-Use, and Perceived Relation Clarity. 
    Each subplot includes three grouped bar clusters labeled Original, Target, and In General. 
    Within each cluster, there are three colored bars representing different XAI types (for the Target Domain): light gray for None, light blue for Target, and dark blue for Transferable. The y-axis ranges from 0 to 2 for subplot (a) and (b), and from 0 to 1 for panel (c). All bars display mean user ratings with vertical error bars representing 95 percent confidence intervals.
    Subplot a) Perceived Helpfulness. Y-axis label: Perceived Helpfulness (scale 0 to 2). Air Pollution: 
    Original group — None 1.3, Target 1.5, Transferable 1.5. 
    Target group — None 1.0, Target 1.4, Transferable 1.5.
    In General group — None 1.3, Target 1.6, Transferable 1.7.
    Health Risk:
    Original group — None 1.4, Target 1.6, Transferable 0.8.
    Target group — None 1.3, Target 1.6, Transferable 0.9.
    In General group — None 1.4, Target 1.6, Transferable 0.8.
    Error bars are approximately ±0.4 for all bars.
    Subplot b) Perceived Ease-of-Use. Y-axis label: Perceived Ease-of-Use (scale 0 to 2).
    Air Pollution:
    Original group — None 0.8, Target 1.0, Transferable 1.1.
    Target group — None 0.7, Target 0.9, Transferable 1.0.
    In General group — None 0.8, Target 1.0, Transferable 1.1.
    Health Risk:
    Original group — None 0.8, Target 1.0, Transferable 1.1.
    Target group — None 0.6, Target 0.9, Transferable 1.0.
    In General group — None 0.8, Target 1.0, Transferable 1.1. Error bars are approximately ±0.4.
    Subplot c) Perceived Relation Clarity. Y-axis label: Perceived Relation Clarity (scale 0 to 1).
    Air Pollution:
    Original group — None 0.4, Target 0.6, Transferable 0.8.
    Target group — None 0.2, Target 0.6, Transferable 0.8.
    In General group — None 0.3, Target 0.6, Transferable 0.9.
    Health Risk:
    Original group — None 0.3, Target 0.6, Transferable 0.9.
    Target group — None 0.3, Target 0.6, Transferable 0.8.
    In General group — None 0.4, Target 0.6, Transferable 0.8.
    Error bars are approximately ±0.3.}
    \label{fig:summative_attributestransfer_perceivedratings}
\end{figure*}
\subsection{Summative User Study} 
\label{sec:attributes_summative}
We conducted a summative study to evaluate how well participants understand, recall, and apply explanations and their understanding of attributes correlation across domains.

\subsubsection{Participants}
We recruited workers from Prolific.co of which 222 participants passed screening.
Participants who completed the study had a median age of 41 years (range = 19--76, 52\% female). The survey took a median of 70 minutes, with £7.00 base pay and a median bonus of £1.18 (max £2.90).

\subsubsection{Statistical Analysis}
The analysis approach is the same as that used in Task Transfer, described in Section~\ref{sec:task_summative}. We controlled the false discovery rate (FDR) at 0.05 across 70 comparisons using the Benjamini–Hochberg procedure; results with $p \le 0.0114$ remained statistically significant.

\subsubsection{Quantitative Results}
In general, Transferable XAI best improves understanding and appropriate domain alignment. However, performance in the Health Risk application, which involves more complex transformation information, is generally worse than in Air Pollution.
Table~\ref{tab:attributes_transfer_hypo_and_finding} summarizes our hypotheses and corresponding results.
See Table~\ref{tab:attributes_transfer_summative_statistical_analysis} for effect test significances. 
\begin{itemize}
    \item \textit{Decision Faithfulness.}
    Participants with \textit{Transferable} XAI exhibited lower unfaithfulness on estimating the AI Explainer's output (see Fig.~\ref{fig:summativestudy_attributes_forward}a,b).
    \item \textit{Weight of Advice.} For the Air Pollution application, users with \textit{Transferable} explanations exhibited a higher WoA (see Fig.~\ref{fig:summativestudy_attributes_forward}c).
    However, this difference was not significant in the Health Risk (Fig.~\ref{fig:summativestudy_attributes_forward}d).
    % This may be because participants in a lab study might not have sufficient cognitive capacity to apply such complex information carefully.
    \item \textit{Factors Recall.} 
    Participants with \textit{Transferable} explanations exhibited marginally lower Factor Recall errors in the Original domain and significantly lower errors in the Target domain (see Fig.~\ref{fig:summativestudy_attributes_recall}a,b).
    \item \textit{Attributes Correlation.} 
    Participants with \textit{Transferable} explanations exhibited lower ordinal error on Attributes Correlation  (see Fig.~\ref{fig:summativestudy_attributes_recall}c,d). 
        \item \textit{Perceived Helpfulness} and \textit{Ease-of-Use.} 
    We found no significant differences between XAI Types (see Fig.~\ref{fig:summative_attributestransfer_perceivedratings}a,b). 
    In the Health Risk, the \textit{Transferable} explanation received lower subjective ratings, possibly due to increased cognitive load, which may have been challenging to process carefully in a lab study.
    \item \textit{Perceived Relation Clarity.}     
    We found a similar pattern of no significant differences across XAI types (see Fig.~\ref{fig:summative_attributestransfer_perceivedratings}c), and ratings in Health Risk were slightly lower for \textit{Transferable} explanations, likely due to the information complexity.
\end{itemize}

\section{Discussion \label{sec:discussion}}
Our approach for transferable explanations helps reduce the effort of non-technical AI users to understand multiple relevant AI systems and, in the long term, may contribute to enhancing their overall AI literacy~\cite{toker2025investigating}. We extend the work in~\cite{bo2024incremental}, and generalize to different domain types. 
This provides an initial attempt to address the research gap in explaining multiple AI systems across domains. Both qualitative and quantitative evaluations revealed users’ overgeneralization across multiple AI systems, shedding new light on the often-overlooked issue of overreliance in XAI.
Here, we discuss comparing and combining domain types, and generalizing explanation transfer.

\subsection{Comparing Transfers Across Domain Types}
Although we examined explanation transfers across domain types in separate experiments, results from our formative and summative studies provide some insights.
We found that Task transfer was more effective and less tedious to learn than Attributes transfer.
Since multi-task machine learning is common, this shows the potential of Task transfer in applications.
However, more work is needed to make Attributes transfer more usable. This may not be necessary if different stakeholders interpreting with different attributes never communicate, but this would limit their shared knowledge and discussion.
Furthermore, although not directly examined in our experiments, we posit that Subspace transfer would be easier to understand than Task transfer since users only need to interpret additive difference ($\bm{x}_\Delta$) instead of multiplicative scale ($\bm{\kappa}_\Delta$)~\cite{ashcraft1992cognitive, imbo2008effects}.

\subsection{Compositing Multiple Explanation Transfers}
\label{sec:discussion-compositing-transfers}

A powerful implication of the affine transformation framework of Transferable XAI is the capability to combine the transfer of multiple domain types.
This could be used to explain the relation between two domains that differ in subspace and task (e.g., osteoporosis risk of underweight patients and diabetes risk of overweight patients),
task and attribute (e.g., cancer risk in terms of age, and diabetes risk in terms of weight),
subspace and attribute (e.g., explain obesity risk with BMI for heavier patients, but with weight for lighter patients).

To composite multiple transfers, we first represent the affine transform in homogeneous coordinates
\begin{equation}
    \begin{pmatrix}\bm{w}_T \\ 1\end{pmatrix} = \mathcal{A} \begin{pmatrix}\bm{w}_O \\ 1\end{pmatrix},
    \qquad 
    \mathcal{A} = \begin{pmatrix}
        A & \bm{b} \\
        \bm{0}^\top & 1
    \end{pmatrix},
\end{equation}
where $\bm{w}_T = A\bm{w}_O + \bm{b}$ is the original form for affine transformation, 
$\bm{O}^\top$ is a vector of 0's with the same length as $\bm{w}_O$, and
$A$ is made a square matrix with zero-padding.
Then multiple transformations (e.g., 
$\mathcal{A}_1 = \big(\begin{smallmatrix} A_1 & \bm{b}_1 \\ \bm{0}^\top & 1 \end{smallmatrix}\big)$, then
$\mathcal{A}_2 = \big(\begin{smallmatrix} A_2 & \bm{b}_2 \\ \bm{0}^\top & 1 \end{smallmatrix}\big)$) can be composited with matrix multiplication, i.e.,
\begin{equation}
    \mathcal{A}_2 \mathcal{A}_1 = \begin{pmatrix}
        A_2 A_1 & A_2 \bm{b}_1 + \bm{b}_2 \\
        \bm{0}^\top & 1
    \end{pmatrix}.
\end{equation}

Note that order matters, since matrix multiplication is non-commutative.
This can be determined by human interpretability or usability needs, e.g., 
1) explain by attributes to contextualize with respect to the new terms,
2) explain by task since it is global, then
3) explain by subspace to drill down into specific subsets of cases.
Alternatively, if no order is preferred, different sequences can be attempted, then choose the composite with best explanation faithfulness and simplicity, i.e., minimal overall loss in Eq.~\ref{eq:loss_functions} across all transformations.

Furthermore, 
although the framework supports compositing multiple explanation transfers, this can be very cognitively demanding, so this needs to be examined in future work. Perhaps, it would suitable for expert analysts using visual analytics tools for XAI~\cite{alicioglu2022survey}.

\subsection{Scope of Transferable XAI}

To identify compelling scenarios for explanation transfer, we had limited domain transfer within the same application (dataset).
There needs to be sufficient similarity between the domains to sustain a sense of familiarity and trust in users.
Therefore, consistent with transfer learning in machine learning~\cite{pan2009survey}, Transferable XAI will not be suitable to apply to applications that are too different, e.g., cannot transfer knowledge about house pricing to health diagnosis, domains that sparsely share too few attributes, domains that do not relate linearly.
 
For brevity in the user study, we had only investigated a single transfer between two domains. However, in practice, users may generalize across multiple domains (e.g., diagnosing a patient for kidney, liver, or heart diseases). Future work can investigate this in a longitudinal study where participants have more time to assimilate the relationships among >2 domains.

We have proposed Transferable XAI for structured tabular data with semantically meaningful attributes.
We do not propose it for \textit{perception} (e.g., vision and audio) or language (NLP) tasks, since they involve innate mental processes due to stimuli or low-level skills rather than deliberate reasoning.
We limit it to interpretable, low-dimensional features to avoid information overload~\cite{sweller1988cognitive}.
It could be extended for such unstructured by first extracting concepts through deep feature learning (e.g., with CNN) or concept-based explanations~\cite{koh2017understanding}, then reasoning over them numerically (e.g., border irregularity in a skin lesion indicating melanoma).

Transferable XAI applies to discriminative AI models for both regression and classification, which constitute a large portion of XAI research~\cite{saarela2024recent}. 
However, it is not currently generalizable to other modeling formulations, such as generative AI and reinforcement learning.
Reinforcement learning (RL) models utility-based reward and non-independent events, which discriminative AI does not. Consequently, explainable RL (XRL) methods are different from XAI, and seldom use global linear or nonlinear explanations~\cite{milani2024explainable}.
Generative AI (GAI) is also beyond the scope of our work due to the high-dimensionality of representing generative content, the stochasticity of generation, and different explanations techniques:
i) the high number of non-interpretable features makes the tabular UI non-applicable
ii) our transformation approach does not account for probabilistic distributions that are fundamental to GAI models;
iii) linear or non-linear explanations are not popular for explaining GAI based on Large Multimodal Models (LMM) due to the focus on concept-based explanations with concept bottleneck~\cite{parekh2024concept}, sparse autoencoders (SAE)~\cite{bricken2023towards}, or circuits~\cite{olah2020zoom}.

\subsection{Limitations of Experiment Design and Apparatus}
\rev{We had evaluated Transferable XAI with an experiment design focusing on the efficacy of transfer explanations.
We made design choices, on the experiment apparatus UI and participant incentives, that were sufficient to observe significant effects, but these embedded assumptions that affect the scope of our results.
Although we had presented the AI explanation and AI system predictions side-by-side to support comparison, some users could find their discrepancy confusing. This difference is due to XAI unfaithfulness. An alternative UI layout is to add a row for a ``faithfulness correction'' term to add to the linear weighted sum explanation, resulting in a match with the AI prediction (see Fig.~\ref{fig:faithfulness_correction}).
Nonetheless, participants did not report being confused with the two predictions.

We focused on deliberative, analytical usage by domain users with sufficient domain knowledge and numeracy to adequately interpret the tabular and matrix UIs with weighted sum, though we did not require knowledge of linear regression.
This was suitable for our analytical use case applications focusing on healthcare and pollution.
However, this leaves out less mathematically savvy users. 
More accessible explanation design could be developed in the future, such as thresholding factors to show or omit them which does not require multiplication.
Furthermore, our summative user study included bonus incentives to encourage participants to think effortfully to interpret the AI as correctly as they can.
This is a form of cognitive forcing~\cite{buccinca2021trust}, which may be absent in ordinary usage. 
In such situations, users struggle even more to interpret Transferable XAI, so it may be less effective.
Nevertheless, Transferable XAI provides a new capability for explaining across AI domains, which would benefit sufficiently math-literate and motivated users.}

\subsection{Generalizing Transfer of Non-linear AI Explanations}

\rev{With our current technical approach in Transferable XAI, we had assumed that explanations are linear, and can be sufficiently transferred with linear affine transformations.
However, due to the complexity in AI, non-linear explanations may be needed to improve faithfulness.
We discuss three ways to generalize Transferable XAI to non-linear explanations: sequential affine transformations, transferring local linear explanations, and nonlinear mappings.

Incremental XAI leveraged subspace modeling to partition non-linear decisions into piecewise linear explanations~\cite{bo2024incremental}.
As a component of Transferable XAI, the subspace transfer of Incremental XAI can be applied first to transfer from the dominant linear subspace (e.g., middle-class houses) to other subspaces (e.g., expensive mansions), followed by transfer across other domain types. This involves compositing explanation transfers as described in Section~\ref{sec:discussion-compositing-transfers}.

Despite ubiquitous non-linearity in AI, local explanations are popular~\cite{ribeiro2016should}, since they allow linear approximations in regional subspaces of interest.
Hence, where local linear explanations are appropriate, Transferable XAI can be applied linearly to transform across domains.
However, the local neighborhood may have to be reduced in size to accommodate the increased linear assumption between domains, otherwise, the transferred explanations may have lower faithfulness.

To accommodate non-linear subspace transfer, Bo et al. had discussed generalizing Incremental XAI with generalized additive models (GAM)~\cite{hastie2017generalized}, where the original explanation $f_O^{(r)}(x^{(r)})$ and translation $f_\Delta^{(r)}(x^{(r)})$ to the target depend on the attribute value $x^{(r)}$ (see Section 5.2 in \cite{bo2024incremental}), i.e.,}
\begin{equation}
    \tilde y_T = \sum_r \left( f_O^{(r)}(x^{(r)}) + f_\Delta^{(r)}(x^{(r)}) \right)
    % \tilde y_T = \sum_r f_O^{(r)}(x^{(r)}) + f_\Delta^{(r)}(x^{(r)})
\end{equation}
\rev{Similarly, task transfer of non-linear explanations could involve multiplying with a scale $f_\kappa^{(r)}(x^{(r)})$ that also depends on attribute value, i.e.,}
\begin{equation}
    \label{eq:nonlinear-task-transfer}
    \tilde y_T = \sum_r \left( f_O^{(r)}(x^{(r)}) \cdot f_\kappa^{(r)}(x^{(r)}) \right)
    % \tilde y_T = \sum_r f_O^{(r)}(x^{(r)}) \cdot f_\kappa^{(r)}(x^{(r)})
\end{equation}
\rev{Attributes transfer involved a matrix mapping of attributes from the Original domain $x_O^{(r)}$ to the target domain $x_T^{(r)}$. We consider extending the attribute-dependent coefficients for task transfer in Eq.~\ref{eq:nonlinear-task-transfer} to map multiple target attributes, i.e.,}
\begin{equation}
    \tilde y_T = \sum_r \left( f_O^{(r)}(x_O^{(r)}) \cdot \sum_q \left( f_m^{(rq)}(x_T^{(r)}, x_O^{(q)}) \right) \right)
    % \tilde y_T = \sum_r f_O^{(r)}(x_O^{(r)}) \cdot \sum_q  f_m^{(rq)}(x_T^{(r)}, x_O^{(q)})
\end{equation}
\rev{However, future work is needed to validate these non-linear transfers and to mathematically formulate a unified manifold alignment framework (e.g., via matrix-based basis function expansion~\cite{wood2017generalized}).}

\subsection{Generalizing to Other Domain Shifts}
We have developed Transferable XAI to be generalizable across domain types and investigated this for subspace, task, and attributes transfers.
However, there are other dimensions not covered.
Pan et al. summarized approaches of transductive transfer learning (e.g., covariate shift, sample selection bias), and cases of what to transfer (instance, feature representation, parameter, and knowledge representation)~\cite{pan2009survey}.
In our framing, 
subspace transfer relates to addressing the covariate shift of instances with different distributions,
task transfer relates to sharing parameters in multi-task learning, and
attributes transfer is related to feature representation transfer. 
New opportunities to extend Transferable XAI include:
explaining sample selection bias toward AI fairness~\cite{lyu2024if, ahn2019fairsight, schoeffer2024explanations};
instance transfer that weights instances based on relevance to both domains, perhaps like influence functions~\cite{koh2017understanding}; and
relational knowledge transfer that considers entities by their network graph relation to other entities and analogously mapping to entities with similar relational structures~\cite{mihalkova2007mapping}.

\subsection{Explanation Transfer of Spurious AI Explanations}
In this work, we had assumed that the AI and XAI models mimic the real-world process, where real-world relations match the model-learned ones.
However, models are sensitive to issues in data, such as misrepresentative data, imbalanced data, outliers, or incorrect feature selection~\cite{mohammed2025effects}.
This causes them to learn \textit{spurious relations}, leading to misleading explanations.
We argue that this can lead to misleading transfer explanations too.

For our case with linear factor explanations, the spurious weights $\bm{w}$ could be too high or low for various attributes.
This could occur for any domain, i.e., the Original $\bm{w}_O$ or Target $\bm{w}_T$ weights could be spurious.
Either way, this causes the affine transformation $\bm{w}_T = A \bm{w}_O + \bm{b}$ to optimize to fit them, and consequently be misleading too.
Since both XAI models in either domain and the transfer explanation are co-trained, any spuriousness in either domain can lead to the other domain to learn spurious relations too, thus decreasing its accuracy.
Nevertheless, for non-expert users, the mapping is still faithful to the misleading XAI, but users may not learn generalizable real-world knowledge.
In contrast, domain experts may lose trust in Transferable XAI. 
For subspace or task transfers, the addition or scaling modifications may be inconsistent with their domain knowledge.
For attributes transfer, since they would be familiar with the actual mapping of attribute values across domains, and notice the incoherent learned transfer mapping $A$.

\section{Conclusion}
Extending the idea of Incremental XAI, we propose Transferable XAI, a general framework that explains relationships between domain explanations and enables users to transfer understanding across related domains. With an affine transformation, it supports explanation transfer through translation for data subspace, scaling for decision tasks, and mapping for attributes.
In user studies, Transferable XAI helped participants make decisions more consistent with AI systems and reduced cross-domain confusion, while enhancing their recall and understanding of inter-domain relationships. This work offers an initial exploration of relation-aware explanations that guide users toward more appropriate knowledge transfer, mitigating overreliance on prior explanations and improving comprehension of new domains.

\section*{GenAI Usage Disclosure}   % https://iui.acm.org/2026/call-for-papers/
The authors used generative AI tools (ChatGPT, Gemini) to assist with language refinement and clarity.
The authors reviewed and edited all AI-generated content and take full responsibility for the final version of the manuscript.

\begin{acks}
We thank Jing Liu and Hao Pan for their early discussions and preliminary experiments.
This research is supported by the National Research Foundation, Singapore and Infocomm Media Development Authority under its Trust Tech Funding Initiative (Award No: DTC-RGC-09) and the NUS Institute for Health Innovation and Technology (iHealthtech). 
Any opinions, findings and conclusions or recommendations expressed in this material are those of the author(s) and do not reflect the views of National Research Foundation, Singapore and Infocomm Media Development Authority.
\end{acks}

\balance
\bibliographystyle{ACM-Reference-Format}
\bibliography{reference}


\clearpage
\onecolumn
\appendix
\section{Appendix}
\subsection{Modeling Performances for User Studies} \label{sec:app_modeling_performance}
Table \ref{tab:modeling_performance} shows the performances of the AI and XAI models used in the experiments, and the faithfulness of the XAI models to the AI models.
Since Health Risk is a classification task, of which the data labels are binary (i.e., 0 or 1), we accordingly applied logistic activation to the linear regression outputs of XAI models and changed the training objective to the binary cross-entropy loss when evaluating Glass-box AI, and used the same $\lambda$ as evaluating XAI faithfulness.
In summary, the Single-domain XAI shows marginally higher predictive performance than the Transferable XAI, yet both are outperformed by the neural network-based AI model when used as glass-box predictors. 
This underscores the need to adopt AI models for these applications. In terms of faithfulness, the Transferable XAI remains comparable to the Single-domain XAI.

\subsection{User Interface (UI) for Air Pollution datasets}
\begin{figure}[t]
    \centering
    \includegraphics[width=0.9\linewidth]{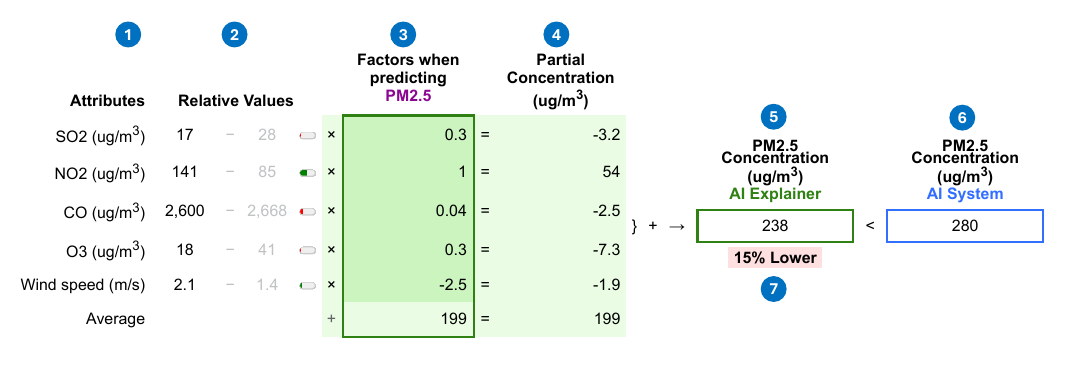}
    \caption{User interface (UI) of AI System with linear factor explanation showing: 1) attributes used in the prediction task, 2) instance relative values, computed as the actual value minus the average value for each attribute, i.e., $x^{(r)}$ {\color{lightgray}- $\bar{x}^{(r)}$}, 3) factors $w^{(r)}$ that the explainer multiplies with values, 4) partial contributions $w^{(r)}x^{(r)}$ of each attributes, 5) estimation $\tilde{y} = \sum_r \tilde{y}^{(r)}$ from the AI Explainer, 6) prediction $\hat{y}$ from the AI System, 7) indicator of how different the AI Explainer estimation is from the AI System.}
    \Description{This figure illustrates the user interface (UI) of the AI System with a linear factor explanation for predicting PM2.5 concentration. 
    The layout is divided into seven labeled components (1--7):
    1. Attributes column lists five environmental features: SO₂ (µg/m³), NO₂ (µg/m³), CO (µg/m³), O₃ (µg/m³), and Wind speed (m/s).  
    2. Relative Values column shows the instance value minus the average for each attribute. 
    For example: SO₂ = 17–28, NO₂ = 141–85, CO = 2 600–2 668, O₃ = 18–41, and Wind speed = 2.1–1.4.  
    3. Factors when predicting PM2.5 (labeled “PM2.5” in magenta) display each attribute’s weight used by the explainer model: SO₂ = 0.3, NO₂ = 1, CO = 0.04, O₃ = 0.3, Wind speed = –2.5, and Average bias term = +199.  4. Partial Concentration (µg/m³) column shows each attribute’s partial contribution obtained by multiplying its relative value with the corresponding factor:  
    SO₂ → –3.2 µg/m³, NO₂ → +54 µg/m³, CO → –2.5 µg/m³, O₃ → –7.3 µg/m³, Wind speed → –1.9 µg/m³, and Average → +199 µg/m³.  
    5. The AI Explainer’s total estimated PM2.5 concentration is 238 µg/m³, enclosed in a green box labeled “AI Explainer.”  
    6. The AI System’s actual predicted PM2.5 concentration is 280 µg/m³, shown in a blue box labeled “AI System.”  
    7. A pink annotation below shows “15 \% Lower,” indicating that the AI Explainer’s estimate (238) is 15 \% lower than the AI System’s prediction (280).
    Overall, the UI visualizes how the explainer’s linear factors and attribute contributions produce a slightly smaller predicted PM2.5 value compared to the AI System output, offering a transparent breakdown of how each attribute contributes to the overall prediction.}
    \label{fig:ui_org_domain_air}
\end{figure}

\begin{figure}[H]
    \centering
    \includegraphics[width=0.9\linewidth]{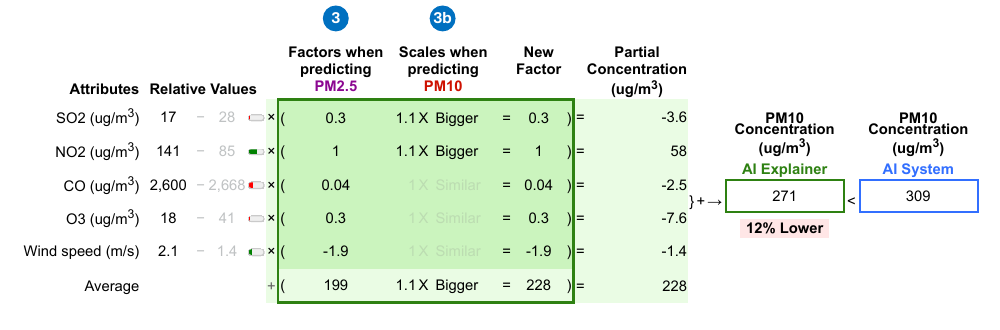}
    \caption{User interface (UI) of the Transferable explanation in \textbf{Task transfer}, showing: 
    3) factors from the Original Domain, $w_O^{(r)}$ (PM2.5 Concentration prediction task), 
    3b) the corresponding scaled factors for the Target domain (PM10 Concentration prediction task). For example, Wind speed (m/s) retains a similar influence across prediction tasks, $\mathrm{SO}_2 (\mathrm{ug/m}^3)$, and $\mathrm{NO}_2 (\mathrm{ug/m}^3)$ becomes 1.1× more influential when transferred to PM10 Concentration prediction. The numbers in the UI are all rounded, and the user can hover to check the exact values.}
    \Description{This figure depicts the user interface (UI) of the Transferable explanation in the Task Transfer condition, illustrating how factors from the original domain (PM2.5 prediction) are adapted to the target domain (PM10 prediction). The interface includes the following components:
    1. Attributes column lists five environmental variables: SO₂ (µg/m³), NO₂ (µg/m³), CO (µg/m³), O₃ (µg/m³), and Wind speed (m/s).  
    2. Relative Values show each instance’s difference from the average: SO₂ = 17 – 28, NO₂ = 141 – 85, CO = 2 600 – 2 668, O₃ = 18 – 41, Wind speed = 2.1 – 1.4, and Average bias term indicated as “+”.  
    3. “Factors when predicting PM2.5” (in magenta) give the original domain weights:  SO₂ = 0.3, NO₂ = 1, CO = 0.04, O₃ = 0.3, Wind speed = –1.9, and Average = 199.  
    3b “Scales when predicting PM10” (in orange text) specify how each factor changes when transferred: SO₂ → 1.1× Bigger, NO₂ → 1.1× Bigger, CO → 1× Similar, O₃ → 1× Similar, Wind speed → 1× Similar, and Average → 1.1× Bigger. The resulting “New Factor” column keeps the same numeric coefficients as the original (e.g., SO₂ = 0.3, NO₂ = 1, etc.), but the average bias increases from 199 to 228 due to the 1.1× scaling.  
    4. The Partial Concentration (µg/m³) column lists each attribute’s contribution: SO₂ = –3.6, NO₂ = +58, CO = –2.5, O₃ = –7.6, Wind speed = –1.4, and Average = +228. 
    5. The AI Explainer’s predicted PM10 concentration is **271 µg/m³**, shown in a green box labeled “AI Explainer.”  
    6. The AI System’s actual PM10 prediction is **309 µg/m³**, shown in a blue box labeled “AI System.”  
    7. A pink annotation reads “12 \% Lower,” indicating that the AI Explainer’s estimate is 12 \% below the AI System’s prediction.
    Overall, the UI demonstrates how scaling factors are scaled when transferring explanatory parameters between related air-quality prediction tasks, with the explainer providing a transparent estimate of PM10 concentration compared with the AI system output.}
    \label{fig:ui_transferable_xai_task_air}
\end{figure}

\begin{figure}[H] 
    \centering
    \includegraphics[width=1\linewidth]{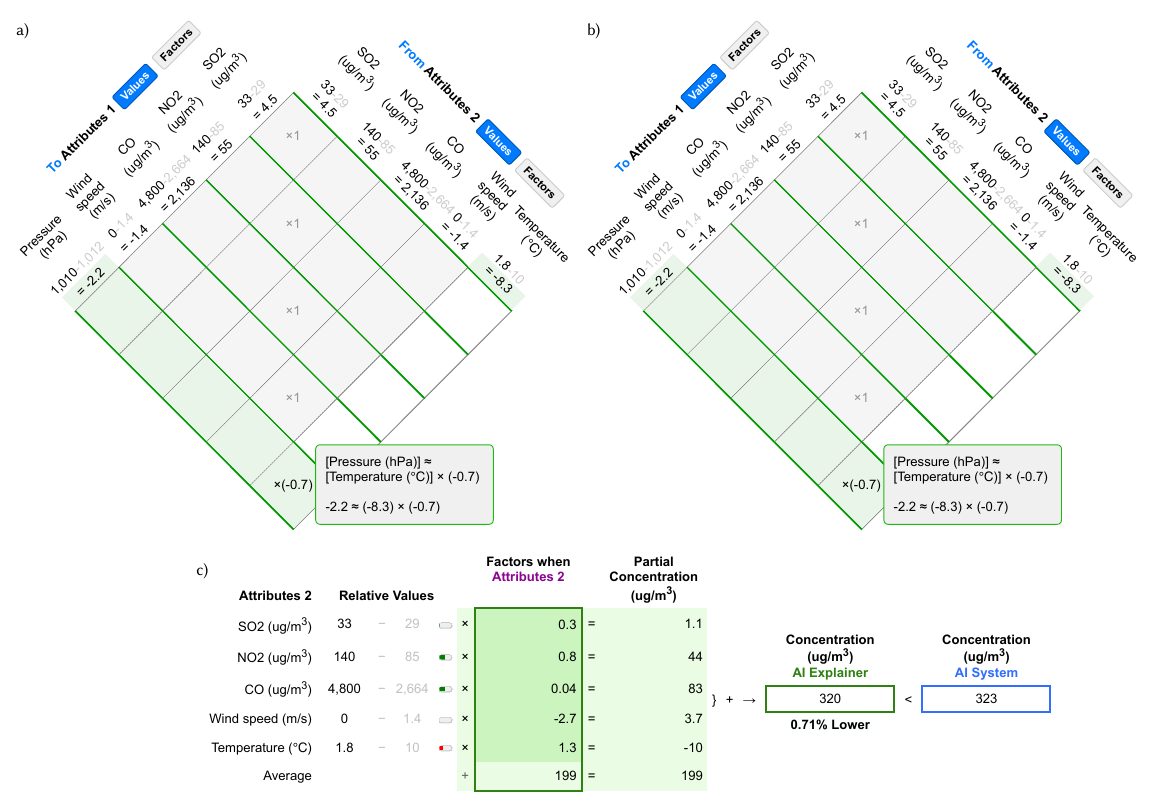}
    \caption{User interface (UI) of the Transferable explanation in \textbf{Attributes transfer}. 
    a) Map Target attribute values to the corresponding Original attribute value. Matching row and attribute values are highlighted when the mouse hovers, with tooltips displaying the transformation formula (Pressure in this case). b) Map Original attribute factors to the corresponding Target attribute factor. Matching column and attribute factors are highlighted when the mouse hovers, with tooltips displaying the transformation formula (Temperature in this case). c) Tabular interface, similar to the one introduced previously for the Target domain.}
    \Description{This figure presents the user interface (UI) of the Transferable explanation in the Attributes Transfer condition, illustrated through three panels (a–c). It demonstrates how attribute mappings and factor transformations occur between two related domains.(a) The first panel maps Target attribute values to the corresponding Original attribute values. 
    Rows and columns correspond to attributes from Original domain and Target domain: Pressure (hPa), Wind speed (m/s), CO (µg/m³), NO₂ (µg/m³), SO₂ (µg/m³), and Temperature (°C).  Each diagonal cell contains an “×1” to indicate direct correspondence between identical attributes.  
    When the mouse hovers over a matching pair (highlighted in green), a tooltip displays the transformation formula.  
    In this example, Pressure (hPa) is related to Temperature (°C) with a factor of –0.7, as shown in the annotation:  
    [Pressure (hPa)] = [Temperature (°C)] × (–0.7), and numerically, –2.2 = (–8.3) × (–0.7).  
    (b) The second panel mirrors this process but maps Original attribute factors to their corresponding Target attribute factors.  
    The same six attributes are displayed, with matching pairs highlighted upon hover.  
    The tooltip again shows the transformation formula [Pressure (hPa)] = [Temperature (°C)] × (–0.7).
    (c) The third panel displays a tabular explanation interface.  
    The leftmost column (“Attributes 2”) lists SO₂ (µg/m³), NO₂ (µg/m³), CO (µg/m³), Wind speed (m/s), Temperature (°C), and Average.  
    Relative Values show the deviation from the mean:  
    SO₂ = 33 – 29, NO₂ = 140 – 85, CO = 4 800 – 2 664, Wind speed = 0 – 1.4, Temperature = 1.8 – 10.  
    The Factors when Attributes 2 column gives coefficients used in the linear explainer:  
    SO₂ = 0.3, NO₂ = 0.8, CO = 0.04, Wind speed = –2.7, Temperature = 1.3, and Average = 199.  
    Partial Concentrations (µg/m³) are computed by multiplying each relative value by its factor:  
    SO₂ = 1.1, NO₂ = 44, CO = 83, Wind speed = 3.7, Temperature = –10, and Average = 199.  
    The AI Explainer predicts a total concentration of **320 µg/m³**, while the AI System predicts **323 µg/m³**, 
    as shown in green and blue boxes respectively.  
    A small annotation, “0.71 \% Lower,” indicates that the AI Explainer’s estimate is slightly below the system’s output.}
    \label{fig:ui_transferable_xai_attributes_air}
\end{figure}

Fig.~\ref{fig:ui_org_domain_air} shows the explanation UI for all instances in the Original domain, and for instances in the Target domain with baseline XAI Types (Original, Target). 
Fig.~\ref{fig:ui_transferable_xai_task_air} shows the explanation UI for Transferable XAI with Task transfer. 
Like Incremental XAI~\cite{bo2024incremental}, Transferable XAI has an additional column to show a modification on the factors.
Incremental XAI's column shows the sparse \textit{differences} ($+$) in factors, where unchanged factors have $\Delta\text{factor} = 0$.
Transferable XAI's column shows the sparse \textit{scales} ($\times$) in factors, where unchanged factors have $\kappa_\Delta \text{scale} = 1$.
Fig.~\ref{fig:ui_transferable_xai_attributes_air} shows the explanation UI for Transferable XAI with Attributes transfer. 
The main tabular UI (Fig.~\ref{fig:ui_transferable_xai_attributes_air}c) for the Target attributes is similar to that for the Original attributes, but the factors are different.
Indeed, the factors are derived from the matrix mapping $M_\chi^\top \bm{w}_O$, and is different from the independent Target Attributes $\bm{w}_T$.

\begin{table}[ht]
\caption{Modeling results from 5-fold cross-validation performances of neural network-based AI (AI Model), linear factor models as glass-box AI (Glass-box AI), and the faithfulness of linear factor models against AI Model (XAI). AI Model and Glass-box AI were trained on the data ground truth, and XAI was trained on AI Model predictions. Except for AI Model and Glass-box AI performances on Health Risk that are measured with \% Accuracy, all other metrics are R$^2$. We report mean $\pm$ standard deviation, and larger is better.}
\renewcommand{\arraystretch}{1.2}
\label{tab:modeling_performance}
\begin{tabular}{cllllll}
\hline
\multirow{8}{*}{\begin{tabular}[c]{@{}c@{}}Task\\ Transfer\end{tabular}}    & \multirow{2}{*}{\textbf{Health Risk}}      & \multirow{2}{*}{AI (Acc \%)}    & \multicolumn{2}{c}{Glass-box AI (Acc \%)} & \multicolumn{2}{c}{XAI (R$^2$)}      \\ \cline{4-7} 
    &     &   & Single   & Transferable   & Single & Transferable \\ \cline{2-7} 
    & Diabetes     &  83.1 $\pm$ 1.5   &   79.7 $\pm$ 1.4  &   78.2 $\pm$ 1.9   &   0.72 $\pm$ 0.02   &  0.72 $\pm$ 0.02   \\
    & Heart disease     &   74.3 $\pm$ 1.3  &   73.9 $\pm$ 1.0   &   70.5 $\pm$ 1.8   &   0.74 $\pm$ 0.02   &  0.74 $\pm$ 0.02         \\ \cline{2-7} 
    & \multirow{2}{*}{\textbf{Air Pollution}}    & \multirow{2}{*}{AI (R$^2$)}    & \multicolumn{2}{c}{Glass-box AI (R$^2$)} & \multicolumn{2}{c}{XAI (R$^2$)}      \\ \cline{4-7} 
    &  &  & Single   & Transferable   & Single & Transferable \\ \cline{2-7} 
    & PM2.5     &  0.76 $\pm$ 0.01   &   0.67 $\pm$ 0.01   &   0.66 $\pm$ 0.01   &   0.89 $\pm$ 0.01   &  0.88 $\pm$ 0.01    \\
    & PM10     &  0.72 $\pm$ 0.01   &   0.63 $\pm$ 0.01   &   0.63 $\pm$ 0.01   &   0.89 $\pm$ 0.01   &  0.88 $\pm$ 0.01            \\ \hline
\multirow{8}{*}{\begin{tabular}[c]{@{}c@{}}Attributes\\ Transfer\end{tabular}}   & \multirow{2}{*}{\textbf{Health Risk}}      & \multirow{2}{*}{AI (Acc \%)}    & \multicolumn{2}{c}{Glass-box AI (Acc \%)} & \multicolumn{2}{c}{XAI (R$^2$)}       \\ \cline{4-7} 
    &     &   & Single   & Transferable   & Single & Transferable \\ \cline{2-7} 
    & Attributes w/ BMI     &  83.3 $\pm$ 1.8   &   79.4 $\pm$ 1.1   &   77.3 $\pm$ 7.3   &   0.70 $\pm$ 0.03   &  0.69 $\pm$ 0.02   \\
    & Attributes w/ Height     &  83.3 $\pm$ 1.8   &   77.4 $\pm$ 4.2  &   77.2 $\pm$ 7.0   &   0.69 $\pm$ 0.02   &  0.68 $\pm$ 0.03    \\ \cline{2-7} 
    & \multirow{2}{*}{\textbf{Air Pollution}}    & \multirow{2}{*}{AI (R$^2$)}    & \multicolumn{2}{c}{Glass-box AI (R$^2$)} & \multicolumn{2}{c}{XAI (R$^2$)}      \\ \cline{4-7} 
    &  &  & Single   & Transferable   & Single & Transferable \\ \cline{2-7} 
    & Attributes w/ Pressure     &  0.74 $\pm$ 0.01   &  0.68 $\pm$ 0.01   &   0.68 $\pm$ 0.01   &   0.97 $\pm$ 0.01   &  0.97 $\pm$ 0.01           \\
    & Attributes w/ Temperature     &  0.73 $\pm$ 0.02   &   0.67 $\pm$ 0.01   &   0.67 $\pm$ 0.01   &  0.99 $\pm$ \textless{}0.01  & 0.99 $\pm$ \textless{}0.01  \\ \hline
\end{tabular}
\end{table}

\clearpage

\newpage

\begin{figure}[H]
    \centering
    \includegraphics[width=0.7\linewidth]{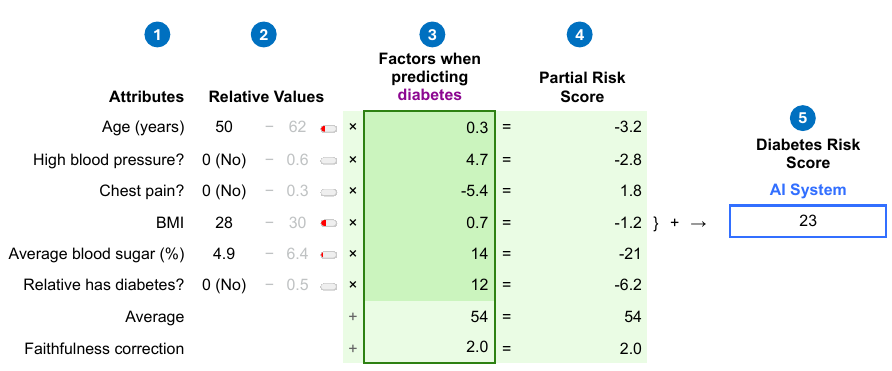}
    \caption{An alternative UI layout by adding a \textit{faithfulness correction} row, such that the sum of the explainer outputs match the AI system predictions.}
    \label{fig:faithfulness_correction}
    \Description{The figure shows a single table-based interface with numbered markers labeled 1 through 5 above different sections. 
    The first column, labeled ``Attributes'', lists Age (years), High blood pressure? (No), Chest pain? (No), BMI, Average blood sugar (\%), Relative has diabetes? (No), an ``Average'' row, and a ``Faithfulness correction'' row. 
    The second column, labeled ``Relative Values'', shows numeric values for each attribute, including 50 for Age, 0 (No) for High blood pressure, 0 (No) for Chest pain, 28 for BMI, 4.9 for Average blood sugar, and 0 (No) for Relative has diabetes, with small slider indicators next to each value. 
    The third column, highlighted in green and labeled ``Factors when predicting diabetes'', displays numeric factor values for each attribute, including 0.3, 4.7, –5.4, 0.7, 14, and 12, followed by 54 in the Average row and 2.0 in the Faithfulness correction row. 
    The fourth column, also shaded in green and labeled ``Partial Risk Score'', shows the computed partial scores –3.2, –2.8, 1.8, –1.2, –21, –6.2, 54, and 2.0 corresponding to each row. 
    Mathematical symbols between the columns indicate multiplication and summation of values. 
    On the right side of the figure, a blue box labeled ``Diabetes Risk Score'' and ``AI System'' displays the final numeric value 23. 
    Curly braces and an arrow connect the summed partial risk scores to the final risk score output.}
\end{figure}

\newpage
\subsection{Qualtrics survey for the user study} \label{app:task_transfer_survey}
\begin{figure*}[htp]
    \centering
    \includegraphics[width=0.7\linewidth]{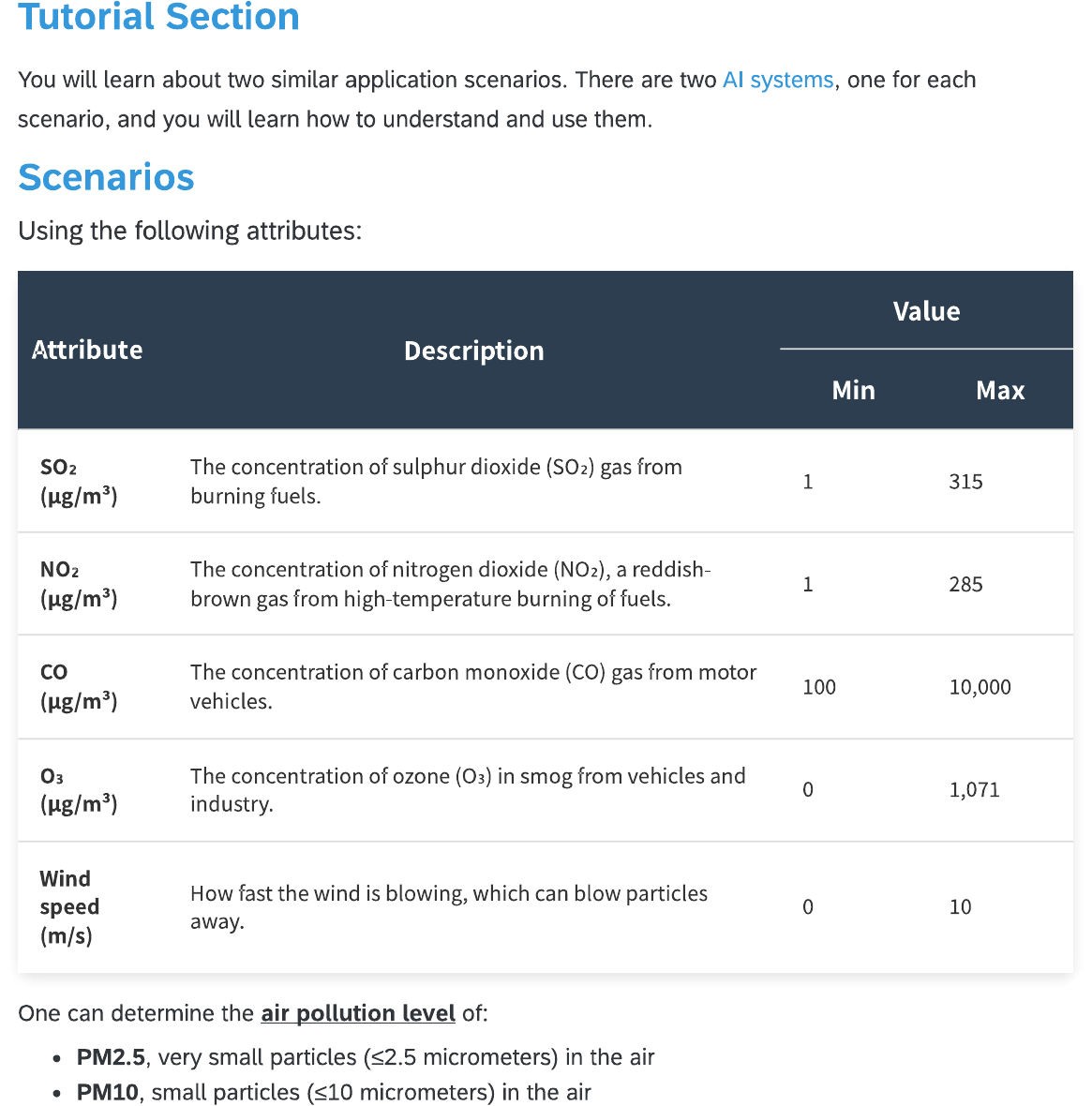}
    \caption{Introduction of attributes and scenarios on air pollution.}
    \Description{The figure shows a tutorial page titled “Tutorial Section.” 
    A paragraph states that the user will learn about two similar application scenarios, each with an AI system, and that the user will learn how to understand and use them. 
    A heading labeled “Scenarios” appears below, followed by the sentence “Using the following attributes:”. 
    Under it is a table with three columns labeled “Attribute,” “Description,” and “Value.” 
    The “Value” column has subheadings ``Min'' and ``Max''. 
    The table contains five rows:
    (1) SO₂ (µg/m³) – ``The concentration of sulphur dioxide (SO₂) gas from burning fuels'', with min value 1 and max value 315.
    (2) NO₂ (µg/m³) – ``The concentration of nitrogen dioxide (NO₂), a reddish-brown gas from high-temperature burning of fuels'', with min value 1 and max value 285.
    (3) CO (µg/m³) – ``The concentration of carbon monoxide (CO) gas from motor vehicles'', with min value 100 and max value 10,000.
    (4) O₃ (µg/m³) – ``The concentration of ozone (O₃) in smog from vehicles and industry'', with min value 0 and max value 1,071.
    (5) Wind speed (m/s) – ``How fast the wind is blowing, which can blow particles away'', with min value 0 and max value 10. 
    Below the table, a paragraph begins with ``One can determine the air pollution level of:'', and lists two bullet points: 
    PM2.5, very small particles (≤2.5 micrometers) in the air;
    and PM10, small particles (≤10 micrometers) in the air.}
    \label{fig:task_transfer_survey_scenarios}
\end{figure*}

\begin{figure*}[t]
    \centering
    \includegraphics[width=0.8\linewidth]{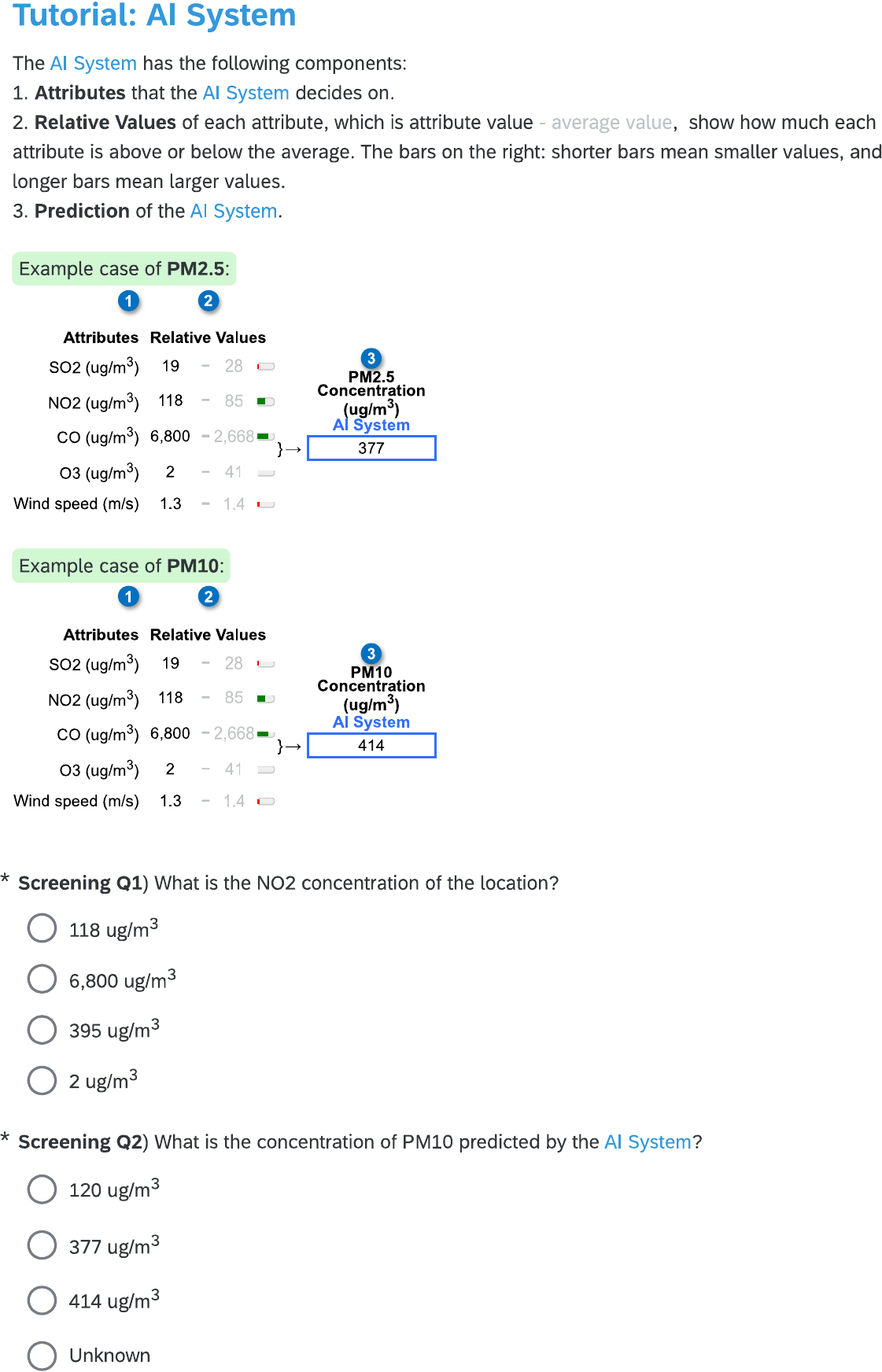}
    \caption{Task transfer tutorial about the AI Systems for two scenarios. The displayed UI is the air pollution application.}
    \Description{The figure shows a tutorial page titled “Tutorial: AI System.” 
    The text explains three components of the AI System: 
    (1) Attributes that the AI System decides on, 
    (2) Relative Values of each attribute, described as the difference between the attribute value and the average value, 
    and (3) Prediction of the AI System. 
    Below, there are two example cases labeled “Example case of PM2.5” and “Example case of PM10”, each containing a numbered sequence “1 2 3.” 
    In both cases, a table lists five rows of attributes: SO2 (µg/m³), NO2 (µg/m³), CO (µg/m³), O3 (µg/m³), and Wind speed (m/s). 
    Each attribute has a numeric value on the left and a smaller number with a short horizontal bar on the right. 
    In the PM2.5 example, the prediction area on the right shows “PM2.5 Concentration (µg/m³)” and a blue box with “AI System” and the number “377.” 
    In the PM10 example, the corresponding area shows “PM10 Concentration (µg/m³)” and a blue box with “AI System” and the number “414.” 
    Below the examples are two questions labeled “Screening Q1” and “Screening Q2”, 
    each followed by four circular multiple-choice options with numeric answers. 
    Q1 asks for the NO2 concentration, and Q2 asks for the PM10 concentration predicted by the AI System.}
    \label{fig:task_transfer_survey_tutorial1_AI}
\end{figure*}

\begin{figure*}[t]
    \centering
    \includegraphics[width=0.8\linewidth]{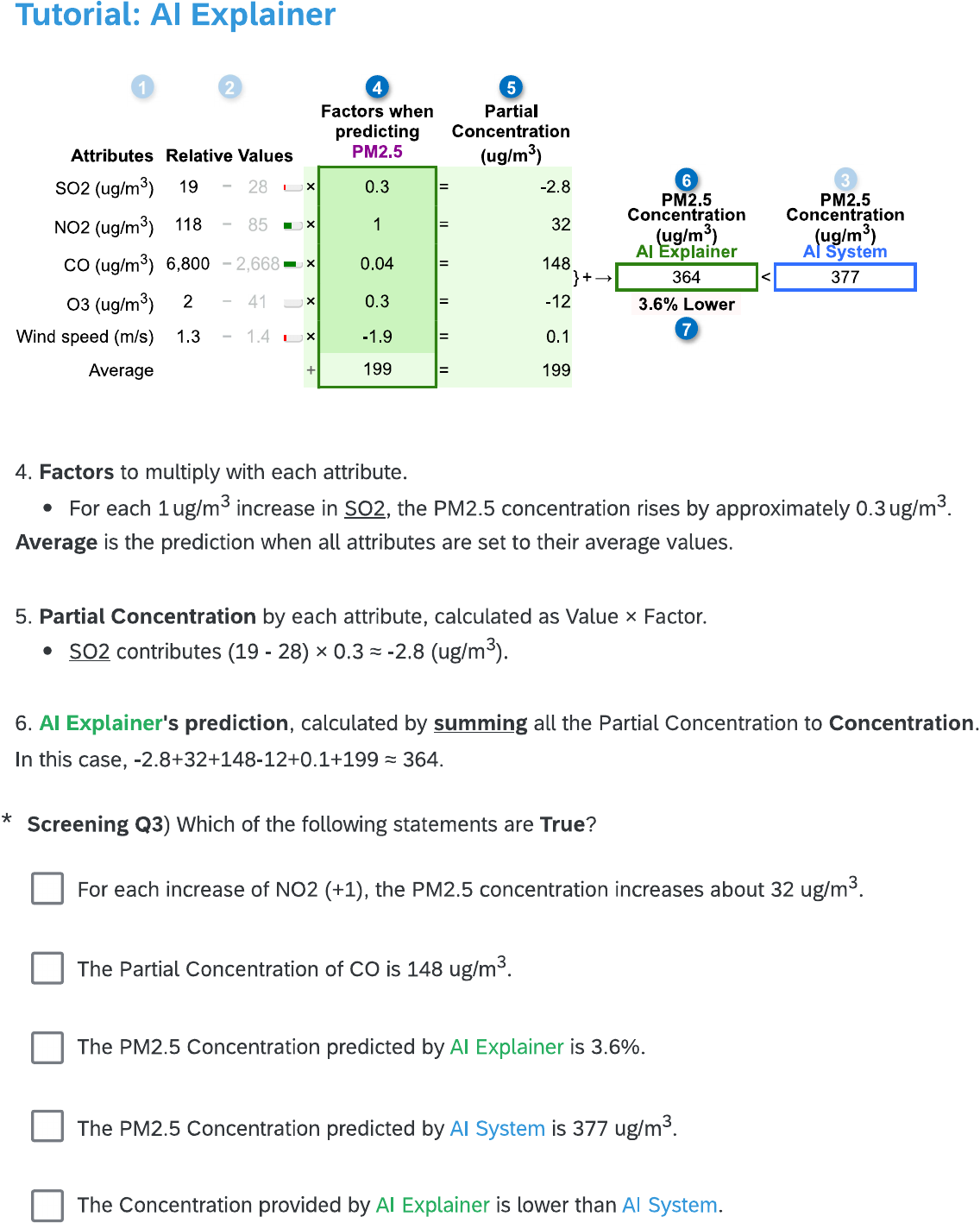}
    \caption{Task transfer tutorial about the AI Explainer. The displayed UI is the Transferable condition on the air pollution application.}
    \Description{The figure shows a tutorial page titled ``Tutorial: AI Explainer''.
    The top section displays a diagram divided into several labeled parts. 
    On the left, a table lists five rows of attributes: SO₂ (µg/m³), NO₂ (µg/m³), CO (µg/m³), O₃ (µg/m³), and Wind speed (m/s), along with their relative values. 
    To the right, a green-shaded column labeled ``Factors when predicting PM2.5'' contains numeric factors for each attribute: 
    0.3 for SO₂, 1 for NO₂, 0.04 for CO, 0.3 for O₃, and –1.9 for Wind speed, followed by an “Average” row with 199. 
    Next to it, another column labeled ``Partial Concentration (µg/m³)'' shows calculated values: –2.8, 32, 148, –12, 0.1, and 199 for the average. 
    On the right side, a section labeled ``PM2.5 Concentration (µg/m³)'' shows two boxes: one labeled “AI Explainer” with the value 364, and another labeled “AI System” with the value 377. 
    A small arrow and label between them indicate ``3.6\% Lower''.
    Below the figure, numbered paragraphs 4 through 6 describe ``Factors'', ``Partial Concentration'', and ``AI Explainer’s prediction'', each with one or more bullet points and mathematical expressions. 
    At the bottom, ``Screening Q3'' asks ``Which of the following statements are True?'' followed by five checkbox options, each containing a short sentence about PM2.5 concentration, partial contribution, or prediction comparison.}
    \label{fig:task_transfer_survey_tutorial2_XAI}
\end{figure*}

\begin{figure*}[t]
    \centering
    \includegraphics[width=0.82\linewidth]{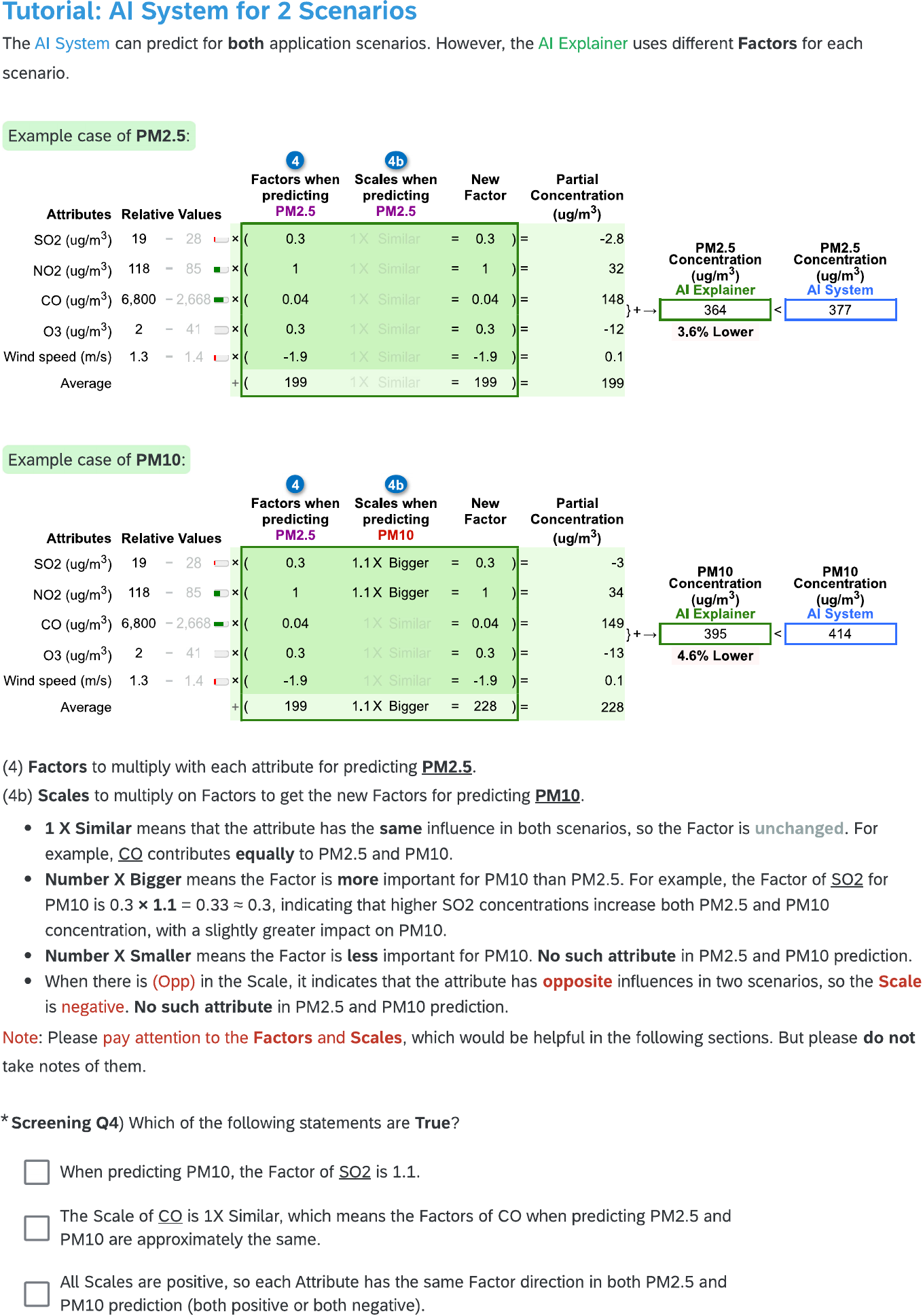}
    \caption{Task transfer tutorial about AI Explainers in two scenarios. The displayed UI is the Transferable condition on the air pollution application.}
    \Description{The figure shows a tutorial page titled ``Tutorial: AI System for 2 Scenarios''. 
    A paragraph explains that the AI System can predict for both application scenarios, while the AI Explainer uses different Factors for each scenario. 
    Below this, two example cases are shown.
    The first section is labeled ``Example case of PM2.5''. A table lists five attributes: SO₂ (µg/m³), NO₂ (µg/m³), CO (µg/m³), O₃ (µg/m³), and Wind speed (m/s), with their relative values next to them. 
    To the right, green-shaded columns are labeled ``Factors when predicting PM2.5'', ``Scales when predicting PM2.5'', ``New Factor'', and ``Partial Concentration (µg/m³)''. 
    In the Factors column, the values are 0.3, 1, 0.04, 0.3, and –1.9, followed by an ``Average'' row with 199. In the Scales column, each row reads ``1× Similar''. In the New Factor column, the same numerical values as the Factors column are repeated. In the Partial Concentration column, the values are –2.8, 32, 148, –12, 0.1, and 199 for the average. 
    On the right, two boxes show PM2.5 concentration results: one labeled ``AI Explainer'' with 364 and one labeled ``AI System'' with 377. 
    A label between them reads ``3.6\% Lower'', with the symbol ``<'' between the two boxes.
    The second section is labeled ``Example case of PM10''. 
    A similar table lists the same five attributes and relative values. To the right, green-shaded columns are labeled ``Factors when predicting PM2.5'', ``Scales when predicting PM10'', ``New Factor'', and ``Partial Concentration (µg/m³)''. 
    In the Factors column, the values are 0.3, 1, 0.04, 0.3, and –1.9, followed by an ``Average'' row with 228. In the Scales column, the entries are ``1.1× Bigger'', ``1.1× Bigger'', ``1× Similar'', ``1× Similar'', and ``1× Similar'', followed by ``1.1× Bigger'' in the Average row. 
    In the New Factor column, the values are 0.3, 1.1, 0.04, 0.3, and –1.9, followed by 228 in the Average row. 
    In the Partial Concentration column, the values are –3, 34, 149, –13, 0.1, and 228 for the average. 
    On the right, two boxes show PM10 concentration results: one labeled ``AI Explainer'' with 395 and one labeled ``AI System'' with 414. 
    A label between them reads ``4.6\% Lower'', with the symbol ``<'' between the two boxes.
    Below the examples, numbered section (4) describes ``Factors to multiply with each attribute for predicting PM2.5'', and section (4b) describes ``Scales to multiply on Factors to get the new Factors for predicting PM10''. Four bullet points follow, defining the meanings of ``1× Similar'', ``Number × Bigger'', ``Number × Smaller'', and ``(Opp)''. 
    Each bullet contains an explanatory sentence and examples with mathematical expressions. 
    A red note at the end says ``Please pay attention to the Factors and Scales, which would be helpful in the following sections. But please do not take notes of them''.
    At the bottom, a question labeled ``Screening Q4'' asks ``Which of the following statements are True?''. Three checkbox options are shown. The first states that when predicting PM10, the Factor of SO₂ is 1.1. The second states that the Scale of CO is 1× Similar, meaning the Factors of CO when predicting PM2.5 and PM10 are approximately the same. The third states that all Scales are positive, so each Attribute has the same Factor direction in both PM2.5 and PM10 prediction, either positive or negative.}
    \label{fig:task_transfer_survey_tutorial3_Transferable}
\end{figure*}

\begin{figure*}[t]
    \centering
    \includegraphics[width=0.9\linewidth]{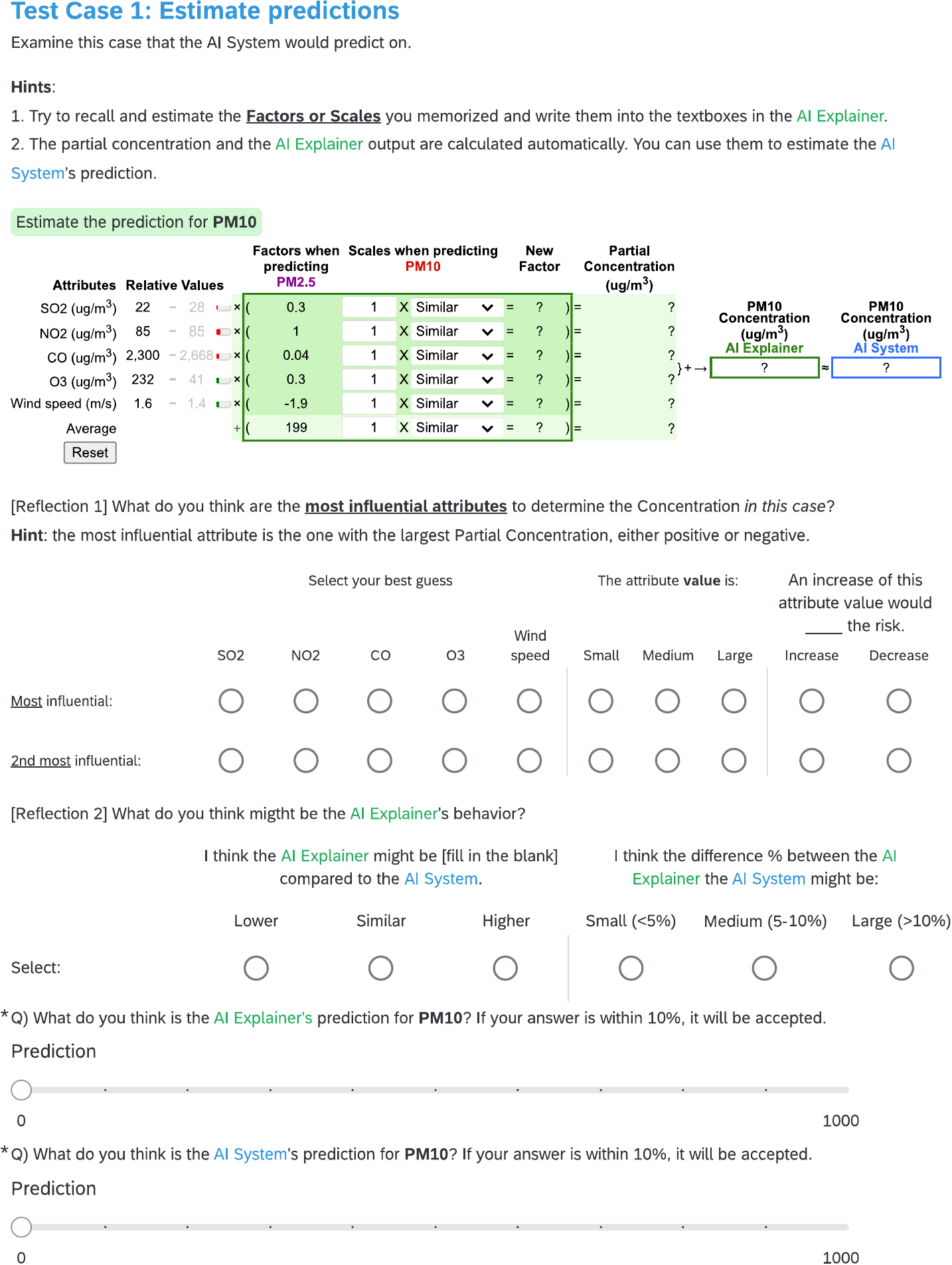}
    \caption{Sample of the forward simulation trial (test page) in the Task transfer, where participants are asked to estimate the explanation and AI system outputs without seeing the factors. The displayed UI is the Transferable condition on the air pollution application.}
    \Description{The figure shows a tutorial page titled ``Test Case 1: Estimate predictions''. 
    A paragraph instructs the user to examine a case that the AI System would predict on. Below, a section titled ``Hints'' lists two numbered instructions: 
    (1) to recall and estimate the Factors or Scales memorized and write them into the textboxes in the AI Explainer, 
    and (2) to use the partial concentration and the AI Explainer output to estimate the AI System's prediction. 
    A title box below reads ``Estimate the prediction for PM10''.
    A diagram follows, containing a table with columns labeled ``Attributes'', ``Relative Values'', ``Factors when predicting PM2.5'', ``Scales when predicting PM10'', ``New Factor'', and ``Partial Concentration (µg/m³)''. 
    The rows list five attributes: SO₂ (µg/m³), NO₂ (µg/m³), CO (µg/m³), O₃ (µg/m³), and Wind speed (m/s), followed by an ``Average'' row. 
    Each row has numeric values in the Relative Values column, and textboxes with ``1× Similar'' in the Scales column, while other cells show question marks (``?'') indicating missing values. 
    At the right of the table, two blue boxes show placeholders labeled ``AI Explainer'' and ``AI System'', both under ``PM10 Concentration (µg/m³)'', each containing a question mark. 
    Buttons labeled ``Average'', ``Reset'', and various calculation symbols are visible within the figure.
    Below the diagram, there are several sections titled ``[Reflection 1]'' and ``[Reflection 2]''. 
    Reflection 1 asks, ``What do you think are the most influential attributes to determine the Concentration in this case?'' A hint below states that the most influential attribute is the one with the largest Partial Concentration, either positive or negative. 
    A selection grid allows the user to choose the most and second most influential attributes among SO₂, NO₂, CO, O₃, and Wind speed. To the right, additional columns labeled ``The attribute value is:'' and ``An increase of this attribute value would ____ the risk.'' contain radio button options labeled ``Small'', ``Medium'', ``Large'', ``Increase'', and ``Decrease''.
    Reflection 2 asks, ``What do you think might be the AI Explainer's behavior?'' Two sentences allow the user to complete and select statements: 
    ``I think the AI Explainer might be [fill in the blank] compared to the AI System.'' with options ``Lower'', ``Similar'', and ``Higher'', and ``I think the difference \% between the AI Explainer and the AI System might be:'' with options ``Small (<5\%)'', ``Medium (5–10\%)'', and ``Large (>10\%)''.
    At the bottom, two starred questions are shown. 
    The first asks, ``What do you think is the AI Explainer's prediction for PM10? If your answer is within 10\%, it will be accepted.'' 
    A horizontal slider labeled ``Prediction'' ranges from 0 to 1000. 
    The second asks, ``What do you think is the AI System's prediction for PM10? If your answer is within 10\%, it will be accepted.'' 
    It also includes a horizontal slider labeled ``Prediction'' ranging from 0 to 1000.}

    \label{fig:task_transfer_survey_test_page1}
\end{figure*}

\begin{figure*}[t]
    \centering
    \includegraphics[width=0.9\linewidth]{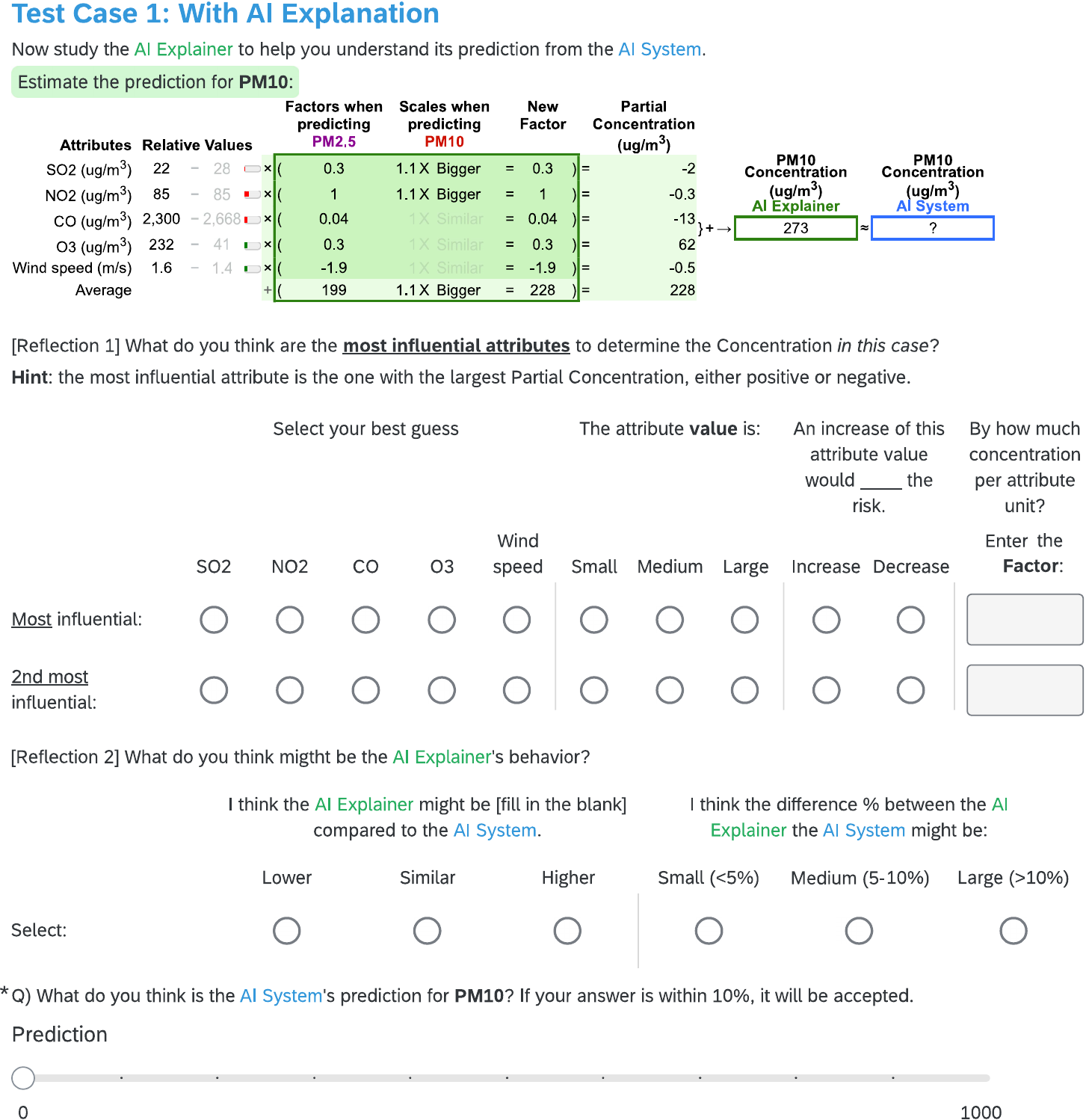}
    \caption{Sample of the forward simulation trial (explanation page) in the Task transfer, where participants are asked to estimate the AI system outputs with the factors shown. The displayed UI is the Transferable condition on the air pollution application.}
    \Description{The figure shows a tutorial page titled ``Test Case 1: With AI Explanation''. A paragraph instructs the user to study the AI Explainer to understand its prediction from the AI System. Below, a box reads ``Estimate the prediction for PM10''.
    A diagram is displayed showing a table with columns labeled ``Attributes'', ``Relative Values'', ``Factors when predicting PM2.5'', ``Scales when predicting PM10'', ``New Factor'', and ``Partial Concentration (µg/m³)''. 
    The table lists five attributes: SO₂ (µg/m³), NO₂ (µg/m³), CO (µg/m³), O₃ (µg/m³), and Wind speed (m/s), followed by an ``Average'' row. 
    In the Factors column, the values are 0.3, 1, 0.04, 0.3, and –1.9, with 199 in the Average row. 
    In the Scales column, the first two rows show ``1.1× Bigger'', the next three rows show ``1× Similar'', and the Average row shows ``1.1× Bigger''. 
    In the New Factor column, the corresponding values are 0.3, 1.1, 0.04, 0.3, and –1.9, followed by 228 in the Average row. 
    In the Partial Concentration column, the values are –2, –0.3, –13, 62, 0.5, and 228 for the average. 
    To the right of the table, two boxes represent the predicted PM10 concentration. The left box, labeled ``AI Explainer'', displays the value 273, and the right box, labeled ``AI System'', contains a question mark. 
    Below the diagram, a section labeled ``[Reflection 1]'' asks ``What do you think are the most influential attributes to determine the Concentration in this case?'' 
    A hint below states that the most influential attribute is the one with the largest Partial Concentration, either positive or negative. 
    A grid allows selection for ``Most influential'' and ``2nd most influential'' among the five attributes: SO₂, NO₂, CO, O₃, and Wind speed. 
    To the right, three additional columns are labeled ``The attribute value is:'', ``An increase of this attribute value would ____ the risk:'', and ``By how much concentration per attribute unit?''. 
    Under these columns are radio buttons labeled ``Small'', ``Medium'', ``Large'', ``Increase'', and ``Decrease'', followed by two empty textboxes labeled ``Enter the Factor''.
    Below, another section labeled ``[Reflection 2]'' asks ``What do you think might be the AI Explainer's behavior?''. Two statements are provided with selection options. The first reads ``I think the AI Explainer might be [fill in the blank] compared to the AI System'', followed by three radio buttons labeled ``Lower'', ``Similar'', and ``Higher''. 
    The second reads ``I think the difference \% between the AI Explainer and the AI System might be:'', followed by three radio buttons labeled ``Small (<5\%)'', ``Medium (5–10\%)'', and ``Large (>10\%)''.
    At the bottom, a question asks ``What do you think is the AI System's prediction for PM10? If your answer is within 10\%, it will be accepted.'' 
    A horizontal slider labeled ``Prediction'' ranges from 0 to 1000.}

    \label{fig:task_transfer_survey_test_page2}
\end{figure*}

\begin{figure*}[t]
    \centering
    \includegraphics[width=0.9\linewidth]{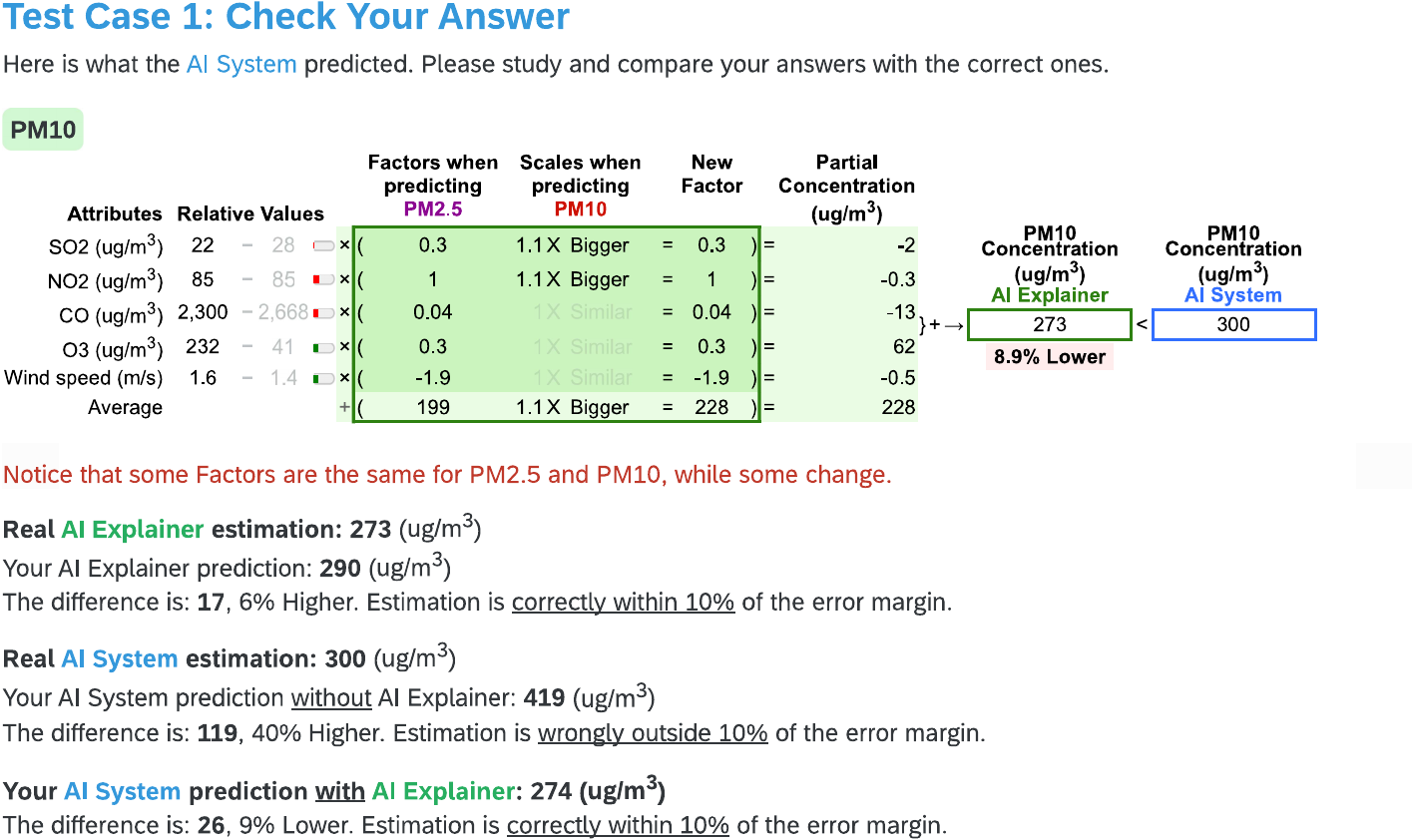}
    \caption{Sample of the forward simulation trial (learning page) in the Task transfer, where participants review their performance on this trial to strengthen their understanding. The displayed UI is the Transferable condition on the air pollution application.}
    \Description{The figure shows a tutorial page titled ``Test Case 1: Check Your Answer''. 
    A short paragraph below instructs the user to see what the AI System predicted and to compare their answers with the correct ones. 
    A green label reads ``PM10''.
    A table is displayed in the center, with columns labeled ``Attributes'', ``Relative Values'', ``Factors when predicting PM2.5'', ``Scales when predicting PM10'', ``New Factor'', and ``Partial Concentration (µg/m³)''. 
    The rows list five attributes: SO₂ (µg/m³), NO₂ (µg/m³), CO (µg/m³), O₃ (µg/m³), and Wind speed (m/s), followed by an ``Average'' row. 
    In the Factors column, the values are 0.3, 1, 0.04, 0.3, and –1.9, with 199 in the Average row. 
    In the Scales column, the entries are ``1.1× Bigger'', ``1.1× Bigger'', ``1× Similar'', ``1× Similar'', and ``1× Similar'', with ``1.1× Bigger'' in the Average row. 
    In the New Factor column, the corresponding values are 0.3, 1, 0.04, 0.3, and –1.9, followed by 228 in the Average row. 
    In the Partial Concentration column, the values are –2, –0.3, –13, 62, –0.5, and 228 for the average. 
    To the right, two blue boxes represent PM10 concentration predictions. 
    The left box, labeled ``AI Explainer'', shows the value 273, and the right box, labeled ``AI System'', shows 300. 
    Between them, a label reads ``8.9\% Lower'', with the symbol ``<'' between the two boxes.
    Below the table, a red sentence states, ``Notice that some Factors are the same for PM2.5 and PM10, while some change.'' 
    The next section contains three paragraphs with numerical results. The first begins with ``Real AI Explainer estimation: 273 (µg/m³)'' followed by lines showing ``Your AI Explainer prediction: 373 (µg/m³)'', ``The difference is: 100, 36\% Higher'', and ``Estimation is wrongly outside 10\% of the error margin''. 
    The second paragraph begins with ``Real AI System estimation: 300 (µg/m³)'' followed by ``Your AI System prediction without AI Explainer: 381 (µg/m³)'', ``The difference is: 81, 27\% Higher'', and ``Estimation is wrongly outside 10\% of the error margin''. 
    The third paragraph begins with ``Your AI System prediction with AI Explainer: 268 (µg/m³)'' followed by ``The difference is: 32, 11\% Lower'' and ``Estimation is wrongly outside 10\% of the error margin''.}
    \label{fig:task_transfer_survey_test_page3}
\end{figure*}

\begin{figure*}[t]
    \centering
    \includegraphics[width=0.9\linewidth]{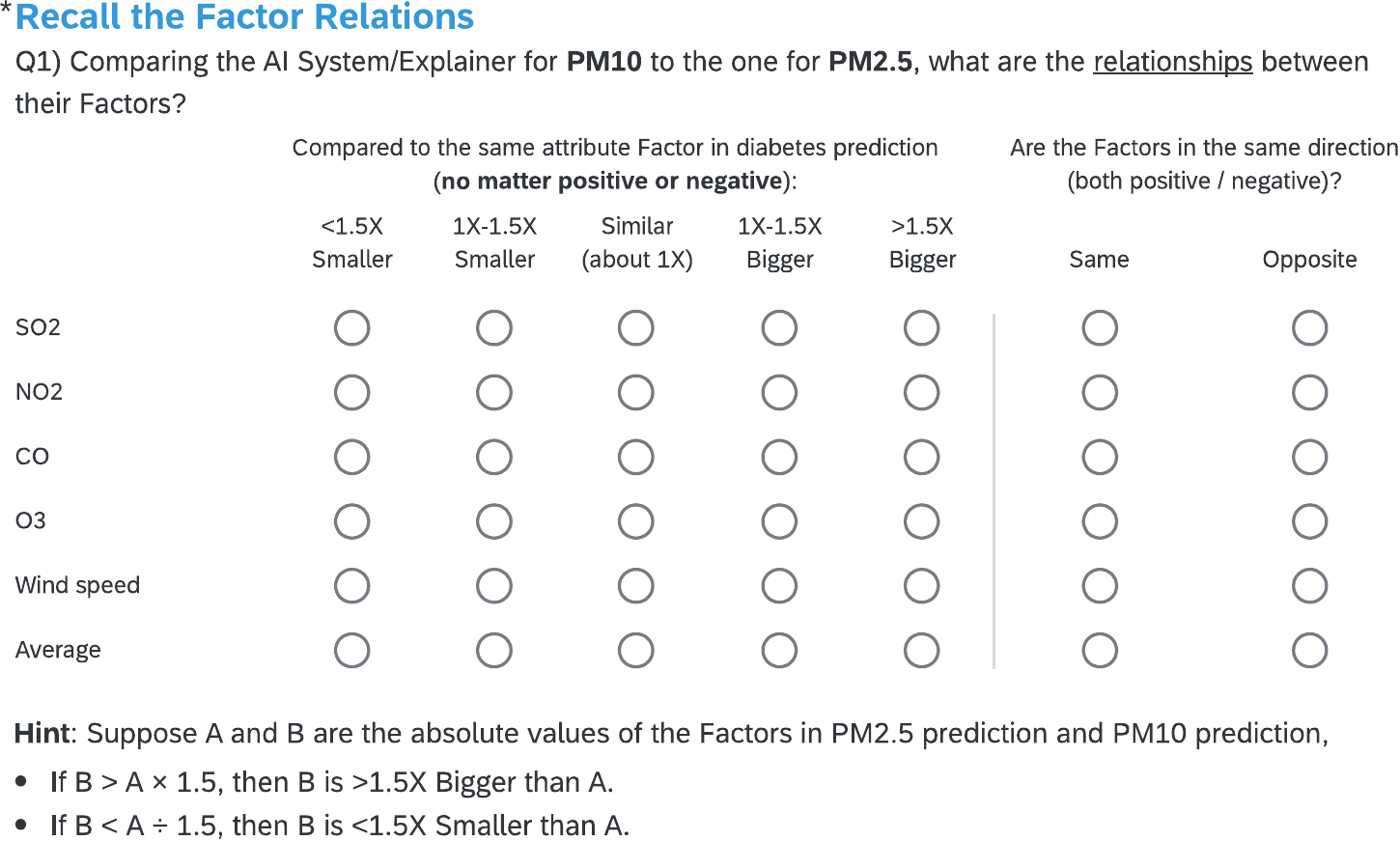}
    \caption{The question about Domain relationships in the Task transfer. The displayed UI is for the air pollution application.}
    \Description{The figure shows a section titled ``Recall the Factor Relations''. 
    A question labeled ``Q1)'' asks ``Comparing the AI System/Explainer for PM10 to the one for PM2.5, what are the relationships between their Factors?''. 
    Below, a table contains six rows labeled SO₂, NO₂, CO, O₃, Wind speed, and Average. 
    The table has two grouped parts of columns. 
    The first group of five columns is labeled ``Compared to the same attribute Factor in diabetes prediction (no matter positive or negative):'' with column headers ``<1.5× Smaller'', ``1×–1.5× Smaller'', ``Similar (about 1×)'', ``1×–1.5× Bigger'', and ``>1.5× Bigger''. 
    The second group of two columns is labeled ``Are the Factors in the same direction (both positive / negative)?'' with column headers ``Same'' and ``Opposite''. 
    Each cell in the table contains a circular radio button for selection.
    Below the table, a hint section provides two bullet points describing how to interpret the comparisons. 
    It begins with the sentence ``Hint: Suppose A and B are the absolute values of the Factors in PM2.5 prediction and PM10 prediction,'' followed by two bullet points: 
    ``If B > A × 1.5, then B is >1.5× Bigger than A.'' and 
    ``If B < A ÷ 1.5, then B is <1.5× Smaller than A.''} 
    \label{fig:task_transfer_survey_relation_question}
\end{figure*}

\begin{figure*}[t]
    \centering
    \includegraphics[width=0.8\linewidth]{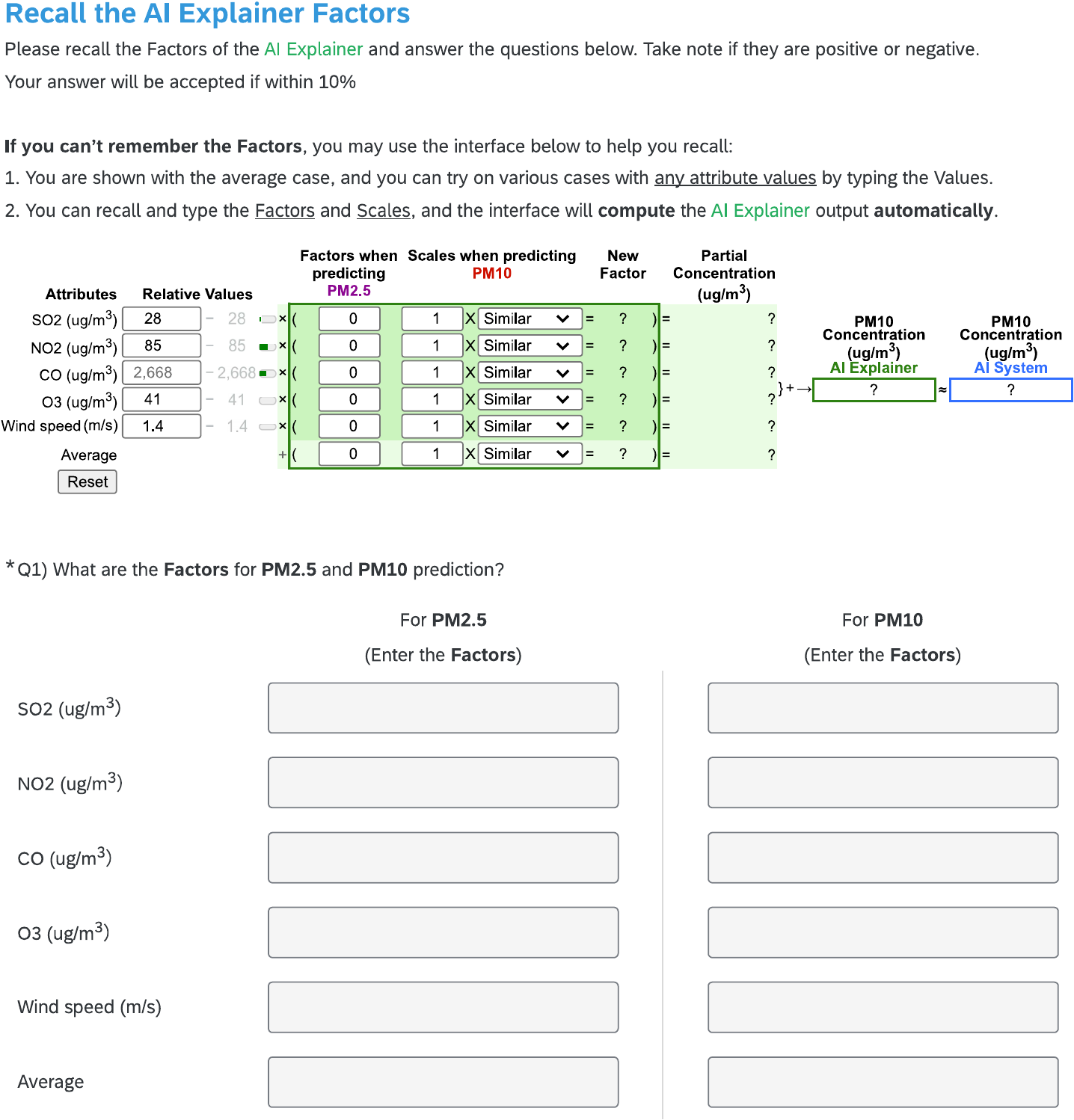}
    \caption{The question about factor recall in the Task transfer. The displayed UI is for the air pollution application.}
    \Description{The figure shows a section titled ``Recall the AI Explainer Factors''. 
    A paragraph instructs users to recall the Factors of the AI Explainer and answer the questions below, noting whether they are positive or negative. 
    It also mentions that the answer will be accepted if within 10\%. 
    A bolded paragraph beginning with ``If you can’t remember the Factors'' explains that users may use the interface below to help recall. 
    Two numbered points follow. 
    Point 1 states that users are shown with the average case and can try various cases with any attribute values by typing the Values. 
    Point 2 states that users can recall and type the Factors and Scales, and the interface will compute the AI Explainer output automatically.
    Below this text, a table is shown with six labeled columns: ``Attributes'', ``Relative Values'', ``Factors when predicting PM2.5'', ``Scales when predicting PM10'', ``New Factor'', and ``Partial Concentration (µg/m³)''. 
    The rows list SO₂ (µg/m³), NO₂ (µg/m³), CO (µg/m³), O₃ (µg/m³), and Wind speed (m/s), followed by an ``Average'' row. 
    Each row has numeric values and dropdown boxes with ``1× Similar'' selected, while the New Factor and Partial Concentration columns contain question marks (``?''). 
    To the right of the table, two blue boxes labeled ``AI Explainer'' and ``AI System'' show the PM10 Concentration (µg/m³), both containing question marks. 
    Two buttons labeled ``Average'' and ``Reset'' appear below the table.
    The section below, labeled ``*Q1) What are the Factors for PM2.5 and PM10 prediction?'', contains two sub-columns. 
    The left side is labeled ``For PM2.5 (Enter the Factors)'' and the right side is labeled ``For PM10 (Enter the Factors)''. 
    Each side lists SO₂ (µg/m³), NO₂ (µg/m³), CO (µg/m³), O₃ (µg/m³), Wind speed (m/s), and Average, each followed by an empty rectangular input box for user entry.}

    \label{fig:task_transfer_survey_factor_recall}
\end{figure*}

\begin{figure*}[t]
    \centering
    \includegraphics[width=0.61\linewidth]{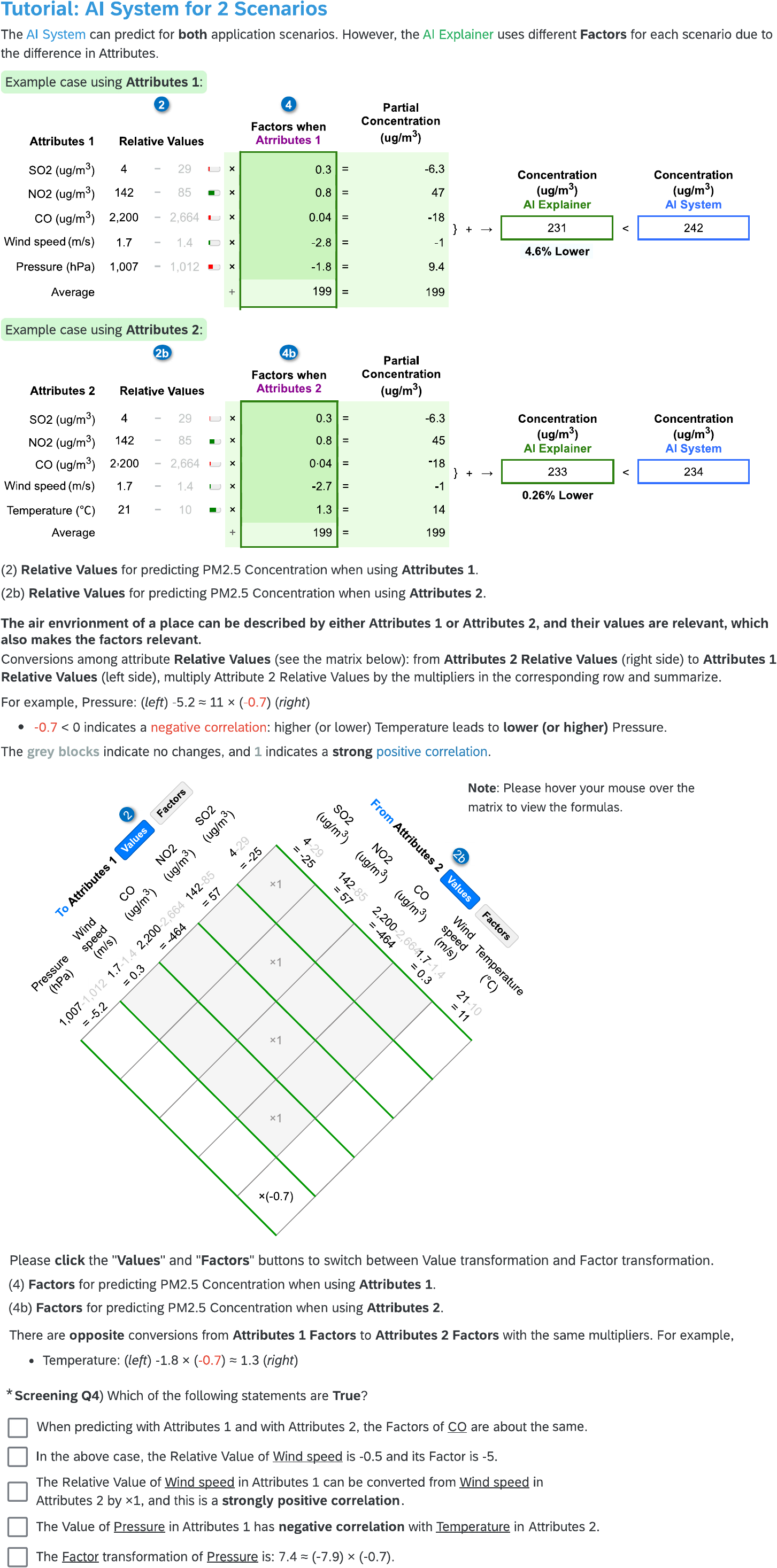}
    \caption{Attributes transfer tutorial about AI Explainers in two scenarios. The displayed UI is the Transferable condition on the air pollution application.}
    \Description{The figure shows a tutorial page titled ``Tutorial: AI System for 2 Scenarios''. 
    A paragraph explains that the AI System can predict for both application scenarios, while the AI Explainer uses different Factors for each scenario due to the difference in Attributes. 
    Two example cases are shown. 
    The first section, labeled ``Example case using Attributes 1'', displays a table with columns labeled ``Attributes 1'', ``Relative Values'', ``Factors when Attributes 1'', and ``Partial Concentration (µg/m³)''. 
    The rows list SO₂ (µg/m³), NO₂ (µg/m³), CO (µg/m³), Wind speed (m/s), and Pressure (hPa), followed by an ``Average'' row. 
    The Factors column contains 0.3, 0.8, 0.04, –2.8, and –1.8, with 199 in the Average row. 
    The Partial Concentration column shows –6.3, 47, –18, –1, and 9.4, followed by 199 in the Average row. 
    On the right, two blue boxes show Concentration results: one labeled ``AI Explainer'' with 231 and one labeled ``AI System'' with 242. 
    A green label between them reads ``4.6\% Lower'', with the symbol ``<'' between the two boxes. 
    The second section, labeled ``Example case using Attributes 2'', displays a similar table with columns labeled ``Attributes 2'', ``Relative Values'', ``Factors when Attributes 2'', and ``Partial Concentration (µg/m³)''. 
    The rows list SO₂ (µg/m³), NO₂ (µg/m³), CO (µg/m³), Wind speed (m/s), and Temperature (°C), followed by an ``Average'' row. 
    The Factors column contains 0.3, 0.8, 0.04, –2.7, and 1.3, with 199 in the Average row. 
    The Partial Concentration column shows –6.3, 45, –18, –1, and 14, followed by 199 in the Average row. 
    To the right, two blue boxes show Concentration results: one labeled ``AI Explainer'' with 233 and one labeled ``AI System'' with 234. 
    A green label between them reads ``0.26\% Lower'', with the symbol ``<'' between the two boxes. 
    Below the examples, a section explains Relative Values for predicting PM2.5 Concentration when using Attributes 1 and Attributes 2. 
    It states that the air environment of a place can be described by either Attributes 1 or Attributes 2, and that conversions among attribute Relative Values can be represented by a matrix. 
    A formula example is given: ``Pressure: (left) –5.2 = 11 × (–0.7) (right)''. 
    A bullet point explains that –0.7 < 0 indicates a negative correlation and that higher Temperature leads to lower Pressure. 
    Grey blocks in the matrix indicate no change, and 1 indicates a strong positive correlation. 
    A note below states, ``Please hover your mouse over the matrix to view the formulas.'' 
    A diamond-shaped matrix diagram is shown below. 
    The left side is labeled ``To Attributes 1'', and the right side is labeled ``From Attributes 2''. 
    The diagonal cells contain grey blocks with ``×1'', and one cell near the bottom shows ``×(–0.7)''. 
    The labels around the matrix include Pressure (hPa), Wind speed (m/s), CO (µg/m³), NO₂ (µg/m³), SO₂ (µg/m³), and Temperature (°C). 
    Below the matrix, text instructs users to click the ``Values'' and ``Factors'' buttons to switch between Value transformation and Factor transformation. 
    Section (4) is titled ``Factors for predicting PM2.5 Concentration when using Attributes 1'', and section (4b) is titled ``Factors for predicting PM2.5 Concentration when using Attributes 2''. 
    A paragraph explains that there are opposite conversions from Attributes 1 Factors to Attributes 2 Factors with the same multipliers. 
    An example shows ``Temperature: (left) –1.8 × (–0.7) = 1.3 (right)''. 
    A red note emphasizes paying attention to the Factors and Correlations in the matrix but not taking notes of them. 
    At the bottom, a question labeled ``*Screening Q4) Which of the following statements are True?'' appears with five checkbox options: 
    (1) When predicting with Attributes 1 and with Attributes 2, the Factors of CO are about the same. 
    (2) In the above case, the Relative Value of Wind speed is –0.5 and its Factor is –5. 
    (3) The Relative Value of Wind speed in Attributes 1 can be converted from Wind speed in Attributes 2 by ×1, indicating a strongly positive correlation. 
    (4) The Value of Pressure in Attributes 1 has negative correlation with Temperature in Attributes 2. 
    (5) The Factor transformation of Pressure is: 7.4 = (–7.9) × (–0.7).}

    \label{fig:attributes_transfer_survey_tutorial3_transferable}
\end{figure*}

\begin{figure*}[t]
    \centering
    \includegraphics[width=0.68\linewidth]{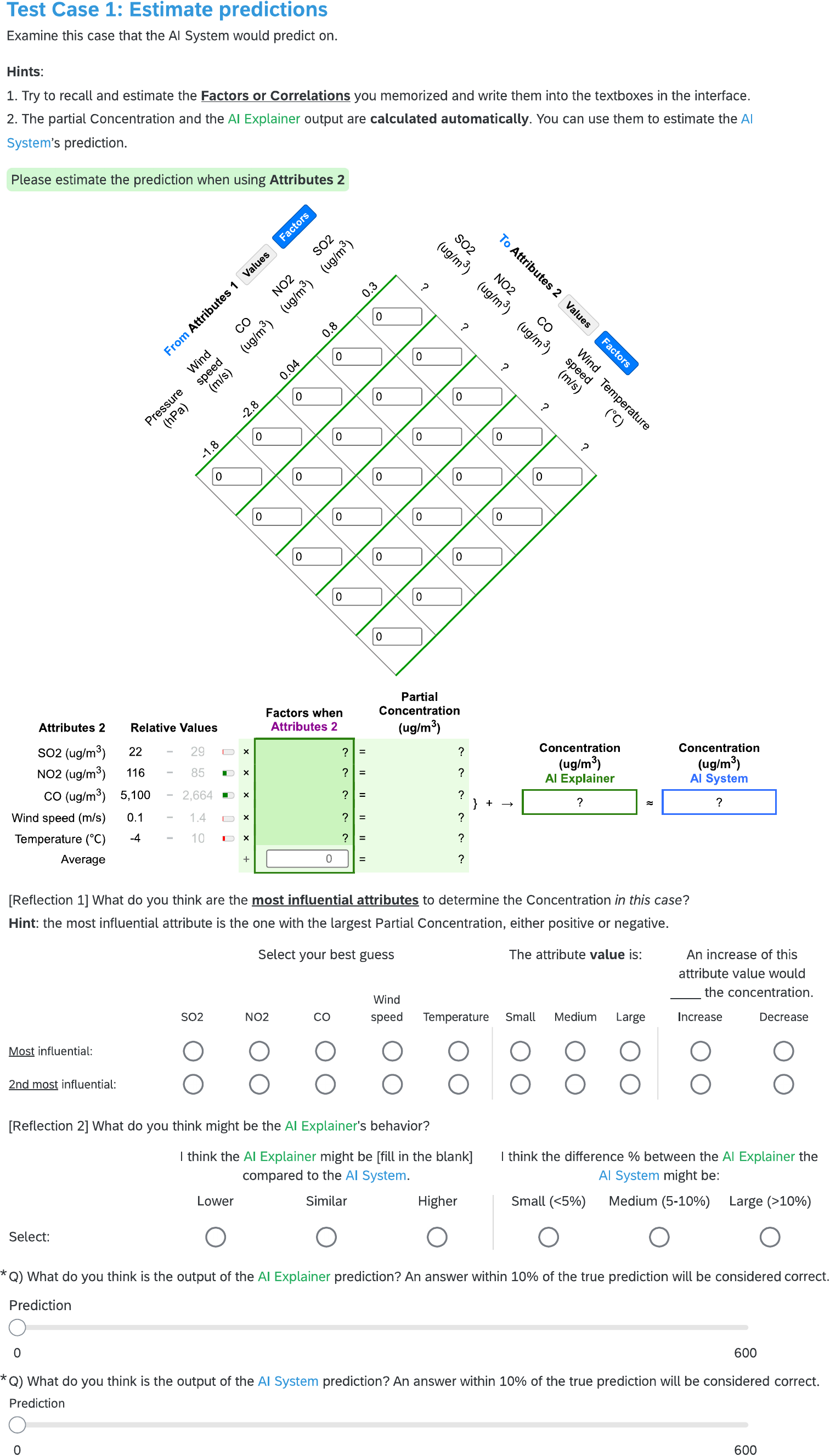}
    \caption{Sample of the forward simulation trial (test page) in the Attributes transfer, where participants are asked to estimate the explanation and AI system outputs without seeing the factors. The displayed UI is the Transferable condition on the air pollution application.}
    \Description{The figure shows a tutorial page titled ``Test Case 1: Estimate predictions''. 
    A paragraph instructs the user to examine a case that the AI System would predict on. 
    Below it, a section labeled ``Hints:'' contains two numbered points. 
    Point 1 asks users to recall and estimate the Factors or Correlations they memorized and write them into the textboxes in the interface. 
    Point 2 states that the partial Concentration and the AI Explainer output are calculated automatically and can be used to estimate the AI System’s prediction. 
    A green box below reads ``Please estimate the prediction when using Attributes 2''.
    A diamond-shaped matrix diagram is shown, labeled on the left as ``From Attributes 1'' and on the right as ``To Attributes 2''. 
    The top of the matrix shows headers ``Values'' and ``Factors''. 
    The matrix includes diagonal and triangular cells containing numeric values such as 0.3, 0.8, 0.04, –2.8, and –1.8, as well as zeros and question marks (``?''). 
    The attributes listed around the matrix include Pressure (hPa), Wind speed (m/s), CO (µg/m³), NO₂ (µg/m³), SO₂ (µg/m³), and Temperature (°C).
    Below the matrix, a green-shaded table lists five rows of attributes under the heading ``Attributes 2'': SO₂ (µg/m³), NO₂ (µg/m³), CO (µg/m³), Wind speed (m/s), and Temperature (°C), followed by an ``Average'' row. 
    The columns are labeled ``Relative Values'', ``Factors when Attributes 2'', and ``Partial Concentration (µg/m³)''. 
    Each row contains numeric values and question marks. 
    On the right, two blue boxes display the predicted Concentration (µg/m³). 
    The left box, labeled ``AI Explainer'', shows a blank space with a question mark, and the right box, labeled ``AI System'', also contains a question mark. 
    A green button labeled ``Reset'' appears below the table.
    The next section, labeled ``[Reflection 1]'', asks ``What do you think are the most influential attributes to determine the Concentration in this case?''. 
    A hint explains that the most influential attribute is the one with the largest Partial Concentration, either positive or negative. 
    Below, two rows of radio button options allow users to select the ``Most influential'' and ``2nd most influential'' attributes from SO₂, NO₂, CO, Wind speed, and Temperature. 
    Additional columns to the right are labeled ``The attribute value is:'' and ``An increase of this attribute value would ____ the concentration'', with radio button options ``Small'', ``Medium'', ``Large'', ``Increase'', and ``Decrease''.
    Another section, labeled ``[Reflection 2]'', asks ``What do you think might be the AI Explainer’s behavior?''. 
    Two sentences allow the user to select options: 
    ``I think the AI Explainer might be [fill in the blank] compared to the AI System'' with radio buttons labeled ``Lower'', ``Similar'', and ``Higher'', 
    and ``I think the difference \% between the AI Explainer and the AI System might be:'' with options labeled ``Small (<5\%)'', ``Medium (5–10\%)'', and ``Large (>10\%)''.
    At the bottom, two starred questions are presented. 
    The first asks, ``What do you think is the output of the AI Explainer prediction? An answer within 10\% of the true prediction will be considered correct.'' 
    Below it, a horizontal slider labeled ``Prediction'' ranges from 0 to 600. 
    The second question asks, ``What do you think is the output of the AI System prediction? An answer within 10\% of the true prediction will be considered correct.'' 
    It also includes a horizontal slider labeled ``Prediction'' ranging from 0 to 600.}
    \label{fig:attributes_transfer_survey_test_page1}
\end{figure*}

\begin{figure*}[t]
    \centering
    \includegraphics[width=0.75\linewidth]{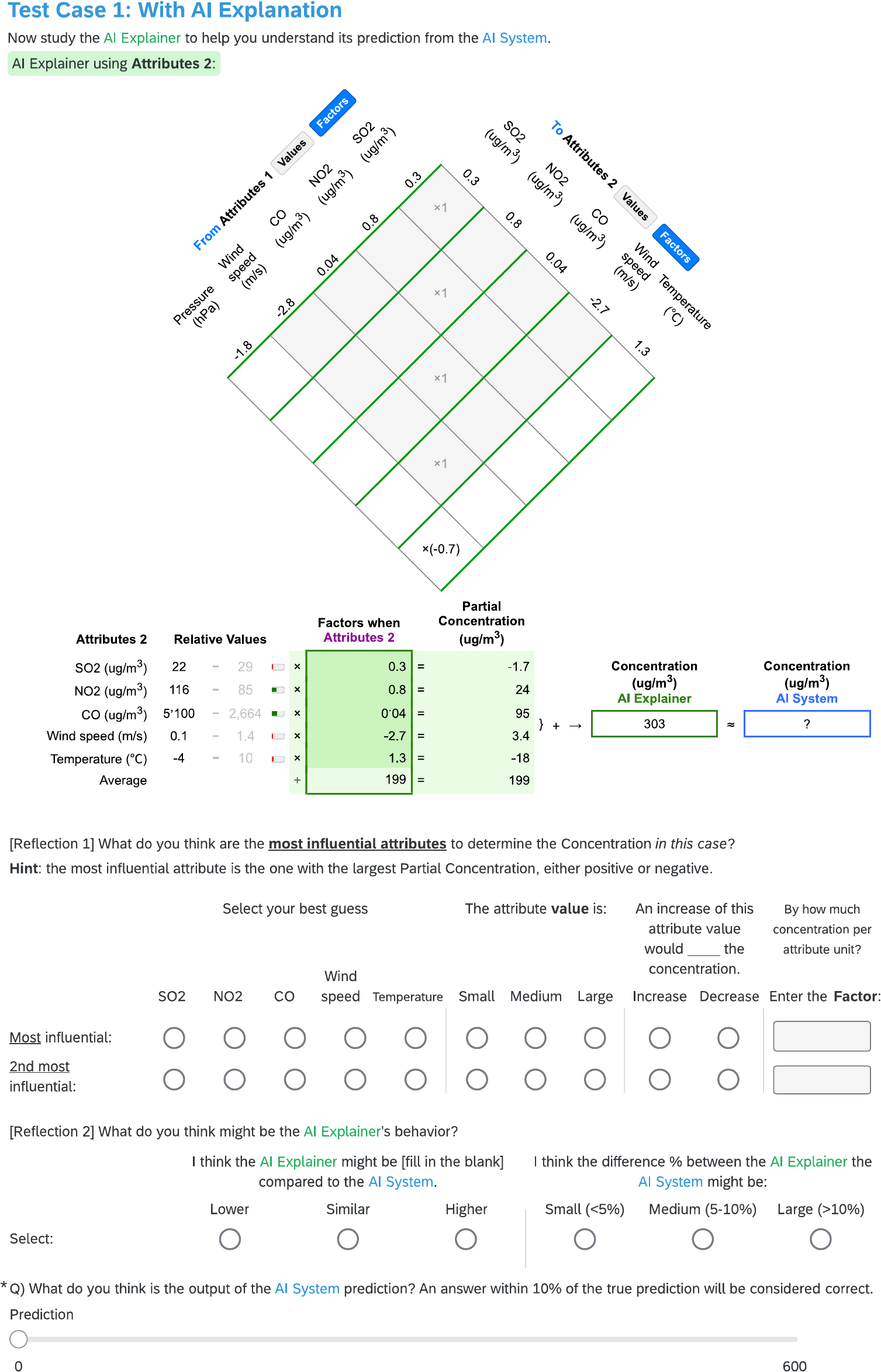}
    \caption{Sample of the forward simulation trial (explanation page) in the Attributes transfer, where participants are asked to estimate the AI system outputs with the factors shown. The displayed UI is the Transferable condition on the air pollution application.}
    \Description{The figure shows a tutorial page titled ``Test Case 1: With AI Explanation''. 
    A paragraph explains that users should now study the AI Explainer to help understand its prediction from the AI System. 
    A green label below reads ``AI Explainer using Attributes 2''.
    At the top, a diamond-shaped matrix diagram is displayed. 
    The left side is labeled ``To Attributes 1'', and the right side is labeled ``From Attributes 2''. 
    The top of the matrix contains headers ``Values'' and ``Factors''. 
    The matrix includes diagonal and triangular cells containing values such as 0.3, 0.8, 0.04, –2.8, and –1.8, as well as several grey blocks with ``×1''. 
    A cell at the bottom right shows ``×(–0.7)''. 
    The attributes listed around the matrix include Pressure (hPa), Wind speed (m/s), CO (µg/m³), NO₂ (µg/m³), SO₂ (µg/m³), and Temperature (°C).
    Below the matrix, a green table lists five rows of attributes under the heading ``Attributes 2'': SO₂ (µg/m³), NO₂ (µg/m³), CO (µg/m³), Wind speed (m/s), and Temperature (°C), followed by an ``Average'' row. 
    The columns are labeled ``Relative Values'', ``Factors when Attributes 2'', and ``Partial Concentration (µg/m³)''. 
    Each row shows numeric values for Relative Values (22, 116, 5,100, 0.1, –4), Factors (0.3, 0.8, 0.04, –2.7, 1.3), and Partial Concentrations (–1.7, 24, 95, 3.4, –18), followed by 199 in the Average row. 
    On the right, two blue boxes display Concentration (µg/m³). 
    The left box, labeled ``AI Explainer'', shows the value 303, and the right box, labeled ``AI System'', contains a question mark (``?'').
    The next section, labeled ``[Reflection 1]'', asks ``What do you think are the most influential attributes to determine the Concentration in this case?''. 
    A hint below states that the most influential attribute is the one with the largest Partial Concentration, either positive or negative. 
    Below the hint, there are two rows of radio button selections labeled ``Most influential'' and ``2nd most influential''. 
    The selectable attributes are SO₂, NO₂, CO, Wind speed, and Temperature. 
    Additional columns to the right are labeled ``The attribute value is:'' and ``An increase of this attribute value would ____ the concentration'', with options ``Small'', ``Medium'', ``Large'', ``Increase'', and ``Decrease''. 
    A final column on the right labeled ``By how much concentration per attribute unit?'' contains two empty textboxes titled ``Enter the Factor''.
    Another section, labeled ``[Reflection 2]'', asks ``What do you think might be the AI Explainer’s behavior?''. 
    It contains two statements for selection. 
    The first reads ``I think the AI Explainer might be [fill in the blank] compared to the AI System'', followed by radio buttons labeled ``Lower'', ``Similar'', and ``Higher''. 
    The second reads ``I think the difference \% between the AI Explainer and the AI System might be:'' followed by options ``Small (<5\%)'', ``Medium (5–10\%)'', and ``Large (>10\%)''. 
    A ``Select:'' label appears above the radio buttons.
    At the bottom, a starred question asks ``What do you think is the output of the AI System prediction? An answer within 10\% of the true prediction will be considered correct.'' 
    Below it, a horizontal slider labeled ``Prediction'' is displayed, ranging from 0 to 600.}
    \label{fig:attributes_transfer_survey_test_page2}
\end{figure*}

\begin{figure*}[t]
    \centering
    \includegraphics[width=0.8\linewidth]{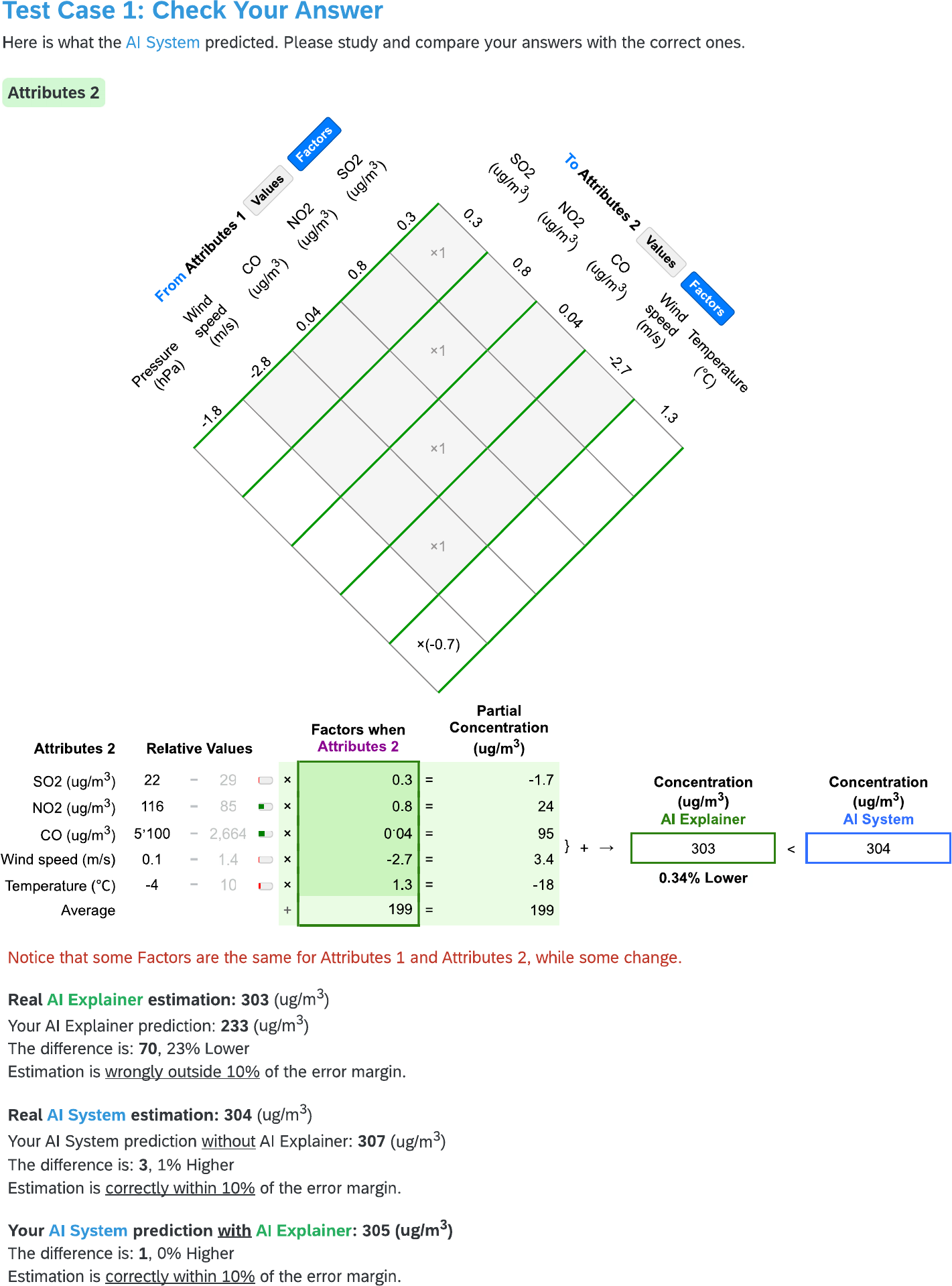}
    \caption{Sample of the forward simulation trial (learning page) in the Attributes transfer, where participants review their performance on this trial to strengthen their understanding. The displayed UI is the Transferable condition on the air pollution application.}
    \Description{The figure shows a results page titled ``Test Case 1: Check Your Answer''. 
    A sentence below states that this page shows what the AI System predicted and asks users to study and compare their answers with the correct ones. 
    A green label near the top reads ``Attributes 2''. 
    At the top center, a diamond-shaped matrix diagram is displayed. 
    The left side of the matrix is labeled ``To Attributes 1'', and the right side is labeled ``From Attributes 2''. 
    The matrix headers include ``Values'' and ``Factors''. 
    Numeric values appear along the top and left edges, including SO₂ (µg/m³), NO₂ (µg/m³), CO (µg/m³), Wind speed (m/s), Pressure (hPa), and Temperature (°C). 
    Inside the matrix, most diagonal cells contain grey blocks labeled ``×1'', and one cell near the bottom shows ``×(–0.7)''. 
    Below the matrix, a green-shaded table labeled ``Attributes 2'' is shown. 
    The table columns are labeled ``Attributes 2'', ``Relative Values'', ``Factors when Attributes 2'', and ``Partial Concentration (µg/m³)''. 
    The rows list SO₂ (µg/m³), NO₂ (µg/m³), CO (µg/m³), Wind speed (m/s), and Temperature (°C), followed by an ``Average'' row. 
    The Factors column shows values 0.3, 0.8, 0.04, –2.7, and 1.3, with 199 in the Average row. 
    The Partial Concentration column shows –1.7, 24, 95, 3.4, –18, and 199. 
    To the right of the table, two blue boxes display Concentration (µg/m³). 
    The left box, labeled ``AI Explainer'', shows 303, and the right box, labeled ``AI System'', shows 304. 
    A label between them reads ``0.34\% Lower'', with the symbol ``<'' between the two boxes. 
    Below the table, a red sentence states that some Factors are the same for Attributes 1 and Attributes 2, while some change. 
    Three result summaries are shown below. 
    The first begins with ``Real AI Explainer estimation: 303 (µg/m³)'', followed by ``Your AI Explainer prediction: 248 (µg/m³)'', ``The difference is: 55, 18\% Lower'', and ``Estimation is wrongly outside 10\% of the error margin''. 
    The second begins with ``Real AI System estimation: 304 (µg/m³)'', followed by ``Your AI System prediction without AI Explainer: 355 (µg/m³)'', ``The difference is: 51, 17\% Higher'', and ``Estimation is wrongly outside 10\% of the error margin''. 
    The third begins with ``Your AI System prediction with AI Explainer: 307 (µg/m³)'', followed by ``The difference is: 3, 1\% Higher'', and ``Estimation is correctly within 10\% of the error margin''. }
    \label{fig:attributes_transfer_survey_test_page3}
\end{figure*}

\begin{figure*}[t]
    \centering
    \includegraphics[width=0.9\linewidth]{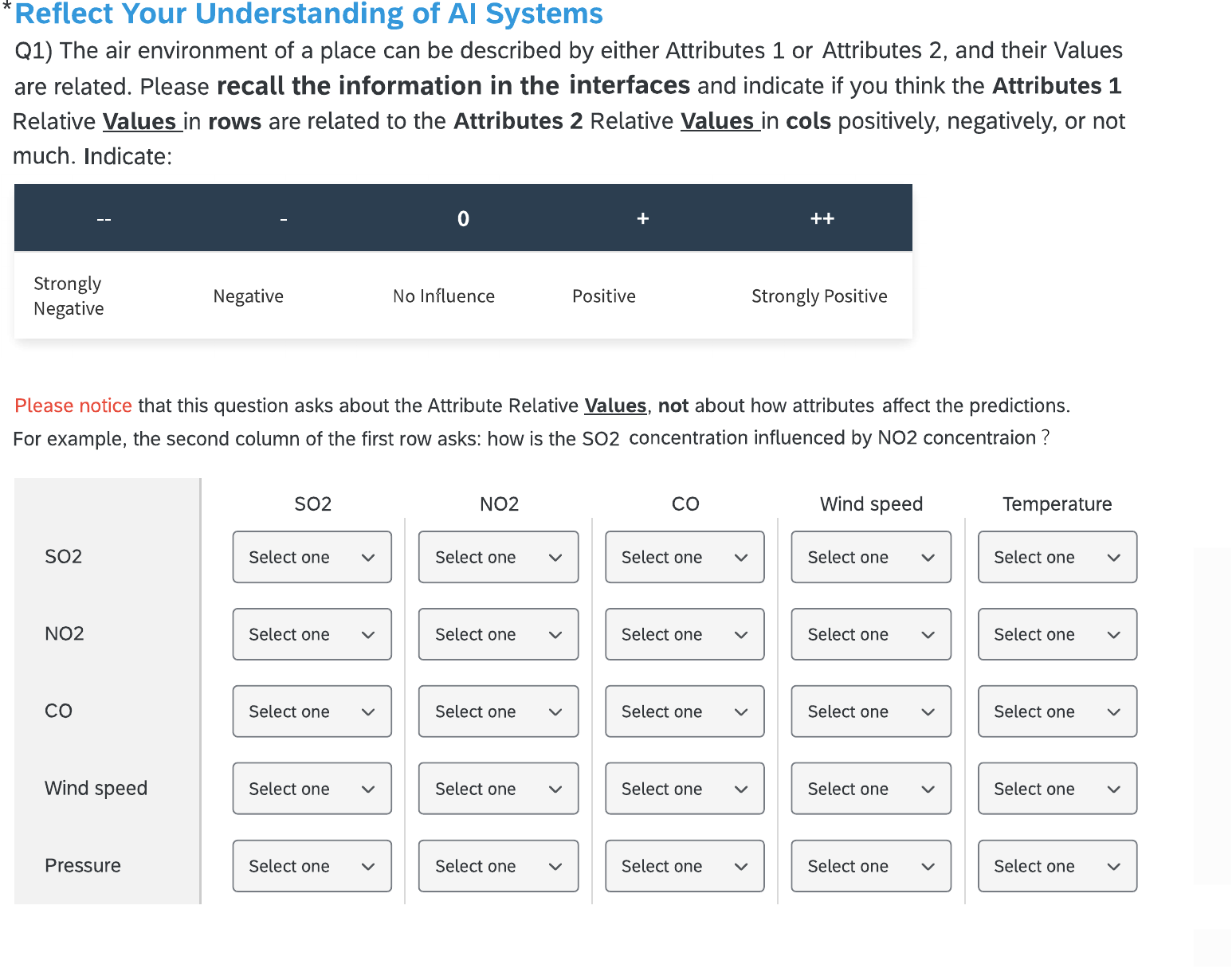}
    \caption{The question about Domain relationships in the Attributes transfer. The displayed UI is for the air pollution application.}
    \Description{The figure shows a survey interface titled ``Reflect Your Understanding of AI Systems''. 
    The question, labeled ``Q1)'', asks participants to consider that the air environment of a place can be described by either Attributes 1 or Attributes 2, and their Values are related. 
    The instructions ask users to recall the information in the interfaces and indicate whether they think the Attributes 1 Relative Values in rows are related to the Attributes 2 Relative Values in columns positively, negatively, or not much. 
    The instruction ends with the word ``Indicate:''.
    A horizontal rating scale is displayed across the top, labeled with five categories from left to right: ``--'', ``-'', ``0'', ``+'', and ``++''. 
    Below each label, descriptive terms are shown: ``Strongly Negative'', ``Negative'', ``No Influence'', ``Positive'', and ``Strongly Positive''. 
    Beneath the scale, a paragraph in red text reads: 
    ``Please notice that this question asks about the Attribute Relative Values, not about how attributes affect the predictions. For example, the second column of the first row asks: how is the SO₂ concentration influenced by NO₂ concentration.'' 
    Below this explanation, a table is presented. 
    The first column lists the row attributes: SO₂, NO₂, CO, Wind speed, and Pressure. 
    The column headers, aligned horizontally across the top, are labeled SO₂, NO₂, CO, Wind speed, and Temperature. 
    Each cell in the table contains a dropdown menu labeled ``Select one''. 
    The layout forms a 5×5 grid of selectable dropdowns, allowing users to specify how each pair of attributes is related.}

    \label{fig:attributes_transfer_survey_relation_question}
\end{figure*}

Fig.~\ref{fig:task_transfer_survey_scenarios}--\ref{fig:task_transfer_survey_factor_recall} depict the Qualtrics workflow with each of the four XAI conditions for the Task transfer user study, including introduction of the survey (Fig.~\ref{fig:task_transfer_survey_scenarios}), tutorials (Figs.~\ref{fig:task_transfer_survey_tutorial1_AI}--~\ref{fig:task_transfer_survey_tutorial3_Transferable}), forward simulation session (Figs.~\ref{fig:task_transfer_survey_test_page1}--\ref{fig:task_transfer_survey_test_page3}), factor relation recall (Fig.~\ref{fig:task_transfer_survey_relation_question}) and factor recall (Fig.~\ref{fig:task_transfer_survey_factor_recall}). 
The Attributes Transfer survey follows the same structure, with the main difference in Transferable XAI and the question on attribute relationships, which are shown in Fig.~\ref{fig:attributes_transfer_survey_tutorial3_transferable}–\ref{fig:attributes_transfer_survey_relation_question} for brevity.

\begin{figure*}[t]
    \centering
    \includegraphics[width=0.9\linewidth]{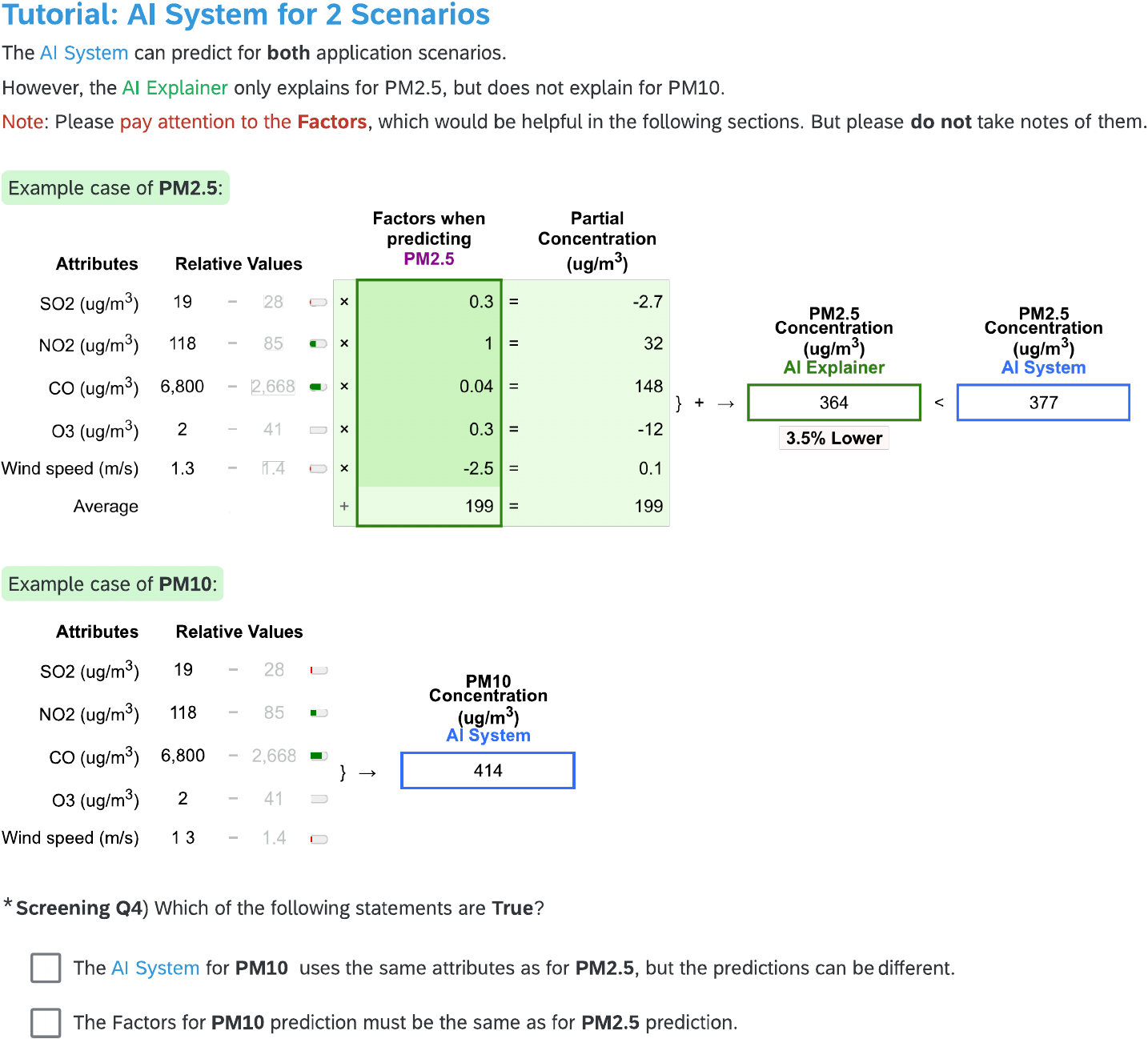}
    \caption{Task transfer tutorial about AI Explainers in two scenarios. The displayed UI is the None condition on the air pollution application.}
    \Description{The figure shows a tutorial page titled ``Tutorial: AI System for 2 Scenarios''. 
    A paragraph explains that the AI System can predict for both application scenarios. 
    It also states that the AI Explainer only explains for PM2.5 but not for PM10. 
    A red note says ``Please pay attention to the Factors, which would be helpful in the following sections. But please do not take notes of them''. 
    Below, two example cases are shown.
    The first section is labeled ``Example case of PM2.5''. 
    A table lists five attributes: SO₂ (µg/m³), NO₂ (µg/m³), CO (µg/m³), O₃ (µg/m³), and Wind speed (m/s), with relative values next to them. 
    To the right, a green-shaded column labeled ``Factors when predicting PM2.5'' includes the values 0.3, 1, 0.04, 0.3, and –2.5, followed by an ``Average'' row with 199. 
    Next to it, a column labeled ``Partial Concentration (µg/m³)'' shows –2.7, 32, 148, –12, 0.1, and 199 for the average. 
    On the far right, two boxes show PM2.5 concentration results: one labeled ``AI Explainer'' with 364 and one labeled ``AI System'' with 377. 
    A label between them reads ``3.5\% Lower'', with the symbol ``<'' pointing from 364 to 377.
    The second section is labeled ``Example case of PM10''. 
    A table lists the same five attributes and relative values as above. 
    To the right, a blue box labeled ``PM10 Concentration (µg/m³) AI System'' shows the value 414.
    At the bottom, a question labeled ``Screening Q4'' asks ``Which of the following statements are True?''. 
    Two checkbox options are shown. 
    The first option states that the AI System for PM10 uses the same attributes as for PM2.5 but the predictions can be different. 
    The second option states that the Factors for PM10 prediction must be the same as for PM2.5 prediction.}
    \label{fig:task_transfer_survey_tutorial3_None}
\end{figure*}

\begin{figure*}[t]
    \centering
    \includegraphics[width=0.9\linewidth]{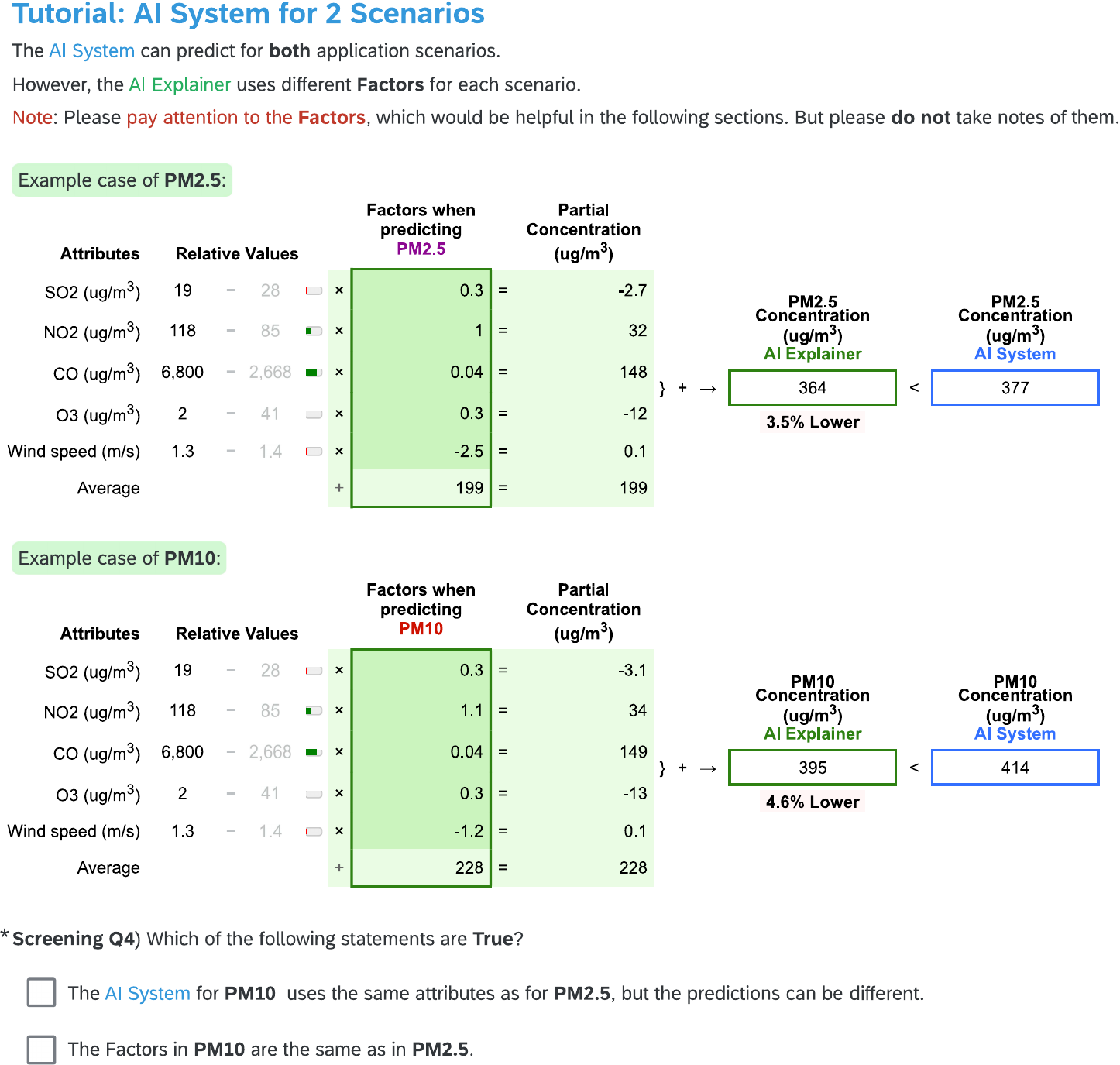}
    \caption{Task transfer tutorial about AI Explainers in two scenarios. The displayed UI is about the air pollution application. The Original condition and Target condition share this same page in the tutorial, but the Target condition only presents the second table for Target domain cases.}
    \Description{The figure shows a tutorial page titled ``Tutorial: AI System for 2 Scenarios''. A paragraph explains that the AI System can predict for both application scenarios. It also states that the AI Explainer uses different Factors for each scenario. A red note says ``Please pay attention to the Factors, which would be helpful in the following sections. But please do not take notes of them''. 
        Below, two example cases are shown.
    The first section is labeled ``Example case of PM2.5''. A table lists five attributes: SO₂ (µg/m³), NO₂ (µg/m³), CO (µg/m³), O₃ (µg/m³), and Wind speed (m/s), with their relative values displayed next to them. 
    To the right, a green-shaded column labeled ``Factors when predicting PM2.5'' includes the values 0.3, 1, 0.04, 0.3, and –2.5, followed by an ``Average'' row with 199. 
    Next to it, another column labeled ``Partial Concentration (µg/m³)'' shows –2.7, 32, 148, –12, 0.1, and 199 for the average. 
    On the right side, two blue boxes show PM2.5 concentration results: one labeled ``AI Explainer'' with 364 and one labeled ``AI System'' with 377. 
    A label between them reads ``3.5\% Lower'', with the symbol ``<'' between the two boxes.
    The second section is labeled ``Example case of PM10''. A similar table lists the same five attributes and relative values. 
    To the right, a green-shaded column labeled ``Factors when predicting PM10'' includes the values 0.3, 1.1, 0.04, 0.3, and –1.2, followed by an ``Average'' row with 228. 
    Next to it, another column labeled ``Partial Concentration (µg/m³)'' shows –3.1, 34, 149, –13, 0.1, and 228 for the average. 
    On the right side, two blue boxes show PM10 concentration results: one labeled ``AI Explainer'' with 395 and one labeled ``AI System'' with 414. 
    A label between them reads ``4.6\% Lower'', with the symbol ``<'' between the two boxes.
    At the bottom, a question labeled ``Screening Q4'' asks ``Which of the following statements are True?''. Two checkbox options are shown. The first states that the AI System for PM10 uses the same attributes as for PM2.5 but the predictions can be different. 
    The second states that the Factors in PM10 are the same as in PM2.5.}
    \label{fig:task_transfer_survey_tutorial3_None}
\end{figure*}

\end{document}